\documentclass[12pt]{article}
\usepackage{comment}
\usepackage{amsmath}
\usepackage{amssymb}
\usepackage{graphicx}
\usepackage{here}
\usepackage{subcaption}
\usepackage{url}
\usepackage[sort&compress, numbers, merge]{natbib}

\usepackage{physics}
\usepackage[compat=1.1.0]{tikz-feynhand}
\usepackage{slashed}
\usepackage{mathtools}

\usepackage{todonotes}

\numberwithin{equation}{section}

\usepackage[colorlinks=true, linkcolor=blue, citecolor=blue,
urlcolor=black]{hyperref}

\begin{document}
\def\ps{\mathbf{p}}
\def\PS{\mathbf{P}}
\baselineskip 0.6cm
\def\simgt{\mathrel{\lower2.5pt\vbox{\lineskip=0pt\baselineskip=0pt
           \hbox{$>$}\hbox{$\sim$}}}}
\def\simlt{\mathrel{\lower2.5pt\vbox{\lineskip=0pt\baselineskip=0pt
           \hbox{$<$}\hbox{$\sim$}}}}
\def\simprop{\mathrel{\lower3.0pt\vbox{\lineskip=1.0pt\baselineskip=0pt
             \hbox{$\propto$}\hbox{$\sim$}}}}
\def\tr{\mathop{\rm tr}}
\def\SU{\mathop{\rm SU}}

\begin{titlepage}

\begin{flushright}
IPMU21-0087
\end{flushright}

\vskip 1.1cm

\begin{center}

{\Large \bf 
Cosmological Constraints on Dark Scalar
}

\vskip 1.2cm
Masahiro Ibe$^{a,b}$, 
Shin Kobayashi$^{a}$, 
Yuhei Nakayama$^{a}$ and
Satoshi Shirai$^{b}$
\vskip 0.5cm

{\it

$^a$ {ICRR, The University of Tokyo, Kashiwa, Chiba 277-8582, Japan}

$^b$ {Kavli Institute for the Physics and Mathematics of the Universe
 (WPI), \\The University of Tokyo Institutes for Advanced Study, \\ The
 University of Tokyo, Kashiwa 277-8583, Japan}
}

\vskip 1.0cm

\abstract{
We discuss cosmological constraints on a dark scalar particle mixing with the Standard Model Higgs boson.
We pay particular attention to the dark scalar production process when the reheating temperature of the Universe is very low, which allows us to give a conservative limit on the low-mass scalar particle.
We also study the effect of the self-interaction of the dark scalars and find this has a significant impact on the cosmological constraints.
We obtain the most conservative cosmological constraint on the dark scalar, which is complementary to accelerator experiments and astrophysical observations. 
}

\end{center}
\end{titlepage}

\section{Introduction}

It is clear that the Standard Model (SM) is not the ultimate theory for describing nature.
The SM does not have a candidate for dark matter. 
It cannot explain the origin of cosmic inflation.
The neutrino mass also requires the extension of the SM.
The origin of the baryon in the Universe also needs the extension of the SM.
These observations require extensions of the SM.

The minimal extension of the Standard Model is to introduce a neutral scalar.
In fact, the neutral scalar appears in many extensions of the SM to solve the problems described above.
For example, the neutral scalar appears in the extended Higgs sector.
The neutral scalar also appears as the dilaton field in string theory.
The neutral scalar may also play the role of a mediator between the SM sector and the dark sector containing dark matter~\cite{Patt:2006fw,Pospelov:2007mp,March-Russell:2008lng} (for recent works, see Ref.\,\cite{Matsumoto:2018acr} and references therein).
The mediator prevents the excessive energy density in the dark sector which could result in the dark radiation
or in overclosure of the Universe by opening decay/annihilation channels into the SM sector.
The neutral scalar in the MeV range is also motivated to provide self-interaction between the dark matter which can resolve the small scale problems of the collisionless cold dark matter (see e.g., Ref.\,\cite{Tulin:2017ara}).
The neutral scalar can also be a candidate for an inflaton (see e.g., Ref.\,\cite{Bezrukov:2009yw}).
Neutral scalars are expected to provide important clues to the more profound theories behind them, and many experimental searches are underway or proposed to find dark scalars: 
NA62 \cite{NA62:2021zjw}, 
FASER \cite{Feng:2017uoz},
SHiP \cite{Alekhin:2015byh},
MATHUSLA \cite{Chou:2016lxi},
and CODEX-b \cite{Gligorov:2017nwh}.

In this paper, we discuss the cosmological constraints on the dark scalar, focusing on 
the lower limit on the dark scalar mass.
The cosmological constraints of the dark scalar have been discussed in the literature (see e.g., Refs.\,\cite{Berger:2016vxi,Fradette:2018hhl}).
In the previous studies, they mainly considered the feebly coupled dark scalar to the Higgs bosons. 
In this study, we discuss 
more general case for a broad range of the interaction strength,
with which the dark scalar can play the role of the mediator between the 
dark sector and the SM sector in the early Universe.
For a rather strong coupling of the dark scalar,  we need
detailed analyses of the Boltzmann equation including the neutrino decoupling.
We also care the self-interaction of the dark scalar, which affects the time evolution of the dark scalar energy density.  

With these improvements, we derive conservative constraints on the dark scalar mass and the coupling to the Higgs sector.
The results show that the lower the reheating temperature of the Universe is, the weaker the constraint becomes. 
On the other hand, for the Big-Bang Nucleosynthesis (BBN) to be successful, the reheating temperature must be higher than the MeV scale~
\cite{Kawasaki:1999na,Kawasaki:2000en,Ichikawa:2005vw,deSalas:2015glj,Hannestad:2004px,Ichikawa:2006vm,DeBernardis:2008zz,Hasegawa:2019jsa}. 
As a result, a conservative lower bound on the mass of the dark scalar which has a sizable coupling to the electron, is found to be $3.8$\,MeV. 
We also show detailed parameter dependence of the cosmological constraints.

This paper is organized as follows. 
In Sec.~\ref{sec:model}, we summarize the model of dark scalars.
In Sec.~\ref{sec:evolution}, we discuss the mechanism of dark scalar production from the heat bath. 
In Sec.~\ref{sec:Constraints}, we present the cosmological constraints on the dark scalar. 
The final section is devoted to conclusions and discussions.

\section{Model of Dark Scalar}
\label{sec:model}
\subsection{Dark Scalar Interaction with the Standard Model}
We introduce the dark scalar $s$ as a SM gauge singlet field.
In this case, 
the renormalizable scalar potential of the dark scalar and the SM Higgs doublet is given by 
\begin{align}
\label{eq: dark scalar_interaction}  
V_{\rm int} = V_H(H) + V_\mathrm{mix}(H,s) 
+ V_{s}(s)\, ,
\end{align}
where $H$ denotes the Higgs doublet in the SM, and $V_H(H)$ 
is the Higgs potential.
The mixing terms and the self-interaction terms of $s$ are given by,
\begin{align}
V_{\mathrm{mix}}(H,s) &= \frac{\lambda_{Hs}}{2} s^2 |H|^2 + \mu_{Hs}s |H|^2 \ ,\\
\label{eq: dark scalar potential}
V_{s}(s) &= \frac{m_s^2}{2}s^2 + \frac{\mu_s}{6} s^3 + \frac{\lambda_s}{24} s^4\ .
\end{align}
In this paper, we are interested in the cosmological lower limit on the dark scalar mass in around the MeV region, and hence, we assume the dark scalar mass of $\mathcal{O}(1)$\,MeV in the following analysis.

After the electroweak symmetry breaking, the physical SM Higgs $h$ is described by the field
excitation of the modulus around the minimum of $V_H$. 
The dark scalar $s$ mixes with the SM Higgs $h$
with the mixing angle, $\theta$.
In the following, the dark scalar in 
the mass eigenstate is denoted by $\phi$.
The relevant interactions for a low temperature
below the electroweak scale
are those to the fermions induced by the mixing between the dark scalar and the Higgs boson. 
The corresponding Lagrangian is given by
\begin{align}
    \mathcal{L}_{\phi\mbox{-}\mathrm{matter}} =
    - \sum_{f = \mathrm{SM\,fermions} } \frac{m_f s_\theta}{v} \phi \bar{f} f
\ . \label{eq:phi-fermion yukawa}
\end{align}
Here, $m_f$ are the SM fermion masses,
$s_\theta = \sin\theta$ is the sine of the mixing angle, 
and $v = (\sqrt{2} G_{\mathrm{F}})^{-1/2} \simeq 246$\,GeV is the SM Higgs vacuum expectation value.
We can take $s_\theta >0$ without loss of generality, thus we assume $s_\theta > 0$ hereafter.
Due to the experimental constraints, we assume $s_\theta\ll 1$ in the following discussion.
The mixing between the Higgs and the dark scalar also induces the coupling to the weak gauge bosons for $s_\theta\ll 1$,
\begin{align}
    \label{eq:phi-WZ interaction}
    \mathcal{L}_{\phi-WZ} \simeq \left(1 +\frac{h + s_\theta \phi}{v}\right)^2
    \left(m_W^2 W^{\mu +} W_\mu^- + \frac{1}{2}m_Z^2 Z^\mu Z_\mu\right)\ ,
\end{align}
where $m_{W,Z}$ are the masses of the weak gauge bosons.

\subsubsection*{Effective Coupling to Gluon and Photon}
The dark scalar does not couple to the gluon and the photon at the tree-level.
However, the interactions \eqref{eq:phi-fermion yukawa} and \eqref{eq:phi-WZ interaction} induce effective interactions between the dark scalar and the QCD and QED gauge fields by integrating out heavy particles~\cite{Chivukula:1989ze,Leutwyler:1989tn,Leutwyler:1989xj}, 
\begin{align}
\label{eq: scalar-gauge interaction}
    \mathcal{L}_{\phi\mbox{-}\mathrm{gauge}} = - \frac{s_\theta\phi}{v}\left(b_h \frac{\alpha_s}{8\pi}G^{a}_{\mu\nu}G^{a\mu\nu} + \tilde{b}_h\frac{\alpha_{\mathrm{QED}}}{8\pi}F_{\mu\nu}F^{\mu\nu}\right)\, .
\end{align}
Here $G_{\mu\nu}$ ($F_{\mu\nu}$) is the QCD (QED) gauge field strength
and $\alpha_s$ ($\alpha_\mathrm{QED}$) is the QCD (QED) coupling. 
The coefficients $b_h$ and $\tilde{b}_h$ are the contributions of the heavy degrees of freedom to the beta functions
of $\alpha_s$ and $\alpha_\mathrm{QED}$, respectively. 
In theory with $n^{(q)}_h$ heavy quarks and $n^{(l)}_h$ heavy leptons, they are given by
\begin{align}
    b_h &= -\frac{2}{3}n^{(q)}_h\ , \\
    \tilde{b}_h &= 7-4\sum_\mathrm{heavy} Q_q^2-\frac{4}{3}n^{(l)}_h\ ,
\end{align}
where $Q_q$ represents the electromagnetic charge of 
the quark and the symbol $\sum_\mathrm{heavy}$ indicates the summation over heavy quarks.
In the following, we treat $c, b, t$ as the heavy quarks, thus we take $n_h ^ {(q)} = 3$.

\subsubsection*{Effective Coupling to Mesons}
Combining light quark couplings of Eq.~\eqref{eq:phi-fermion yukawa} and the gluon part of Eq.~\eqref{eq: scalar-gauge interaction}, in a low energy effective theory, we obtain interactions between $\phi$ and the light mesons 
via the trace anomaly~\cite{Leutwyler:1989xj}
\begin{align}
\label{eq:scalar-meson interaction}
    \mathcal{L}_{\phi-\mathrm{meson}}
    = \frac{s_\theta\phi}{v}\frac{f_\pi^2}{2}\qty[
    \tr(\partial_\mu U^{-1}\partial^\mu U)
    +(3\kappa+1)B\tr(MU+MU^{-1})]
     + \cdots \ .
\end{align}
Here, $f_\pi = 92.4$\,MeV is the pion decay constant, $M$ is the quark mass matrix and $B$ has mass dimension one set by the quark condensate.
The coefficient $\kappa$ is given by $\kappa = 2n^{(q)}_h/3b$, while $b = 11-(2/3)n^{(q)}_l$ is the QCD beta function coefficient 
with $n^{(q)}_l$ light flavors. 
The field $U$ is the $\mathrm{SU}(n^{(q)}_l)\times\mathrm{SU}(n^{(q)}_l)$ nonlinear meson field corresponding to the 
pseudo Goldstone bosons associated with the chiral symmetry breaking of the QCD. 
The ellipses include $\mathcal{O}(\alpha_s)$ corrections to the effective coupling of $\phi$ and 
the mesons which comes from the anomalous dimension of the light quark masses.
Those corrections are small evaluated at the renormalization scale, $\mu_{\mathrm{RG}} \sim (m_c m_b m_t)^{1/3}$~\cite{Barbieri:1988ct}
(see also Refs.\,\cite{Chetyrkin:1997un,Grozin:2011nk}).
In our case with the $3$ light flavor model, $n^{(q)}_l=3$, effective interactions
between $\phi$ and the charged mesons, i.e., the charged pion and the charged kaon, are given by
\begin{align}
    \label{eq:scalar-pion interaction}
    \mathcal{L}_{\pi\pi\phi} &= \frac{s_\theta\phi}{v}\left(\kappa (2\partial_\mu \pi^+\partial^\mu\pi^- - 3m_{\pi^\pm}^2\pi^+\pi^-) - m_{\pi^\pm}^2 \pi^+\pi^- \right)\ , \\
    \label{eq:scalar-kaon interaction}
     \mathcal{L}_{KK\phi} &= \frac{s_\theta\phi}{v}\left(\kappa (2\partial_\mu K^+\partial^\mu K^- - 3m_{K^\pm}^2K^+K^-) - m_{K^\pm}^2 K^+K^- \right)\ .
\end{align}
Here, $m_{\pi^\pm,K^\pm}$ are the masses of the charged pion and the charged kaon. 
We also have the interaction to the neutral pion and the neutral kaon by replacing $\pi^{\pm}$ ($K^{\pm}$) with $\pi^0$ ($K^0$) and multiplying by $1/2$ in \eqref{eq:scalar-pion interaction} and \eqref{eq:scalar-kaon interaction}.
To take into account of the QED interaction, the 
derivatives on the charged mesons $\partial_\mu$ are replaced by the covariant derivatives, $D_\mu$.

\subsubsection*{Direct Coupling to SM Higgs}
The interactions between the dark scalar and the SM particles are determined only by the mixing angle, $\theta$, except for the interaction with the Higgs boson.
In fact, the contact interactions with the Higgs boson, i.e., $\phi\phi h$,
$\phi h h$ and $\phi\phi h h$ depend on parameters other than the mixing angle $\theta$.%
\footnote{In the model with only superrenormalizable couplings of the dark scalar, the couplings $\phi\phi h$, $\phi h h$ and $\phi\phi h h$ are also determined by the mixing angle, $\theta$~\cite{Fradette:2018hhl}.}
These interactions are of particular importance, when we consider the case that the dark scalar possesses (approximate) $Z_2$ symmetry and is a dark matter candidate with freeze-in productions \cite{Yaguna:2011qn,Chu:2011be,Heeba:2018wtf,Lebedev:2019ton,Bringmann:2021sth}.

\subsection{Decay of Dark Scalar}
The decay modes and rates are the most important parameters for the cosmological constraint 
of the dark scalar.
In this paper, we consider the case that $m_\phi < \mathcal{O}(10)$\,MeV, and hence, the decay modes of the dark scalar are $e^-e^+$ and $\gamma \gamma$.
For $m_\phi > 2m_e$,  the dark scalar $\phi$ decays dominantly into the electron-positron pair
through the Yukawa interaction~\eqref{eq:phi-fermion yukawa}, with the rate
\begin{align}
		\Gamma_{\phi\rightarrow e^- e^+}
		= \frac{s_\theta^2 m_e^2 m_\phi}{8\pi v^2}\qty(1-\qty(\frac{2m_e}{m_\phi})^2)^{3/2}\ .
\end{align}
The dark scalar can also decay into massive neutrinos.
The coupling between the neutrino and dark scalar depends on a model of the neutrino mass generation.
For the  seesaw mechanism \cite{Yanagida:1979as,*Gell-Mann:1979vob,*Minkowski:1977sc},
the rate of the decay into neutrinos is given by
\begin{align}
   		\Gamma_{\phi\rightarrow \nu \nu}
		= \frac{s_\theta^2 m_\nu^2 m_\phi}{4\pi v^2}\ , 
\end{align}
where $m_\nu$ is a Majorana neutrino mass.
In addition, the interactions \eqref{eq:phi-fermion yukawa} and \eqref{eq: scalar-gauge interaction} induce the decay into two photons whose rate is given by
\begin{align}
    \label{eq: two photon decay}
    \Gamma_{\phi\rightarrow\gamma\gamma} 
    = \frac{m_\phi}{\pi}\qty(\frac{s_\theta\alpha_\mathrm{QED}m_\phi}{16\pi v})^2|C(m_\phi)|^2\ , 
\end{align}
where $C(m_\phi)$ is a complex-valued function which includes the effect of lepton loops, light meson loops and the higher dimensional operators in Eq.~\eqref{eq: scalar-gauge interaction}~\cite{Leutwyler:1989tn}.

\begin{figure}[htbp]
\centering{\includegraphics[width=0.7\textwidth]{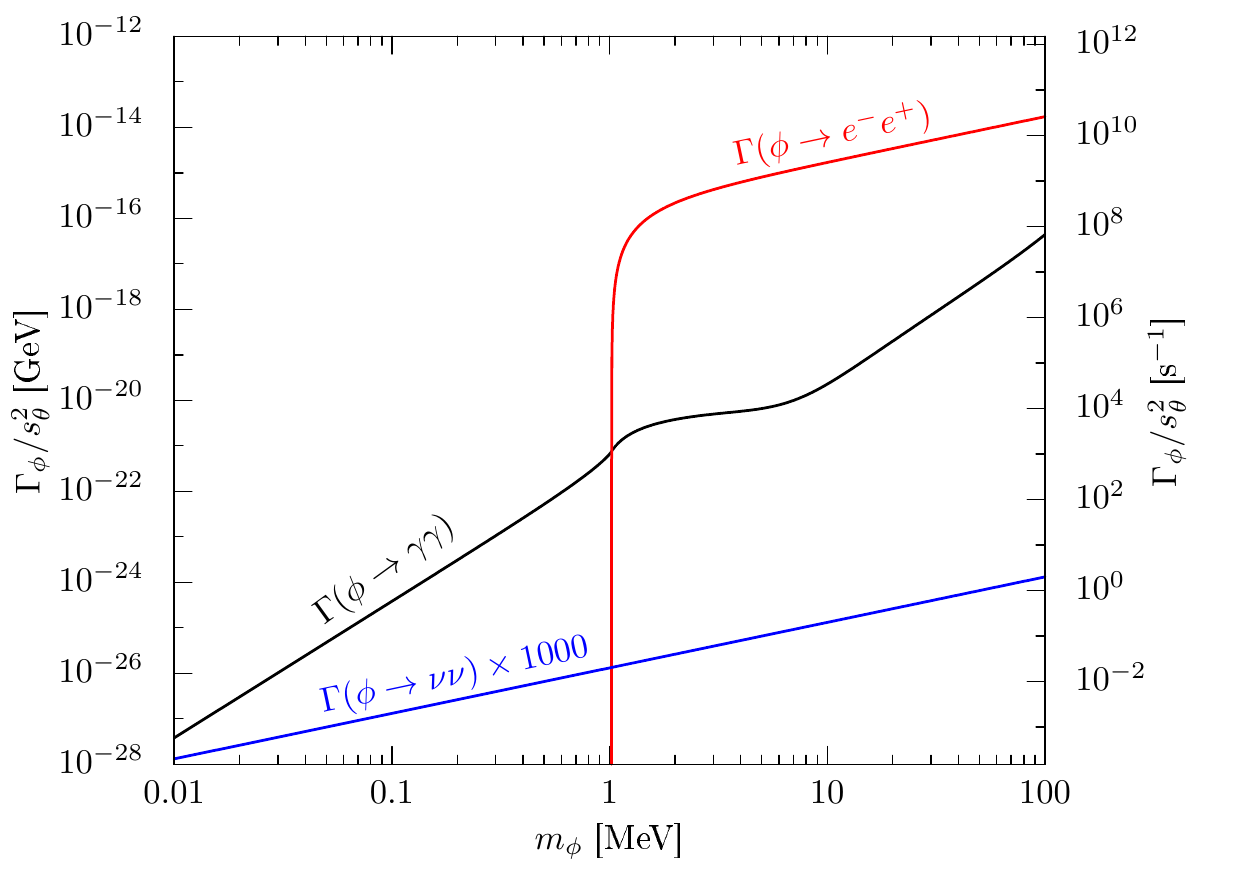}} 
\caption{
The decay rate of the dark scalar $\phi$.
We assume the neutrinos are Majorana fermions where the heaviest mass is $0.1$\,eV.
}
\label{fig:decay}
\end{figure}

In Fig.~\ref{fig:decay}, we show the decay rate of $\phi$ for the mass range we are interested in. 
The figure shows that we can safely neglect the interaction between $\phi$ and the neutrinos for $m_{\phi} > O(10)$\,keV.

\subsection{Self-Interaction of Dark Scalar}
The cubic and quartic terms in the dark scalar potential in Eq.\,\eqref{eq: dark scalar potential} induce the self-interaction of $\phi$.
In the limit of $s_\theta \ll 1$, they are given by,
\begin{align}
    \label{eq:dark scalar self-interaction}
    V_{\mathrm{self-interaction}} = \frac{1}{6}\mu_\phi \phi^3 + \frac{1}{24}\lambda_s \phi^4 + \cdots\ ,
\end{align}
where $\mu_\phi = \mu_s + \lambda_s \langle s \rangle
+ 3 s_\theta  \lambda_{Hs} v$.
These interactions induce $\phi\phi \to \phi\phi$ scattering processes displayed in Fig.\,\ref{fig:diagram}.
In general, $\lambda_s$ and $\mu_\phi/m_\phi$ 
can be of $\mathcal{O}(1)$, even if the mixing parameter $\theta$ is tiny.
In the following analysis, we consider the cases with/without self-interactions.
As we will see, the self-interaction affects the time evolution of the dark scalar.

\subsection{Astrophysical and Collider Constraints}
Let us summarize the constraints other than the cosmological constraints we consider in this work.
\subsubsection*{Supernovae} 
Light particles with mass of $\mathcal{O}(10)\,\mathrm{MeV}$ can be efficiently produced in a core-collapse supernova (for the emissivity of the dark scalar from the supernova, see Refs.~\cite{Ishizuka:1989ts,Krnjaic:2015mbs,Dev:2020eam}).
Such particles produced in a supernova explosion can alter the neutrino cooling rate.
The interaction with nucleons is a dominant source of the dark scalar production. 
The dark scalar with $m_{\phi}\lesssim 100\, \mathrm{MeV}$ can be constrained from the observation of neutrino burst of SN1987A~\cite{Bjorken:2009mm, Dent:2012mx, Kazanas:2014mca, Rrapaj:2015wgs, Chang:2016ntp}.
Note that, however, it is pointed out that there are uncertainties in a model of the neutrino burst and there is a possibility that the constraints from SN1987A are invalidated~\cite{Bar:2019ifz}.

\subsubsection*{Meson Decay}
 In the mass region we are interested in, the most stringent constraint of the accelerator experiments
 comes form the rare decay of  $K^+$.
 The NA62 and E949 experiments put the upper limit
on $\mathrm{Br}(K^+ \to \pi^+ + \phi)= (3\mbox{--}6) \times 10^{-11}$ at the $90\%$\,C.L., which corresponds to $s_\theta \lesssim 2\times 10^{-4}$~\cite{BNL-E949:2009dza,NA62:2021zjw}.

We show these constraints in the summary plot in Fig.\,\ref{fig:constraints} in the final section.

\section{Cosmological Evolution of the Dark Scalar}
\label{sec:evolution}
In this section, we discuss the evolution and the production of the dark scalar in the early Universe.
As the dark scalar couples to all the SM particles, there are many types of production processes at the early Universe, for instance, $f_{\mathrm{SM}}\bar{f}_{\mathrm{SM}} \to \phi V$, where $f_{\mathrm{SM}}$ is an SM fermion and $V$ is a gauge boson.
Depending on the relation of the masses of $\phi$ and the SM particles participating in the production processes, there are two types of contributions to the production processes.
If $m_\phi \ll m_{\mathrm{SM}}$, the production process becomes most efficient when $T_{\mathrm{SM}} \sim m_{\mathrm{SM}}$ and the contribution is suppressed by the Boltzmann factor when the temperature gets much  lower than the mass of the SM particle.
In this sense, the production process behaves like a ``freeze-out" mechanism, and we refer to  this type of contribution as ``UV contribution."

On the other hand, if $m_{\mathrm{SM}} \ll m_\phi$, the production process is most efficient when $T_{\mathrm{SM}} \sim m_\phi$ and we call this type of contribution ``freeze-in contribution."
As we mainly focus on the dark scalar of a mass $\mathcal{O}(10)$ MeV,
the photon and electron contribute
to the production via 
this type of processes.
In particular, the inverse decay $e^- e^+ \to \phi$ provides the largest production among the ``freeze-in" contributions.

In Sec.~\ref{sec: self-thermalization}, 
we discuss how the dark scalar self-interactions affect the time evolution of the dilution after the freeze-out.
In Sec.~\ref{sec: darkscalar-production}, we discuss the production of the dark scalar in detail.
In Sec.~\ref{sec: boltzmann-equation}, we solve the Boltzmann equation to determine 
the abundance of the dark scalar.

\subsection{Effect of Dark Scalar Self-thermalization}\label{sec: self-thermalization}
In previous studies, the impact of the dark scalar self-interactions on the dark scalar constraints is  overlooked.
However, the self-interactions of the dark scalar can affect the evolution of the dark scalar abundance. See e.g., Refs\,\cite{Carlson:1992fn} 
in the context of the dark matter abundance.

In the analysis of the dark scalar abundance,
it is often assumed that the dark scalar does not have self-interaction.
In that case, the dark scalar distribution is
red-shifted after the production.
However, the evolution of the dark scalar is significantly affected by the dark scalar self-interaction. 
The self-scattering $\phi\phi \leftrightarrow \phi\phi$
makes $\phi$ in kinetic equilibrium,
with which the dark scalar can develop its own thermal bath with
the dark sector kinetic 
temperature $T_\phi^{\mathrm{(kin)}}$.
Besides, the number-changing processes such as $\phi \phi \phi \leftrightarrow \phi \phi $ or $\phi \phi \phi \phi \leftrightarrow \phi \phi $ also make $\phi$ in chemical equilibrium, which can be characterized by only one parameter, the temperature $T_\phi$.
We emphasize that the 
temperature $T_\phi$ does not necessarily coincide with the temperature of the SM sector $T_{\mathrm{SM}}$.

\begin{figure}[t]
	\centering
 	\subcaptionbox{ \label{fig:phi4}}
	{\includegraphics[width=0.23\textwidth]{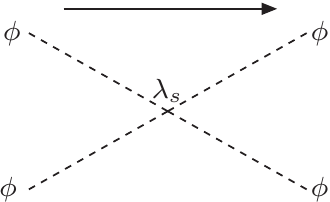}}
 	\subcaptionbox{\label{fig:phi3-t} }
	{\includegraphics[width=0.23\textwidth]{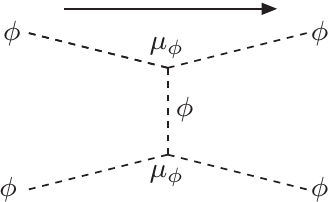}}
 	\subcaptionbox{\label{fig:phi3-s} }
	{\includegraphics[width=0.23\textwidth]{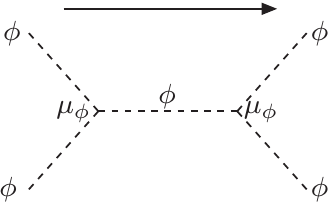}}	
 	\subcaptionbox{\label{fig:split} }
	{\includegraphics[width=0.23\textwidth]{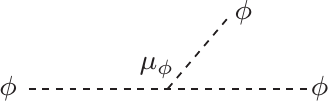}}			

\caption{
Diagrams relevant for the self-scattering.
The diagrams (a)-(c) represent  $\phi \phi \to \phi \phi$ processes, which are relevant for the kinetic equilibrium.
The diagram (d) shows an example diagram for the  collinear contributions, which is relevant for the number-changing processes.
}\label{fig:diagram}
\end{figure}

In the ultra-relativistic limit, the dominant momentum reshuffling comes from the diagram \ref{fig:phi4}.
The cross section $\phi \phi \to \phi \phi$ is approximately given by
\begin{align}
    \sigma_{2 \to 2 } v |_\mathrm{relativistic}  \simeq  \frac{\lambda_s^2}{32 \pi s}\ ,
\end{align}
where $\sqrt{s}$ is the center of mass energy of the collision.
The diagrams \ref{fig:phi3-t} and \ref{fig:phi3-s} provide minor contributions for the momentum reshuffling at high energy.

In the high-energy scattering, there are also number-changing processes  such as
 $\phi \phi \to \phi \phi \phi$ or $\phi \phi \to  \phi \phi  \phi \phi $.
 An important contribution comes from collinear splitting of $\phi$'s.
For instance, the diagram \ref{fig:split} provides the so-called ultra-collinear contribution \cite{Chen:2016wkt}, which provides a large enhancement by $ 1/m_\phi $ to the scattering amplitude. In this limit, by using the splitting function given by Ref.\,\cite{Chen:2016wkt},  we estimate the $\phi \phi \to  \phi \phi \phi  $ cross section as:
\begin{align}
     \sigma_{2 \to 3 } v  |_\mathrm{relativistic}  \simeq  \frac{\mu_\phi^2 \lambda_s^2 (-3 +  \sqrt{3} \pi)}{768\pi^3 m_\phi^2 s} \simeq 0.01 \left(\frac{\mu_\phi}{m_\phi}\right)^2 (\sigma_{2 \to 2 } v) |_\mathrm{relativistic} \ .
\end{align}
Similarly,  the scattering  $\phi \phi \to  \phi \phi \phi \phi$ can be relevant for the number-changing processes.
Moreover, inelastic scatterings of the dark scalar $\phi$ and SM particles $f$, such as $f \phi \to f \phi \phi$, can also contribute to the number-changing processes.

As we will see later, when the dark scalars are produced from the SM plasma of a temperature $T_\mathrm{SM}$,
the typical momentum of the dark scalar is $T_\mathrm{SM}$.
Just after production, the produced dark scalar interacts with other dark scalars.
Through the self-interaction, the dark scalars exchange their momenta 
when $\sigma_{2\rightarrow 2}$ is large enough.
Besides, the comoving number density of the dark scalar increases when $\sigma_{2\rightarrow 3}$ is large enough.
For the efficient momentum exchange, we need
\begin{align}
    \sigma_{2 \to 2 } v|_\mathrm{relativistic}\times n_{\phi} H^{-1} \gg 1\ ,
\end{align}
which indicates 
\begin{align}
    \left.Y_\phi\right|_\mathrm{relativistic} = \frac{n_\phi}{s_\mathrm{SM}} \gg 10^{-18}\lambda_s^{-2} g_*^{-1/2}\times \left(\frac{T_\mathrm{SM}}{1~\mathrm{GeV}}\right)\ .
\end{align}
Here, $H$ denotes the Hubble expansion rate, $s_\mathrm{SM}$ is the entropy density of the SM sector.
The effective degrees of freedom of the SM particles at the temperature $T_\mathrm{SM}$
is given by $g_* \sim g_{*s}$ \cite{Saikawa:2018rcs,*Saikawa:2020swg}, with which the energy and entropy densities are
\begin{align}
    \rho_\mathrm{SM}(T) = \frac{g_{*}(T) \pi^2 T^4}{30}
    \ ,\quad s_\mathrm{SM}(T) = \frac{2\pi^2 g_{*s}(T)  T^3}{45}\ .
\end{align}
If this condition is satisfied, kinetic equilibrium of a kinetic temperature $T^{\mathrm{(kin)}}_\phi$ can be realized.
Similarity, the condition  for the efficient number changing process is
\begin{align}
    \left. Y_\phi\right|_\mathrm{relativistic} \gg 10^{-16}\times\left(\frac{\mu_\phi}{m_\phi}\right)^{-2}\lambda_s^{-2} g_*^{-1/2} \left(\frac{T_\mathrm{SM}}{1~\mathrm{GeV}}\right)
     \left(
    \frac{T^{\mathrm{(kin)}}_\phi}{T_\mathrm{SM}}
    \right)^2\ ,
\end{align}
for $T_\phi^{\mathrm{(kin)}} \gg m_\phi$.
As we will see later, $Y_\phi$ produced from the SM thermal bath is greater than $\mathcal{O}(10^{-15})$ in the parameter region of our interest.
Therefore, for $\lambda_s= \mathcal{O}(1)$ and $\mu_\phi = \mathcal{O}(m_\phi)$, the dark scalar can reach chemical equilibrium.
Hereafter, we refer to this case as ``self-thermalization."

We also need to consider the freeze-out condition of the self-thermalization when the dark scalar temperature gets lower.
As the Universe expands, the number density of the dark scalar is diluted.
In this case, the number changing process $\phi\phi\phi \leftrightarrow \phi \phi$ is decoupled.
The freeze-out condition is
\begin{align}
    \left.\langle\sigma_{3 \to 2 } v^2\rangle\right|_\mathrm{non-relativistic} n^2_{\phi} H^{-1} \ll 1\ .
    \label{eq: 3-2 freeze out}
\end{align}
The perturbative estimation of the cross section $\phi\phi\phi \to \phi \phi$ in the non-relativistic limit is given by
\begin{align}
    \langle \sigma_{3\to2} v^2 \rangle|_{\mathrm{non-relativistic}}  = \frac{25 \sqrt{5} \mu_\phi^2(3\lambda_sm_\phi^2 -  \mu_\phi^2 )^2}{24576 \pi m_\phi^{11}}\ .
    \label{eq:cross_per}
\end{align}
More generally, if the dark scalars are strongly interacting with each other, the pertubative estimation of the cross section may be invalid.
In this case, we use the $s$-wave unitarty bound for the scattering cross section  \cite{Kuflik:2017iqs, Namjoo:2018oyn}: 
\begin{align}
    \langle \sigma_{3\to 2} v^2 \rangle|_\mathrm{non-relativistic}
    <\langle \sigma_{3\to 2} v^2 \rangle|_\mathrm{uni} \equiv \frac{48 \sqrt{3} \pi^2}{m_\phi^3 T_\phi^2}\ ,
    \label{eq:cross_uni}
\end{align}
to estimate the freeze-out condition.

\begin{figure}[tbp]
\centering{\includegraphics[width=0.7\textwidth]{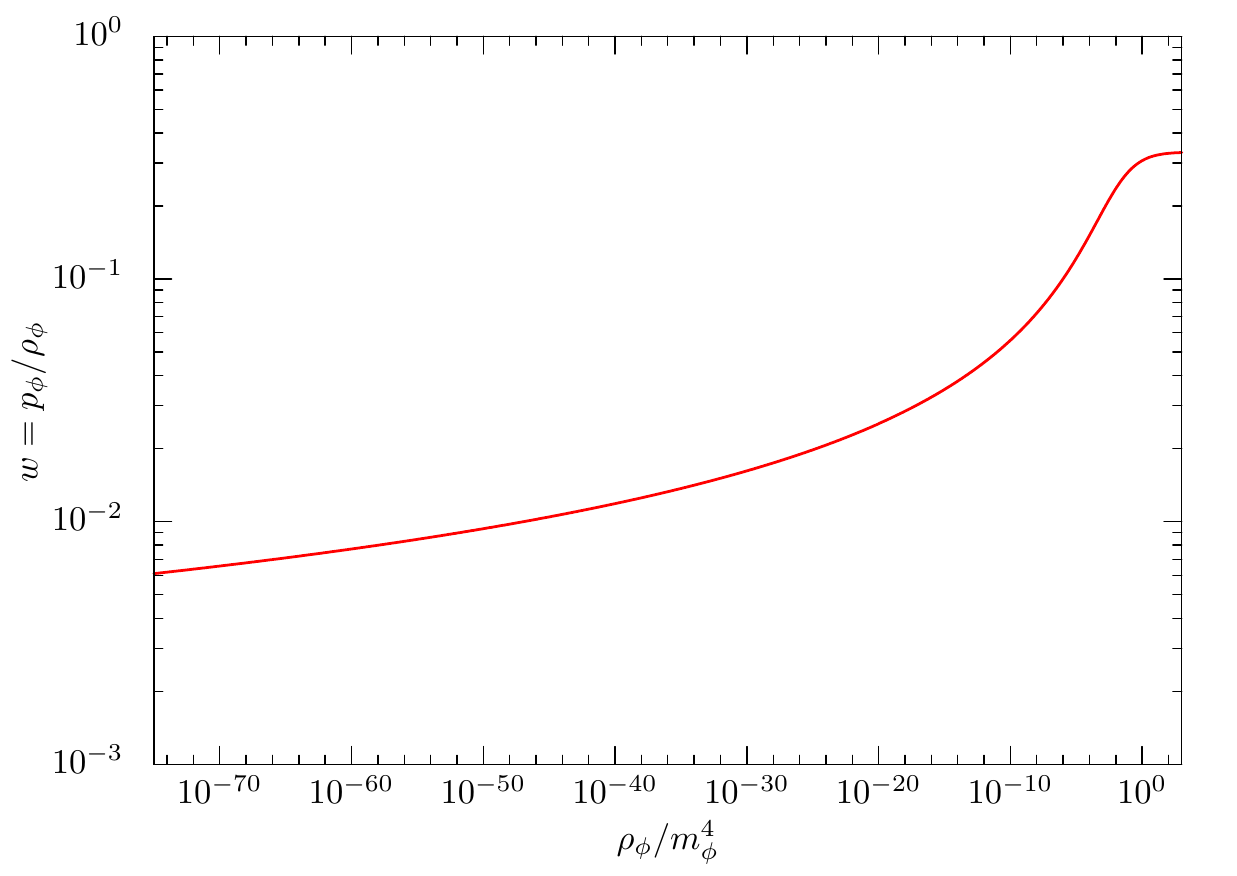}} 
\caption{
Relation between the energy density and the pressure for the self-thermalized $\phi$.
}
\label{fig:w}
\end{figure}

Now, let us discuss the cosmological evolution of the dark scalar with/without self-thermalization.
To focus on the effects of the self-thermalization, let us switch off the interaction with the SM.
The evolution of the energy density of the self-thermalized dark scalar $\rho_\phi$ is given by
\begin{align}
    \frac{d \rho_{\phi}}{ d t} = -3 H (1 + w) \rho_\phi, ~~~~ \text{or}~~~~ a\frac{d \rho_{\phi}}{ d a} = -3  (1 + w) \rho_\phi,\label{eq:self_evolution}
\end{align}
where the pressure $P_\phi$ is given by the parameter for the equation of state (EOS), $P_\phi = w \rho_\phi$, and $a$ is the scale factor of the Universe.
In Fig.\,\ref{fig:w}, we show the EOS parameter $w$.
The $w$ parameter is approximately given by $w \to 1/3$ for $\rho_\phi/m_\phi^4 \gg 1$ and $w \to T_\phi/m_\phi \simeq -1/\log(\rho_\phi/m_\phi^4)$ for $\rho_\phi/m_\phi^4 \ll 1 $.

When the Universe expands and the scale factor goes from $a_\mathrm{ini}$ to $a$, the energy density $\rho_{\phi,\mathrm{ini}}$ approximately changes to
\begin{align}
    \rho_\phi(t) \simeq \rho_{\phi,\mathrm{ini}} \left(\frac{a}{a_\mathrm{ini}}\right)^{-3 (1 + w)}.
\end{align}
Due to $w>0$, the decrease in energy density of self-thermalized $\phi$ is faster than that of non-interacting $\phi$ in the non-relativistic limit.
In this case, the cosmological constraint can be relaxed with the self-thermalization
compared with the non-interacting case.

On the other hand, there is also an effect
of self-scattering that can make the constraint severer.
When the dark scalars are produced from the SM plasma with a temperature of $T_\mathrm{SM}$, the typical momentum of the dark scalars are $p_\phi \sim T_\mathrm{SM}$.
For $T_\mathrm{SM} \gg m_\phi$, the dark scalar is highly relativistic and
the energy density goes $\rho_\phi \propto a^{-4}$ without self-interaction.
However, if the energy density is small $\rho_\phi / m_\phi^4 \ll 1$ and the self-thermalization is efficient,
the dark scalars get immediately non-relativistic through the $\phi\phi \to \phi\phi\phi$ process.
Once the dark scalar becomes non-relativistic, the energy density behaves  $\rho_\phi \propto a^{-3}$.
Therefore the energy density gets larger compared to the case of the non-self-thermalization.

\begin{figure}[tbp]
\centering{\includegraphics[width=0.7\textwidth]{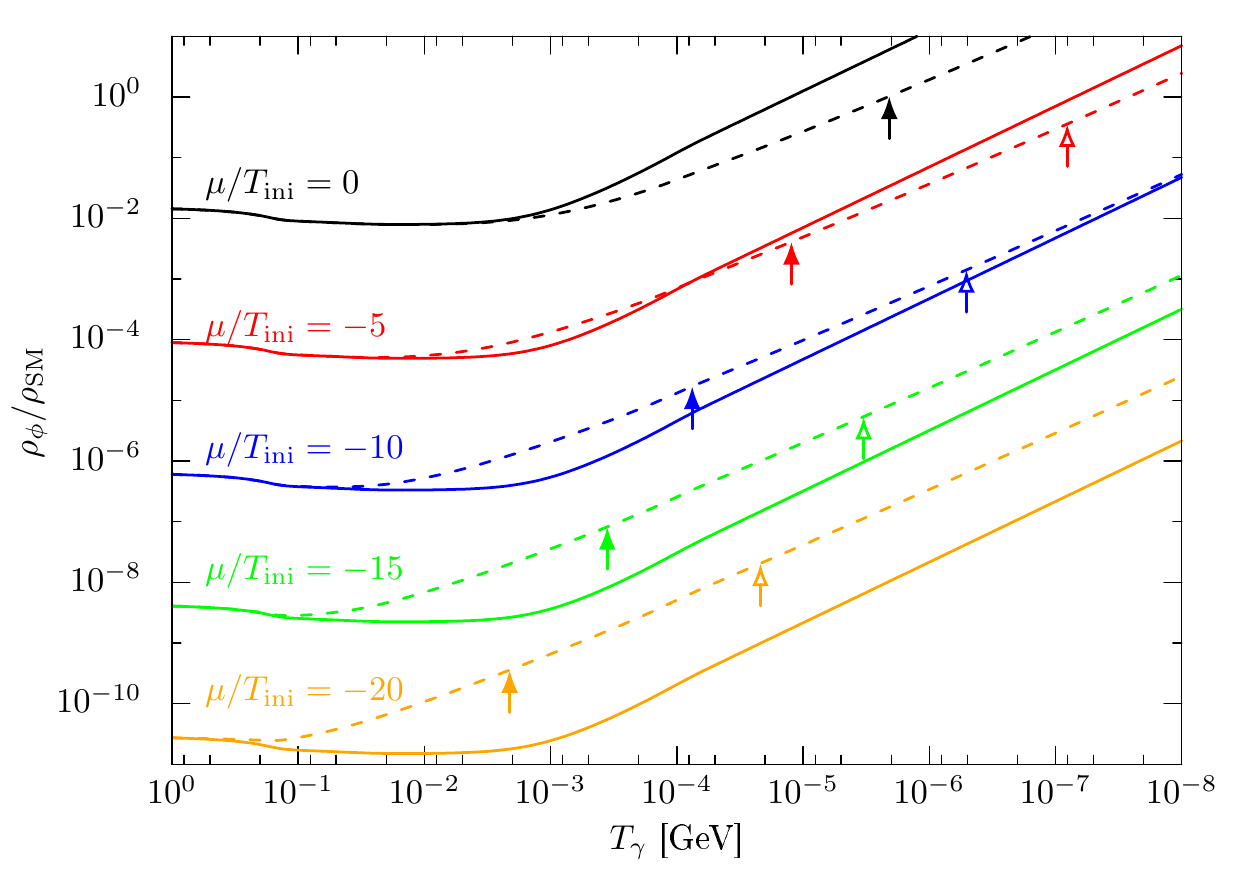}} 
\caption{
The solid (dashed) lines show the cosmological evolution of the energy density of the stable $\phi$ without (with) self-thermalization.
In both cases, we take $m_\phi=2$\,MeV.
The arrows with fill and with empty fill indicate the freeze-out temperature of the $\phi \phi \phi \leftrightarrow \phi \phi $ process with the perturbative cross section in Eq.\,\eqref{eq:cross_per} ($\mu_\phi = m_\phi,\lambda_s = 0$) and the unitarity-bound-saturated cross section in Eq.\,\eqref{eq:cross_uni}, respectively.
}
\label{fig:self}
\end{figure}

In order to see the impact of the self-interaction on the cosmological evolution of the dark scalar energy density, we consider two extreme cases.
One is the case that there is no self-interaction and the phase space density is simply redshifted.
The other is the case that it is always thermalized and in chemical equilibrium.
In the former case, the kinetic equilibrium is no longer maintained.
In Fig.\,\ref{fig:self}, we show some example cases, in which the initial conditions of the phase space density $f_{\phi,i}$ is given by the Bose-Einstein distribution with a chemical potential $\mu$,
\begin{align}
    f_{\phi, \mathrm{ini}}(p) = \frac{1}{\exp((E-\mu)/T_\mathrm{ini})-1 }
\end{align}
at a temperature of the SM plasma $T_\gamma = T_\mathrm{SM} = T_\mathrm{ini} = 1$\,GeV and a scale factor $a_\mathrm{ini}$.
In the case of the self-thermalization, the evolution of $\phi$ is determined by Eq.\,\eqref{eq:self_evolution}.
In the figure, we assume that the
self-thermalization is maintained 
even after the freeze-out of the 
number-changing processes for the 
dashed lines.

In the non-interacting case, the phase space density at a scale factor $a$ is given by
\begin{align}
    f_{\phi}(p) = f_{\phi, \mathrm{ini}}(p \times (a/a_\mathrm{ini})),
\end{align}
with which $\langle p_\phi \rangle \sim T_\gamma$ for $T_\gamma \gg m_\phi$.
On the other hand, the dark scalar becomes non-relativistic in the case of  the self-thermalization.
As we can see from Fig.\,\ref{fig:self}, if the dark scalar energy density is small $\rho/m_\phi^4 \ll1$ but typical momentum is large $\langle p_\phi \rangle \gg m_\phi$, the self-thermalization can relatively enhances the dark scalar abundance.
For $T_\gamma < m_\phi$, the 
self-thermalization induces $w > 0$, which reduces the dark scalar density more rapidly compared with the non-interacting case.

At the very late time, i.e., $T_\gamma \ll m_\phi$, however, we also need to consider the freeze-out of the number-changing self-scattering as discussed at  Eq.\,\eqref{eq: 3-2 freeze out}.
The freeze-out temperature of the number-changing self-scattering are indicated by the filled and the open arrows in Fig.\,\ref{fig:self} 
for perturbative cross section ($\mu_\phi = m_\phi,\lambda_s = 0$) and unitarity-bound-saturated cross section, respectively.
For temperature lower than those temperature, the number-changing-self-scattering is no more effective, and hence, 
the energy density of the dark scalar scales as $\rho_\phi \propto a^{-3}$.

\subsection{
Dark Scalar Production} \label{sec: darkscalar-production}
The dark scalar couplings to the SM in Eq.\,\eqref{eq:phi-fermion yukawa} are renormalizable, and hence, those become more relevant at the lower temperature for each production channel~\cite{Hall:2009bx}.
In fact, if the reheating temperature, $T_R$, is well above the electroweak scale, the abundance of the dark scalar at $T\sim \order{1}$\,MeV is insensitive to $T_R$.
However, as we will see later, it strongly depends on $T_R$ lower than the electroweak scale, because the dark scalar couples to the heavier particles more strongly.
Due to this property, the cosmological evolution of the dark scalar abundance becomes more complicated compared with cases that a dark particle couples to the SM particles with more or less universal couplings such as the dark gauge boson~\cite{Ibe:2019gpv,Ibe:2020dly}.

For example, the cross sections $\sigma_{f}$ 
of the typical production processes such as $f\bar{f} \to \phi V$ with $V$ being either the photon or the gluon
are proportional to $(m_f/v)^2/s$.
Hence, they contribute to the yield of the dark scalar as
\begin{align}
\frac{n_\phi}{s_\mathrm{SM}} \sim \left.\frac{\sigma_f (n_f^\mathrm{eq})^{2} H^{-1}}{s_{\mathrm{SM}}}\right|_{T\sim m_f} \propto \frac{m_f M_\mathrm{Pl}}{v^2} \ , 
\end{align}
where $n^\mathrm{eq}_f$ denotes the thermal number 
density of the fermion $f$ and $M_\mathrm{Pl}$ is the reduced Planck mass.
As the heavy particle contributions for $m_f \gg m_\phi$ are more significant, the  abundance of the dark scalar strongly depends on whether the temperature of the Universe has been higher or lower than the masses of the heavy SM particles.
This is contrary to the conventional freeze-in scenario, in which the most dominant production of the particles comes from the temperature $T_\gamma\simeq m_\phi$ and irrelevant to the high temperature era of the Universe.
In the present dark scalar model, the cosmological constraint has a strong dependence on the reheating temperature
for $T_R \lesssim \order{100}$\,GeV.
In order to get the most conservative constraint, we consider a low reheating temperature scenario.

Depending on the temperature of the Universe, there are various production processes.
In the following, we discuss the main production processes at each temperature.

\subsubsection*{Production for $T\lesssim 10\,$MeV}
\begin{figure}[t]
	\centering
	\subcaptionbox{\label{fig: decay to e}}
	{\includegraphics[width=0.23\textwidth]{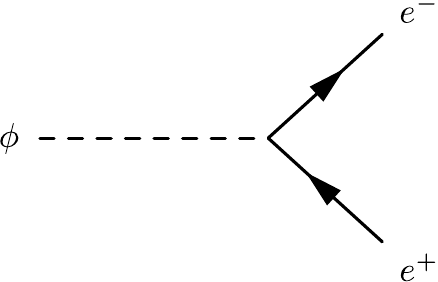}}
	\hspace{50pt}
 	\subcaptionbox{\label{fig: lepton-anni}}
	{\includegraphics[width=0.23\textwidth]{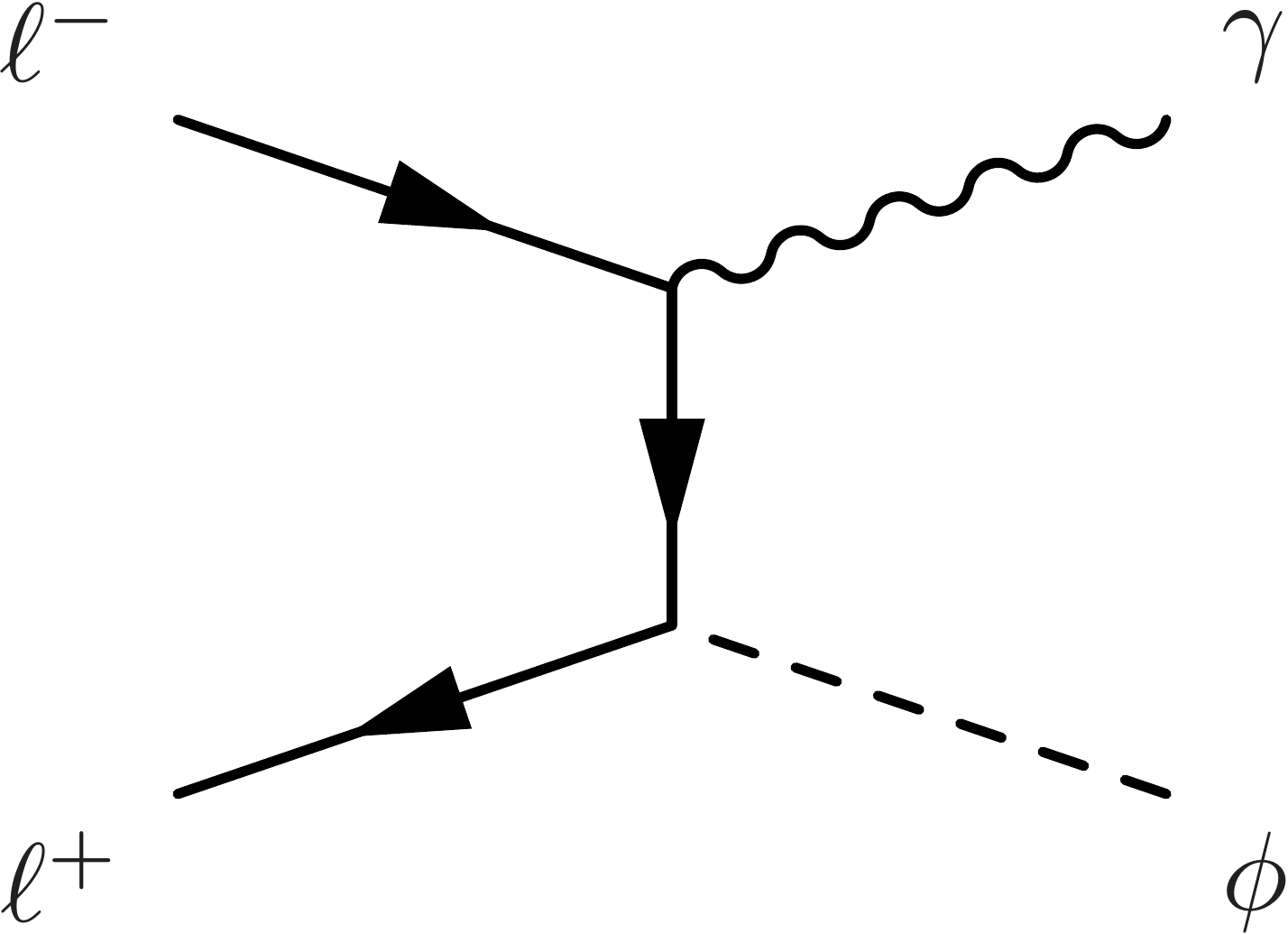}}
	\hspace{50pt}
 	\subcaptionbox{\label{fig: lepton-comp}}
	{\includegraphics[width=0.23\textwidth]{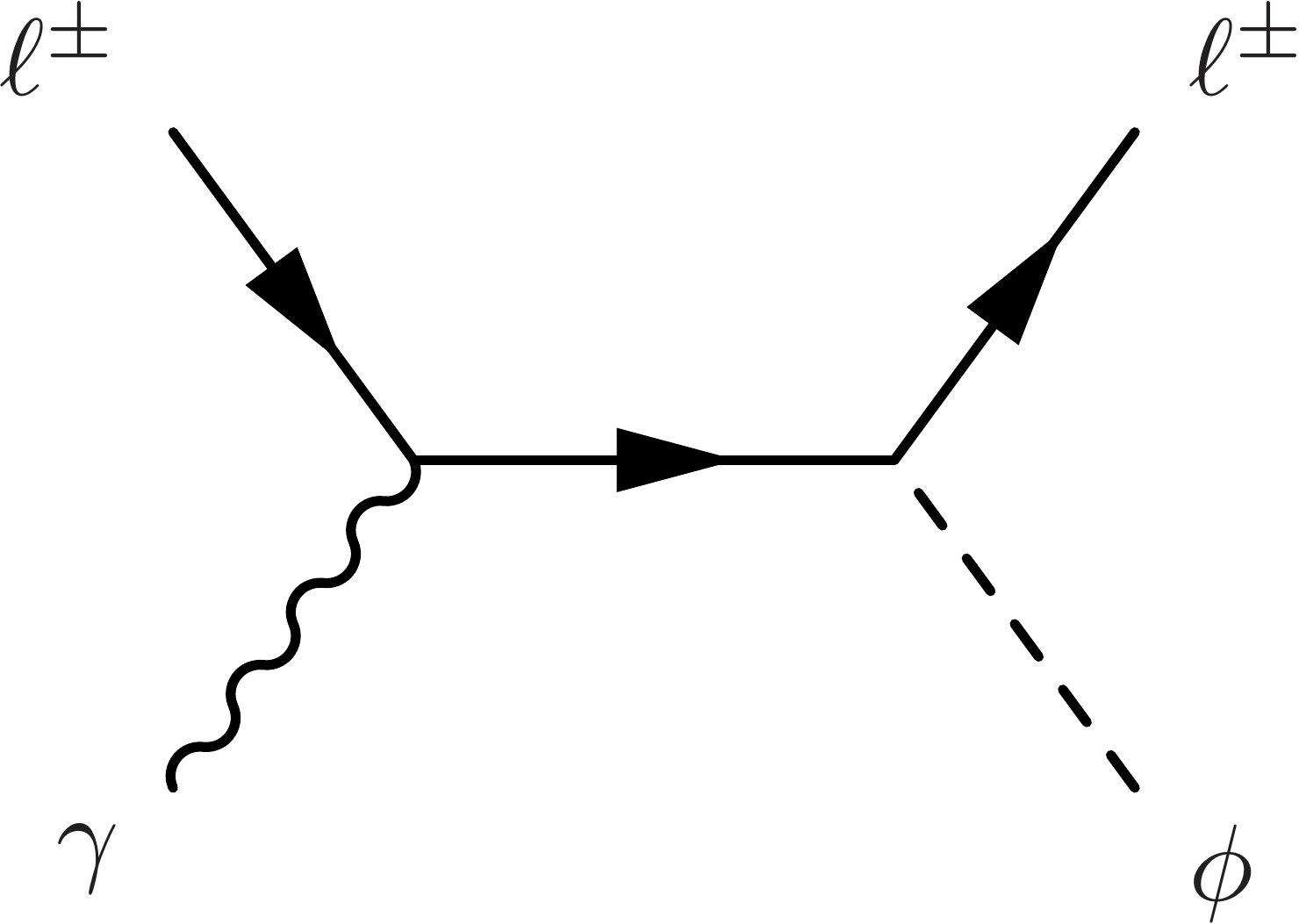}}
\caption{Examples of the production from leptons for $m_\phi < 2m_\mu$.
The inverse decay from the electron positron pair is effective only for $m_\phi > 2 m_e$.}
\label{fig: production from leptons}
\end{figure}

Let us first discuss the production processes relevant for $T\lesssim 10\,$MeV.
In this temperature region, the dark scalar is generated mainly from the electrons. 
The relevant process depends on the dark scalar mass. If $m_\phi > 2m_e$, the inverse decay shown in
Fig.\,\ref{fig: decay to e} is the most important. 
If $m_\phi < 2m_e$, the dark scalar is produced through the annihilation and the Compton-like scattering
shown in Figs.\,\ref{fig: lepton-anni} and~\ref{fig: lepton-comp}.

\subsubsection*{Production for $10$\,MeV$\lesssim T \lesssim 100\,$MeV}
When the temperature is $\mathcal{O}(10)$\,MeV,  the dark scalar is generated mainly
from the muon through the scattering processes shown in Fig.\,\ref{fig: production from leptons}.
Since we are interested in the light dark scalar, $m_\phi<2m_\mu$, it is not produced by the inverse decay from the muons.

\begin{figure}[t]
	\centering
 	\subcaptionbox{\label{fig: meson-anni}}
	{\includegraphics[width=0.23\textwidth]{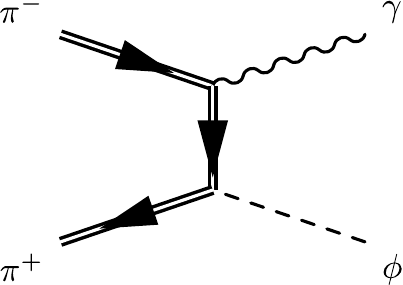}}
	\hspace{50pt}
 	\subcaptionbox{\label{fig: meson-comp}}
	{\includegraphics[width=0.23\textwidth]{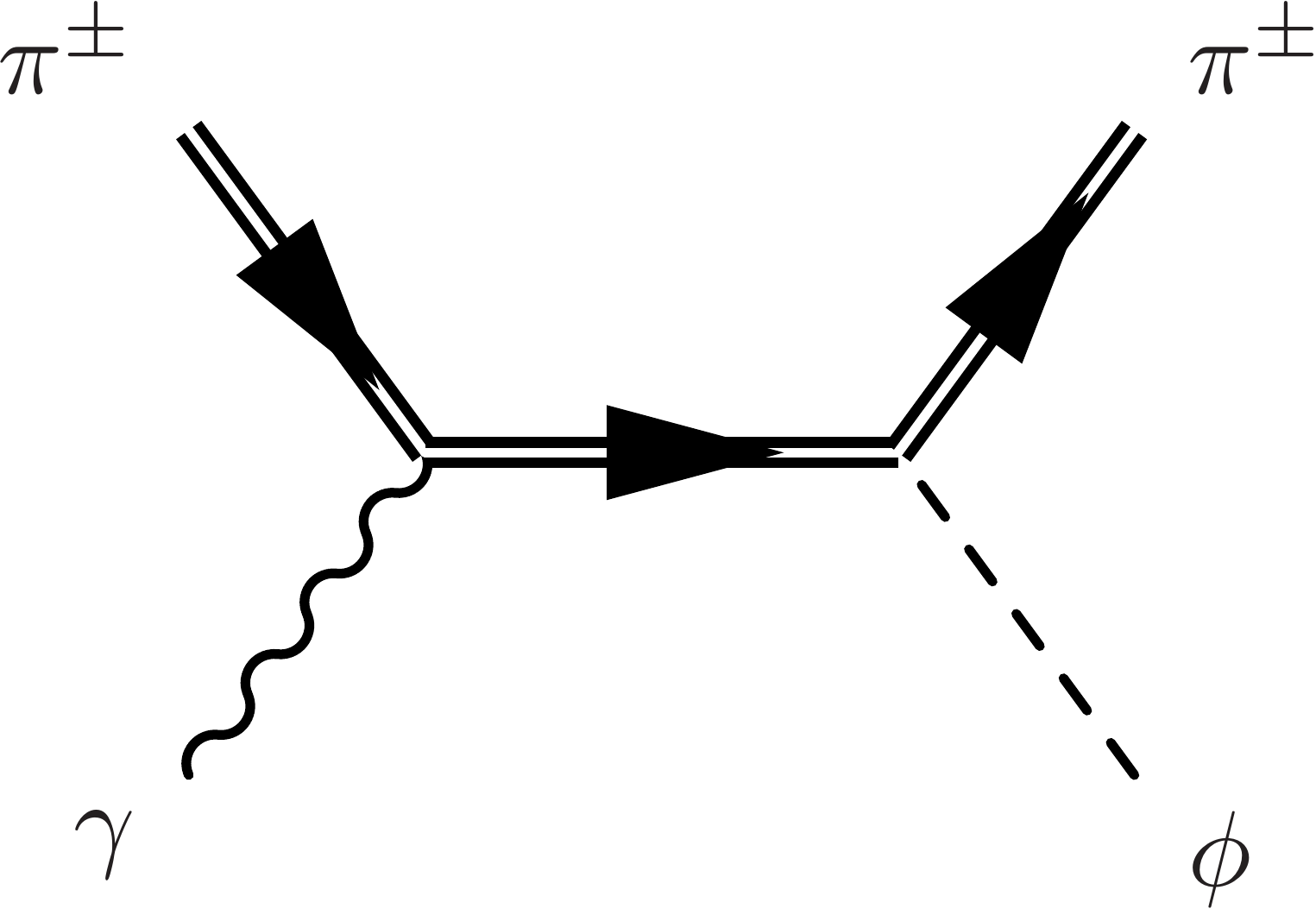}}
	\hspace{50pt}
 	\subcaptionbox{\label{fig: meson-comp}}
	{\includegraphics[width=0.23\textwidth]{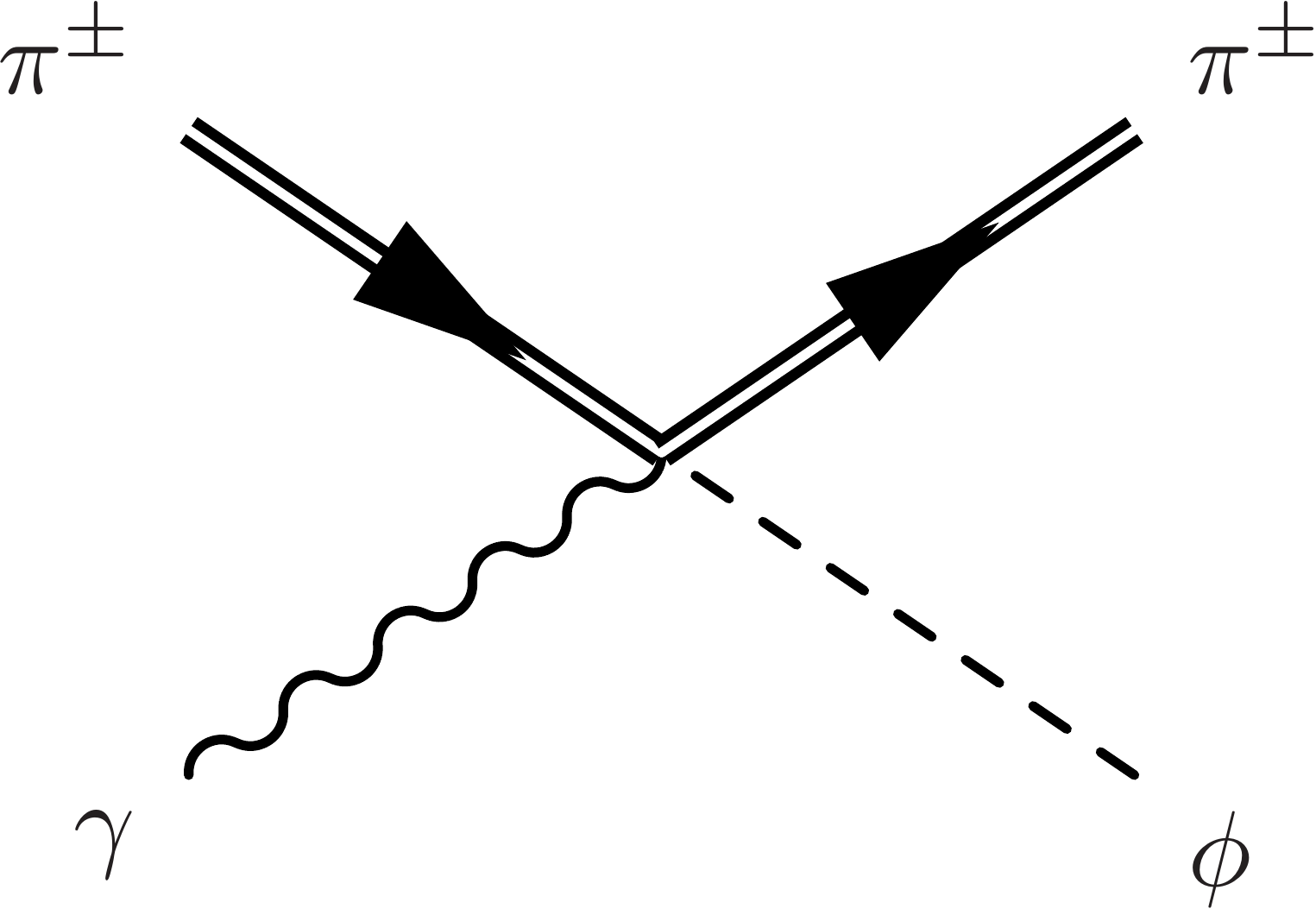}}
\caption{Examples of the production from mesons.}
\label{fig: production from mesons}
\end{figure}

In addition, light quarks ($u$, $d$ and $s$) may also contribute to the dark scalar production.
In this temperature region, we must consider the 
effective theory of the dark scalar couplings to light mesons, in Eqs.\,\eqref{eq:scalar-pion interaction} 
and \eqref{eq:scalar-kaon interaction}.
The effective couplings provide the production channel of the dark scalar $\phi$
shown in Fig.\,\ref{fig: production from mesons}. 
We will find, however, that both the pion and kaon in these channels give 
only subdominant contributions to the dark scalar abundance compared to the muon.
Due to the CP symmetry, the $2 \to 2$ processes involving three pions are forbidden.
Therefore, in the temperature much below the pion mass, the production cross section from light mesons are suppressed by the QED couplings, as in the case of the leptons.

In our analysis, we only include the pion and kaon channels for the hadronic contributions.
We neglect the contributions from heavy hadrons, which results in 
more conservative constraints on the dark scalar.
Note also that the chiral perturbation analysis is useless for the temperature $T \gg 100$\,MeV.
In this case, multiple mesons processes and heavy hadron contributions are more important.
Thus, the treatment to include only the $2 \to 2$ processes is invalid at the higher temperature.
Although our treatments are invalid at the high temperature, 
we naively extrapolate them even for $T_R\gg 100$\,MeV, 
since we will find that heavy quark contributions become dominant in the high temperature region.

\subsubsection*{Production above the QCD Scale}
\begin{figure}[t]
	\centering
 	\subcaptionbox{\label{fig: quark-anni}}
	{\includegraphics[width=0.23\textwidth]{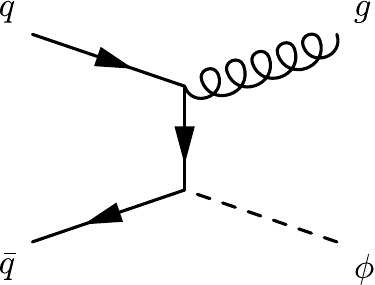}}
	\hspace{50pt}
 	\subcaptionbox{\label{fig: quark-comp}}
	{\includegraphics[width=0.23\textwidth]{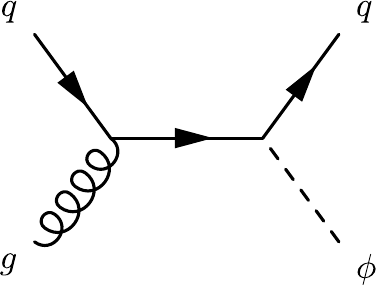}}
\caption{Examples of production from quarks.}
\label{fig: production from quarks}
\end{figure}

In this regime, the heavy $t, b,$ and $c$ quarks give the significant contribution 
through the annihilation $q \bar{q}\rightarrow \phi g$ and the Compton-like scattering 
$q g \rightarrow q\phi$ displayed in Fig.\,\ref{fig: production from quarks}. 
Besides, the $W$ and $Z$ bosons also significantly contribute to the dark scalar abundance through the coupling Eq.\,\eqref{eq:phi-WZ interaction} 
for the temperature $T \gtrsim 10$\,GeV~\cite{Fradette:2018hhl}.
In addition, the Higgs boson contributions
to the dark scalar production are also relevant for $T\gtrsim 10$\,GeV.
In general, the Higgs contributions cannot be parameterized by only the mixing angle $s_\theta$.
As we will see, however, in the parameter region of our interest, $s_\theta \gtrsim 10^{-7}$, 
the cosmological constraints are insensitive to the
details of the production process above $T\gtrsim 10$\,GeV.
Therefore, we do not take into account these weak gauge boson and Higgs contributions.


\begin{figure}[H]
\centering{\includegraphics[width=0.7\textwidth]{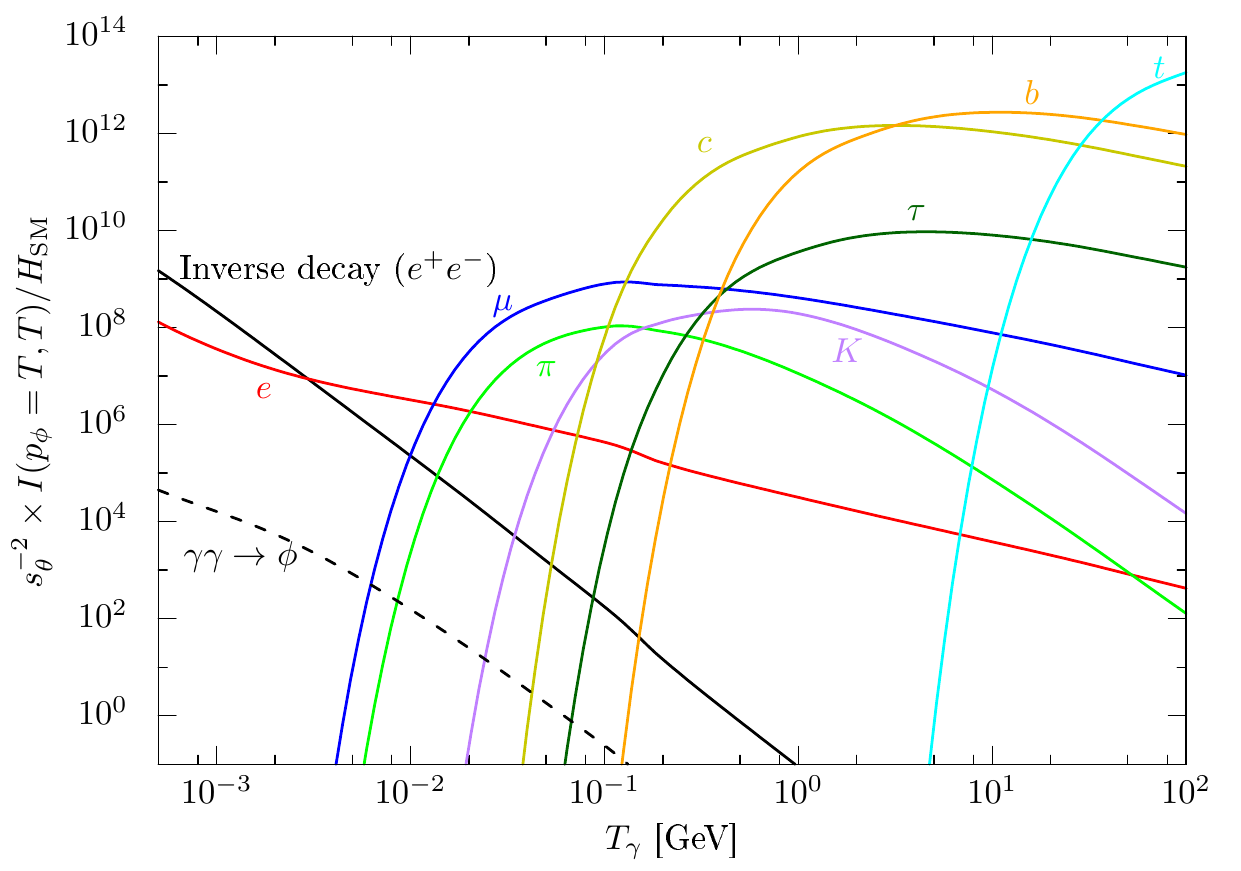}} 
\caption{
Production rates of the dark scalar for $m_{\phi} = 2$\,MeV. 
Each line represents the interaction rate in the right hand side of Eq.\,\eqref{eq:collision intergrals}.
}
\label{fig:rate}
\end{figure}

\subsubsection*{Multiple Dark Scalar Production}
We neglect the multiple $\phi$ production processes which are 
suppressed by more powers of $s_\theta$ since $s_\theta \ll 1$.
In the presence of the self-interaction, multiple $\phi$ production processes such as $f\bar{f} \to \phi \phi$ 
and $f\bar{f} \to \phi \phi\phi$ 
are also possible without further $s_\theta$ suppression, where $\phi$ appears in the $s$-channel and splits into a $\phi$ pair 
and into three $\phi$'s through the $\mu_\phi$ and $\lambda_s$ terms in Eq.\,\eqref{eq:dark scalar self-interaction}.
The cross sections for the multiple productions can be
sizable for $\mu_\phi = \mathcal{O}(m_\phi)$ and $\lambda_s = \mathcal{O}(1)$
when the dark scalar mass is rather large, $m_\phi > \mathcal{O}(10)$\,MeV.
Those contributions enhance the production of the dark scalar
which may make the constraints severer in some parameter region.
However, since our goal is to derive conservative constraints that are as model-independent as possible, we will not incorporate these effects in the following analyses.

Besides, the processes such as $f \phi \to f \phi$ 
and  $f \phi \to f \phi \phi$ which involve the 
dark scalar self-interaction also affect the
evolution of the number density and temperature of the dark scalar. 
These processes cause the energy exchange between the SM and the dark scalar. 
In principle, they can reduce the self-thermalized dark scalar temperature for $T_\phi > T_\mathrm{SM}$.
However, due to the small Yukawa coupling, the rate of the energy exchange is less than the Hubble expansion rate in the parameter region where the self-thermalization relaxes the cosmological constraints. 
Therefore, we also ignore the process $f \phi \to f \phi$ in the following analysis.

\subsection{Boltzmann Equation}\label{sec: boltzmann-equation}
Here, we discuss the cosmological evolution of the 
dark scalar.
It is governed by
the Boltzmann equation, 
\begin{align}
\label{eq:Boltzmann}
&\pdv{f_{\phi}}{t} - Hp_\phi\pdv{f_{\phi}}{p_{\phi}} 
= -\mathcal{C}[f_{\phi}]\ , \\
& H^2 =\frac{\rho_\mathrm{SM} + \rho_{\phi}}{3 M_\mathrm{Pl}^2}\ ,
\end{align}
where $\mathcal{C}[f_\phi]$ denotes 
the collision term of the dark scalar, and $p_\phi = |\mathbf{p}_\phi|$ and $f_\phi$ are the momentum and the distribution function of the dark scalar.
The energy density $\rho_{\mathrm{SM}}$ denotes that of the SM sector, and $\rho_\phi$ is that of the dark scalar.
Since the neutrino decoupling process plays an important role for us to discuss the cosmological constraints on the dark scalar, we decompose the SM energy density into that of neutrinos $\rho_\nu$ and the rest, $\rho_\mathrm{vis}$.
Here, $\rho_\mathrm{vis}$ includes $\gamma$, leptons, the QCD sector, and the Higgs and the weak gauge bosons, which we call the ``visible" sector.
These particles are thermalized, and hence, $\rho_\mathrm{vis}$ is characterized by the 
temperature $T_\gamma$.

The collision term of the dark scalar consists of the self-interaction part discussed in Sec.~\ref{sec: self-thermalization} and the interaction with the SM particles in Sec.~\ref{sec: darkscalar-production}, 
\begin{align}
    \mathcal{C}[f_\phi] = \mathcal{C}_\mathrm{self}[f_\phi] + \mathcal{C}_{\phi \leftrightarrow\mathrm{vis}}[f_\phi]\ .
\end{align}
We reduce $\mathcal{C}_{\phi \leftrightarrow\mathrm{vis}}[f_\phi]$
assuming that all the visible sector particles are in thermal equilibrium, which is approximated by, 
\begin{align}
\label{eq:collision term}
    \mathcal{C}_{\phi \leftrightarrow\mathrm{vis}}[f_{\phi}]
    \simeq & \, 
    I(p_\phi, T_\gamma)\times(
    f_{\phi}(p_{\phi}) - f^{\mathrm{BE}}_{\phi}(p_\phi, T_\gamma)
    ) \ ,
    \end{align}
where, 
    \begin{align}
    \label{eq:collision intergrals}
    I(p_\phi, T_\gamma) =& \sum_{\psi =e,\mu,\tau, \pi, K}\left[I_{\psi^\pm \phi\leftrightarrow \psi^\pm\gamma}(p_\phi, T_\gamma)
    +I_{\gamma \phi\leftrightarrow \psi^-\psi^+}(p_\phi, T_\gamma)\notag \right] \\
    &\left.+  
    \sum_{q =c,b,t}\left[I_{\overset{\scriptscriptstyle(-)}{q}\phi\leftrightarrow \overset{\scriptscriptstyle(-)}{q}g}(p_\phi, T_\gamma)
    +I_{g \phi\leftrightarrow q \bar{q}}(p_\phi, T_\gamma)\notag \right]\right. \notag \\
    &+I_{ \phi \leftrightarrow 2\gamma}(p_\phi, T_\gamma)
    + I_{ \phi \leftrightarrow e^- e^+}(p_\phi, T_\gamma)\    .
\end{align}
Here, $I_\mathrm{process}(p_\phi,T_\gamma)$'s are given in the Appendices~\ref{sec: reducecolint} and \ref{sec: explicit colint}, 
and the superscript ``BE" indicates the Bose-Einstein distribution. 
In Fig.~\ref{fig:rate}, we show the temperature dependence of $I_\mathrm{process}$
for $m_\phi = 2\,$\,MeV at $p_\phi = T_\gamma$.

For the decay and the inverse decay, we can perform the phase space integrals analytically.
For the scattering processes, we can reduce them into the one-dimensional 
integral by using the Maxwell-Boltzmann approximation in part.
We provide the computational details of the reduction of the phase space integrals
in the Appendix~\ref{sec: reducecolint}, and the explicit form of these collision
integrals in the Appendix~\ref{sec: explicit colint}.

\subsubsection{Treatment of 
Collision Term of Self-Interaction}
In general, the dark scalar has self-interactions.
In principle, we need to solve the Boltzmann equation including the collision term of the self-interactions $\mathcal{C}_\mathrm{self}[f_\phi]$.
In the following, however, we consider two extreme cases to demonstrate the effects of the self-interaction.
In the first case, we completely neglect the self-interaction, and solve the Boltzmann equation in Eq.\,\eqref{eq:Boltzmann} with
$\mathcal{C}_\mathrm{self}[f_\phi] = 0$.
In the other extreme case, we assume that the self-interaction is very strong enough so that the self-thermalization of the dark scalar is always kept.
In the self-thermalized case, we assume that the dark scalar reaches the chemical equilibrium and its evolution 
can be traced by the single parameter $\rho_\phi$ (or $T_\phi)$.
Thus, we solve the Boltzmann equation for the energy density of $\phi$, 
\begin{align}
    \frac{d\rho_\phi}{dt} + 3H(\rho_\phi+P_\phi) 
    = -\int\frac{d^3 \mathbf{p}_\phi}{(2\pi)^3}
    \sqrt{p_\phi^2+m_\phi^2}
    \cdot\mathcal{C}_\mathrm{\phi\leftrightarrow \mathrm{vis}}[f_\phi = f_\phi^{\rm BE}(T_\phi)]\ .
\end{align}
As we discussed in Sec.~\ref{sec: self-thermalization}, however, the self-scattering may become inefficient at the lower temperature.
We will address the impact of the freeze-out of the self-scattering on the cosmological constraints later.

\subsubsection{Evolution of SM Sector}

\begin{figure}[t!]
	\centering
 	\subcaptionbox{$T_R = 10$\,MeV \label{fig:10MeV}}
	{\includegraphics[width=0.47\textwidth]{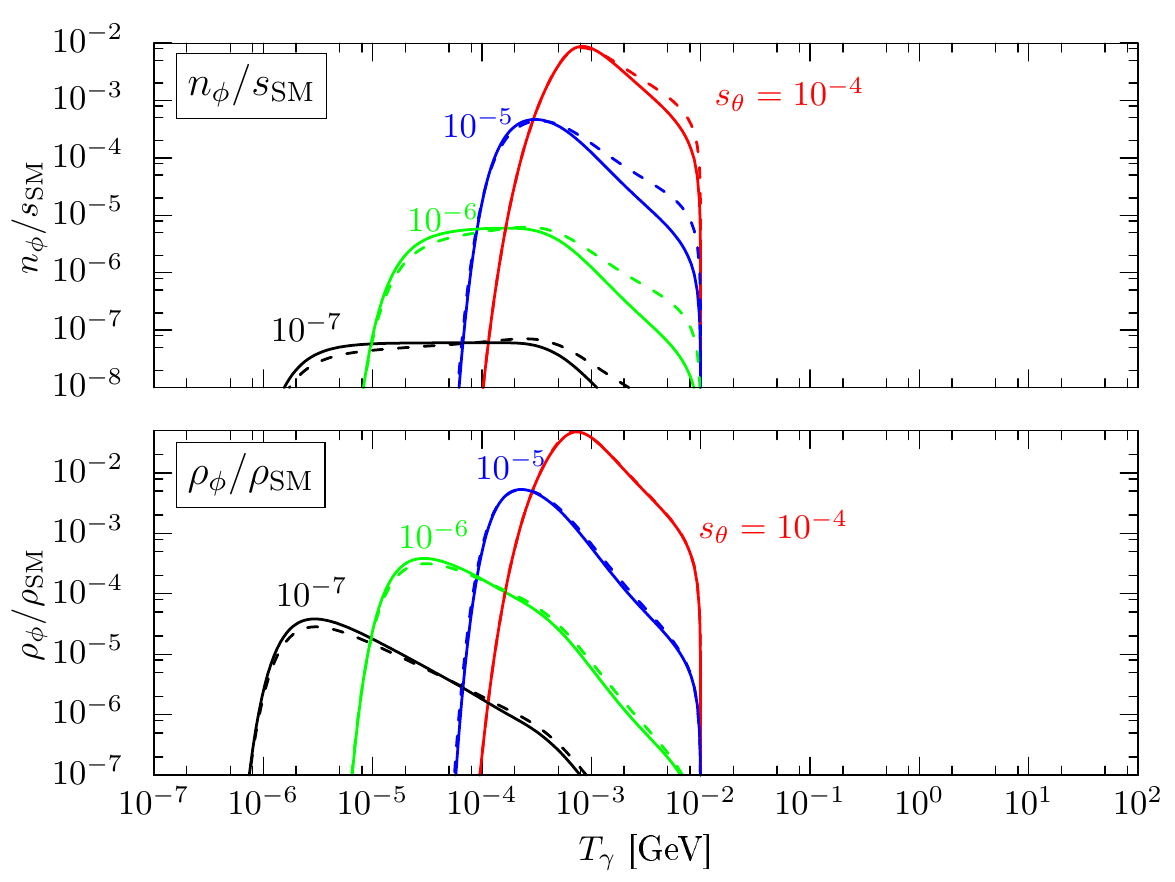}}
 	\subcaptionbox{$T_R = 100$\,MeV\label{fig:100MeV} }
	{\includegraphics[width=0.47\textwidth]{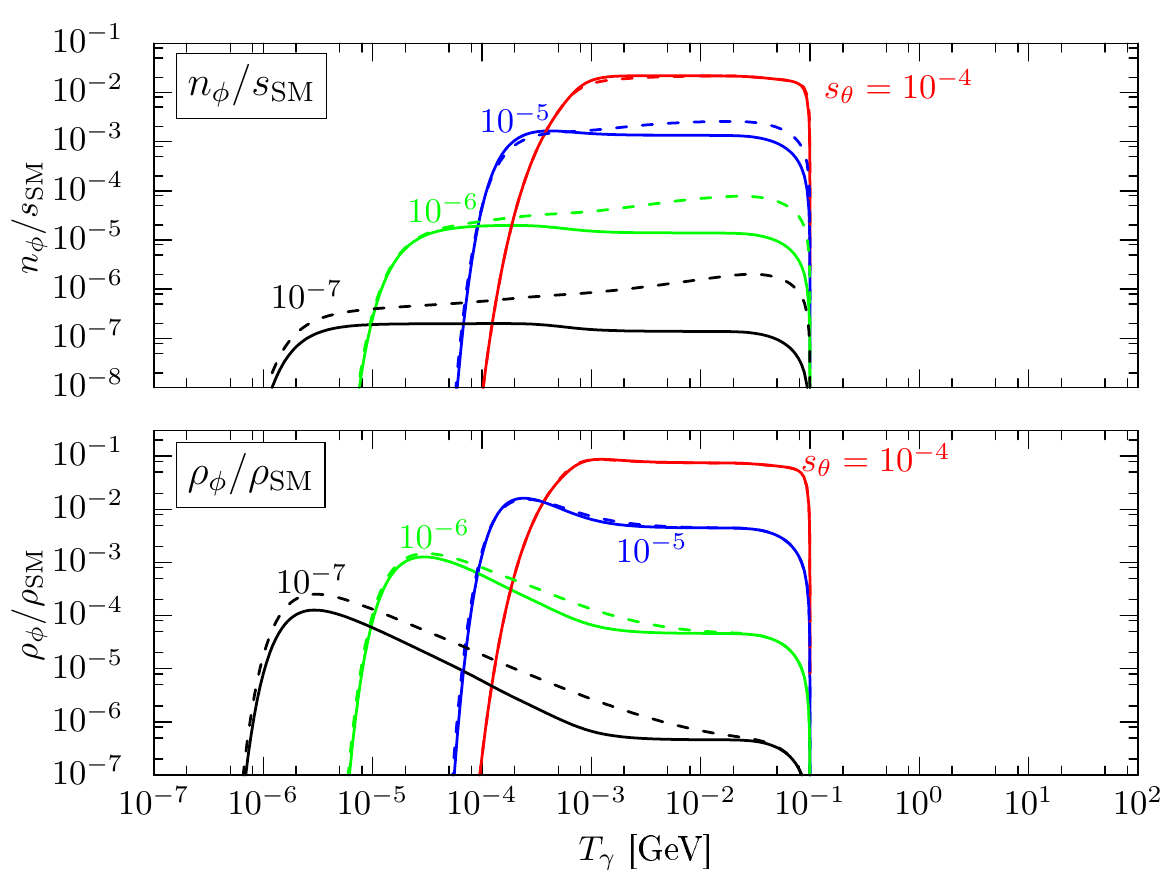}}
	
    \vspace{.4cm}

 	\subcaptionbox{$T_R = 1$\,GeV \label{fig:1GeV} }
	{\includegraphics[width=0.47\textwidth]{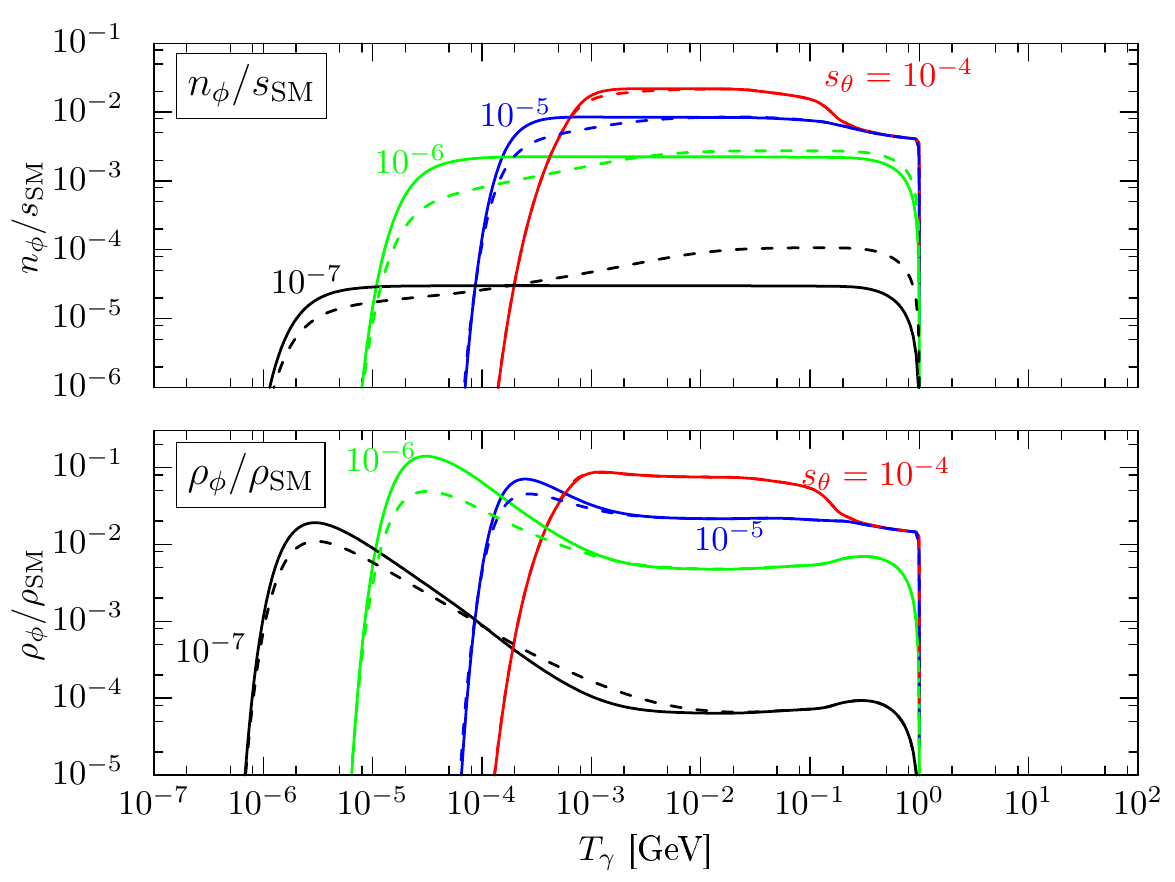}}
 	\subcaptionbox{$T_R = 100$\,GeV \label{fig:100GeV}}
	{\includegraphics[width=0.47\textwidth]{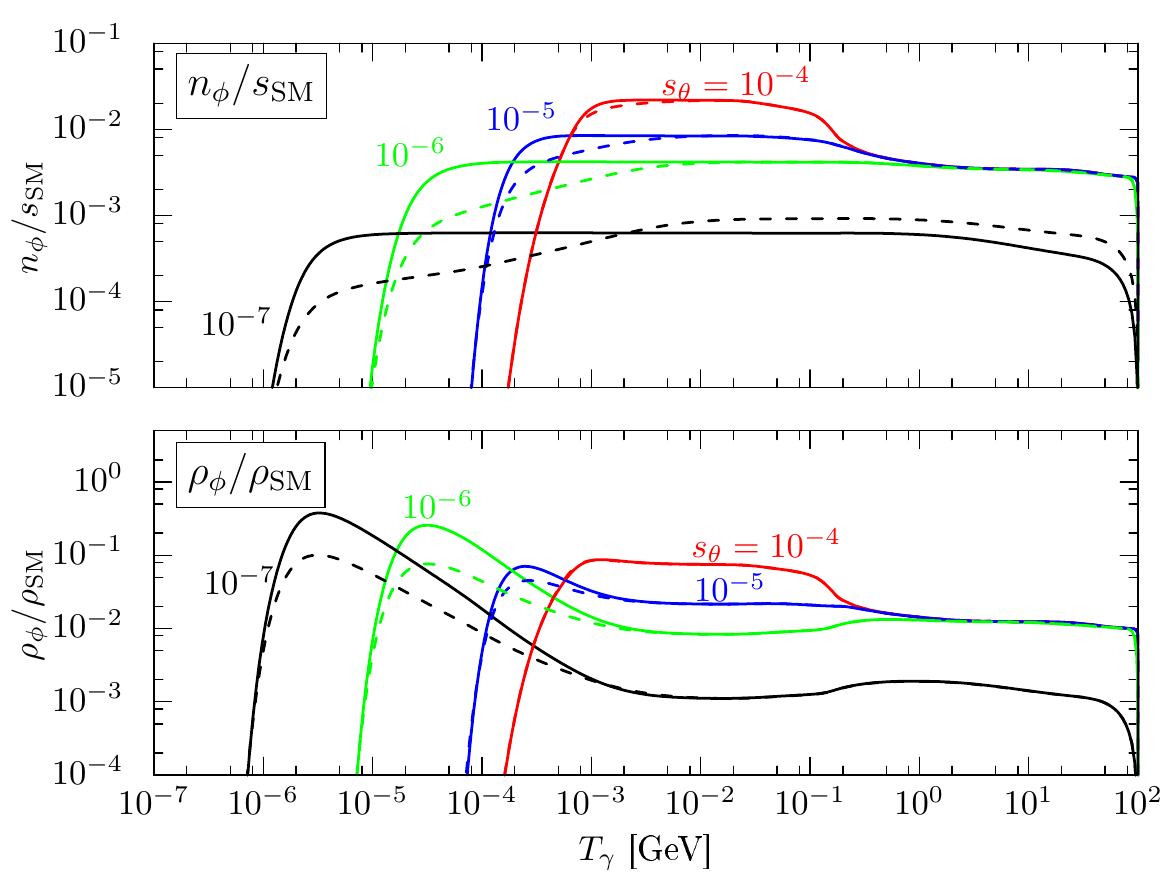}}

\caption{
Time evolution of the number and the energy densities of the dark scalar with $m_{\phi} = 2$\,MeV for the initial temperature $T_R = 10\,\mathrm{MeV}, 100 \,\mathrm{MeV}, 1\,\mathrm{GeV}, 100\,\mathrm{GeV}$.
The dashed (solid) lines show the case of the self-thermalization (no self-interaction). 
}\label{fig:bol}
\end{figure}
The time evolution of the SM sector is given by,
\begin{align}
    \frac{d \rho_\mathrm{vis}}{dt} 
    &= -3H(\rho_\mathrm{vis} + P_\mathrm{vis} ) +   C_{\phi \leftrightarrow \mathrm{vis}
    }(T_{\gamma }) 
    + C^{(1)}_{e\leftrightarrow\nu}
          (T_{\gamma}, T_{\nu}, \mu_{\nu})\  , \\
    \frac{dn_\nu}{dt} &= -3Hn_{\nu} 
         - C^{(0)}_{e\leftrightarrow\nu}(T_{\gamma}, T_{\nu}, \mu_{\nu})
         \ , 
          \label{eq:rho_nu Boltzmann}\\
          \frac{d\rho_\nu}{dt} &= -4H\rho_{\nu} 
          - C^{(1)}_{e\leftrightarrow\nu}
          (T_{\gamma}, T_{\nu}, \mu_{\nu})
         \ ,
      \label{eq:rho_gamma Boltzmann}
\end{align}
where
\begin{align}
     C_{\phi \leftrightarrow \mathrm{vis}
    }(T_{\gamma })
    = \int\frac{d^3 p_\phi}{(2\pi)^3}\sqrt{p_\phi^2+m_\phi^2}
    \cdot\mathcal{C}_\mathrm{\phi\leftrightarrow \mathrm{vis}}[f_\phi]\ .
    \label{eq: energy injection}
\end{align}
Here, $P_\mathrm{vis}$ denotes the pressure of the visible sector.
To treat the neutrino decoupling, we follow Ref.\,\cite{EscuderoAbenza:2020cmq}, which provides the collision integrals, $C_{e\leftrightarrow \nu}^{(0,1)}$.
The neutrino temperature, $T_{\nu}$, is equal to the photon temperature $T_\gamma$ at $T_\gamma \gg \mathcal{O}(1)$\,MeV.
The neutrino chemical potential $\mu_\nu$ vanishes at  $T_\gamma \gg \mathcal{O}(1)$\,MeV.
As for the QED corrections in $\rho_\mathrm{vis}$ and $P_\mathrm{vis}$, 
we use the formula provided in 
Ref.\,\cite{ Heckler:1994tv, Mangano:2001iu, Fornengo:1997wa, Escudero:2018mvt}.
We have used the thermodynamics parameters of Refs.\,\cite{Saikawa:2018rcs,*Saikawa:2020swg} for the electroweak and QCD sectors.

Putting all the Boltzmann equations together, we can compute the cosmological evolution of the dark scalar.
In Fig.\,\ref{fig:bol}, we show the evolution of the number and energy densities of the dark scalar
for different initial temperatures, in the case of $m_\phi = 2\,$MeV. 
The dashed (solid) lines show the case of the self-thermalization (no self-interaction). 
Here we adopt the instantaneous reheating approximation, where the initial condition is that $f_\phi = 0$ and that the SM sector has a temperature $T_R$.

The figure shows that the dark scalar abundance depends on the reheating temperature unlike the case of the simple freeze-in scenario. 
In particular, for a small mixing angle, the abundance becomes larger for the higher reheating temperature.
the abundance is insensitive to $T_R$ for $T_R > \mathcal{O}(100)$\,MeV.
For a large mixing angle, on the other hand,
the dark scalar abundance becomes insensitive to the reheating temperature above some particular temperature.
For instance, for $s_\theta \simeq 10^{-5}$,
This is due to the thermalization with the SM sector at the high temperature.

\begin{figure}[tbp]
\centering{\includegraphics[width=0.7\textwidth]{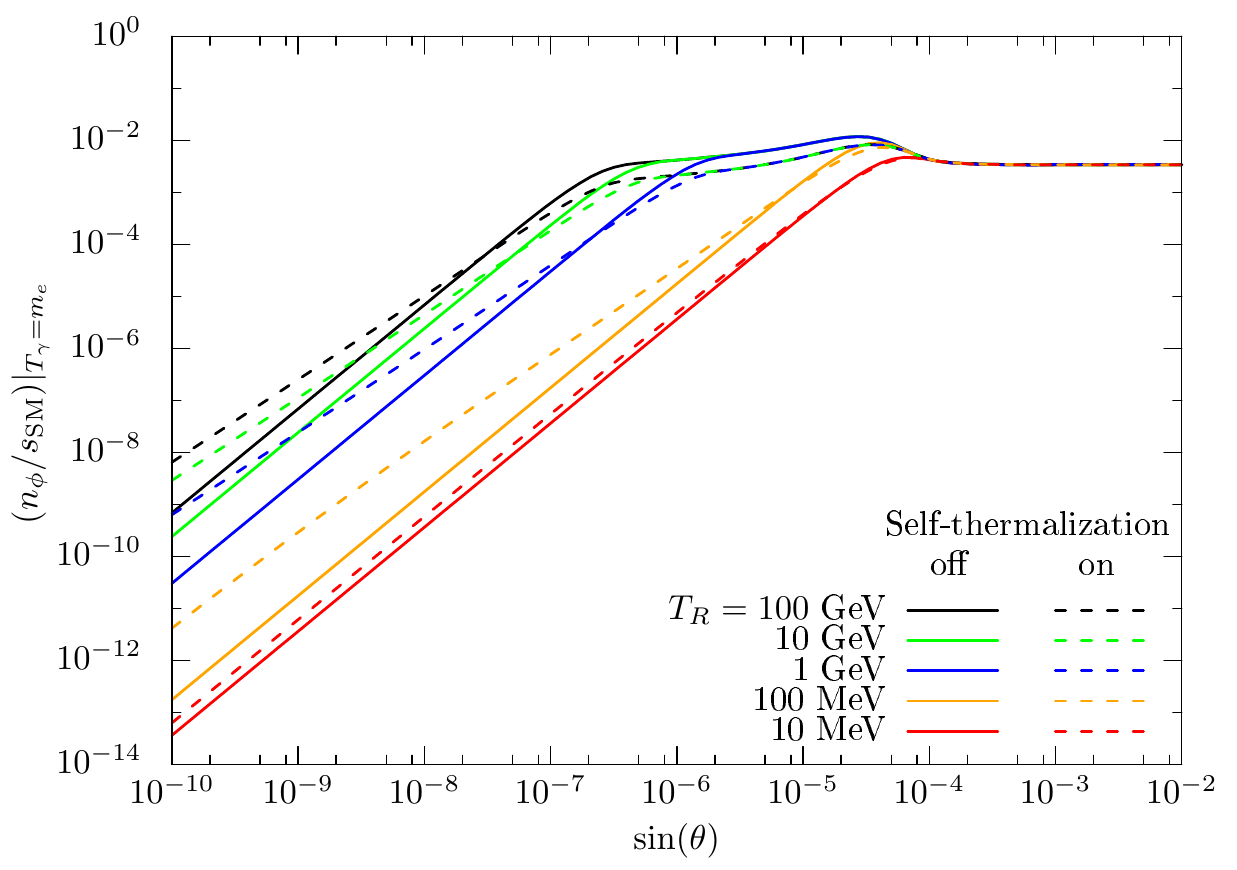}} 
\caption{
The dark scalar number density at $T_\gamma = m_e$ normalized by $s_\mathrm{SM}$.
We take $m_\phi = 2$\,MeV.
}
\label{fig:production}
\end{figure}

In Fig.\,\ref{fig:production}, 
we show the dark scalar number density at $T_\gamma = m_e$ 
 for $m_\phi = 2$\,MeV and various reheating temperature.
The density is normalized by the SM entropy density  $s_\mathrm{SM}$.
The solid lines correspond to the non-interacting cases while 
the dashed ones to the self-thermalized cases.
The figure shows that the dark scalar abundance depends on 
the reheating temperature for a small mixing parameter, $s_\theta$.

\subsubsection{Thermalization and Decoupling of Dark Scalar}
Here, we discuss the thermalization of the dark scalar with the SM sector.
For this purpose, we introduce the following quantity,
\begin{align}
     R(T) \equiv  \frac{ \int\frac{d^3\mathbf{p}_\phi}{(2\pi)^3} I(p_\phi,T) f^{\mathrm{BE}}_\phi(p_\phi,T)}
   { H|_{\rho_\phi=0,\,T_\gamma = T}\times  n^\mathrm{eq}_{\phi} (T) } 
   \ .
    \label{eq: decoupling temperature}
\end{align}
Here, $n^\mathrm{eq}_\phi$ denotes the thermal number density of the dark scalar and $I(p_\phi, T)$ is the summation of the collision integrals 
over all the processes provided in Eq.~\eqref{eq:collision intergrals}.
If $R(T) \gg 1$, the dark scalar $\phi$ is thermalized with the SM sector, and if $R(T) \ll 1$, it is decoupled from the SM.
At the temperature where $R(T)\sim 1$, the dark scalar $\phi$ starts to be either decoupled or thermalized with the SM sector.
We define the UV decoupling temperature, $T_{\mathrm{dec}}$, where $R(T_\mathrm{dec}) = 1$ and $dR(T)/dT|_{T=T_{\mathrm{dec}}} > 0$ are satisfied.
If the temperature is much above the UV decoupling temperature, the dark scalar is, as a rule, thermalized with the SM sector.
This temperature indicates the decoupling of production processes involving heavy particles, such as muons and pions.
Similarly, the freeze-in temperature, $T_{\mathrm{FI}}$, is defined as the temperature where $R(T_{\mathrm{FI}})=1$ and $dR(T)/dT|_{T=T_{\mathrm{FI}}} < 0$ are satisfied.
This indicates the freeze-in of the processes involving light particles, i.e., electrons and photons.\footnote{When this relation has multiple solutions, we take the lowest values 
for $T_\mathrm{dec}$ and $T_\mathrm{FI}$
within $T_\mathrm{dec} > 10$\,MeV and $T_\mathrm{FI} < 10$\,MeV, respectively.}

\begin{figure}[t]
	\centering
	\subcaptionbox{\label{fig:dec_temp} UV decoupling temperature for $m_\phi=2$\,MeV}{\includegraphics[width=0.47\textwidth]{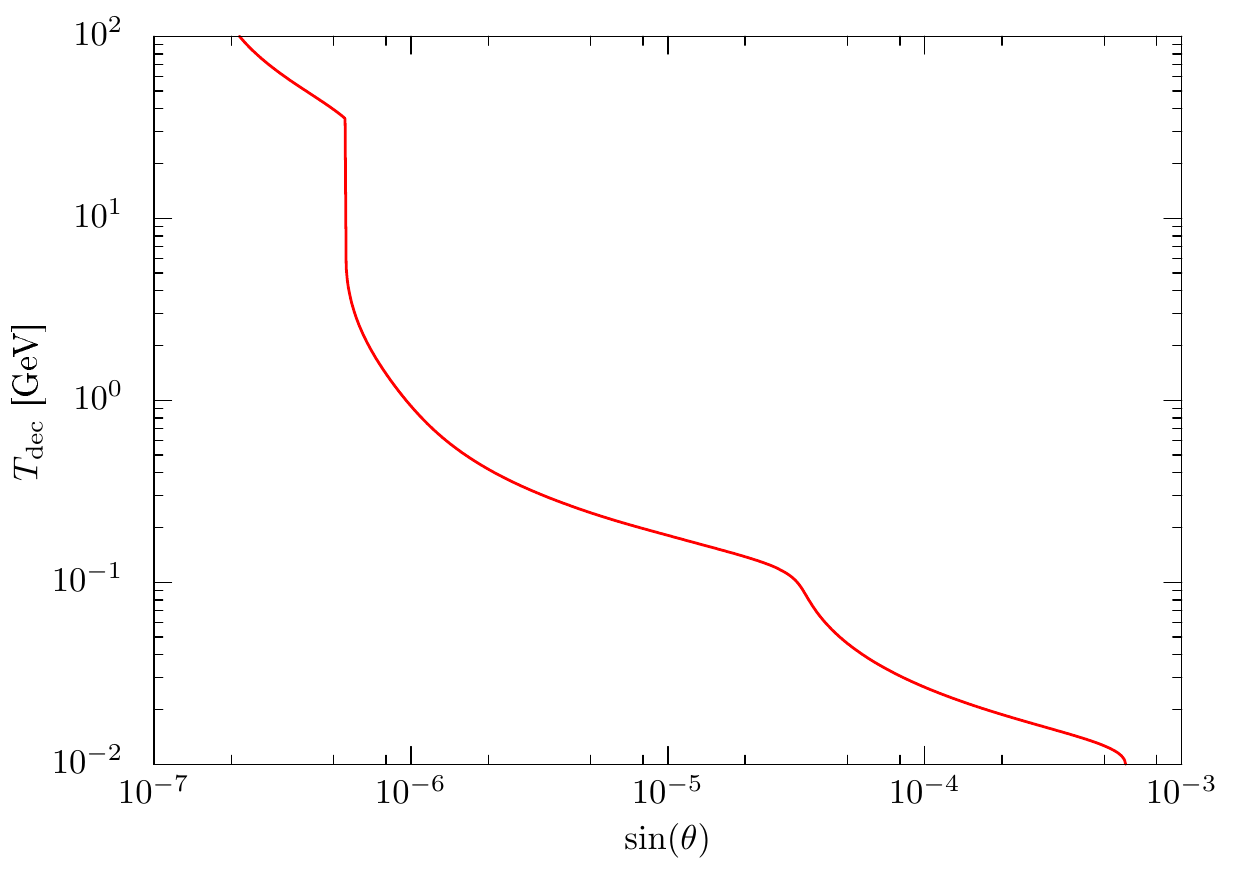}}
	\qquad
	\subcaptionbox{\label{fig:FI_temp}Freeze-in temperature}{\includegraphics[width=0.47\textwidth]{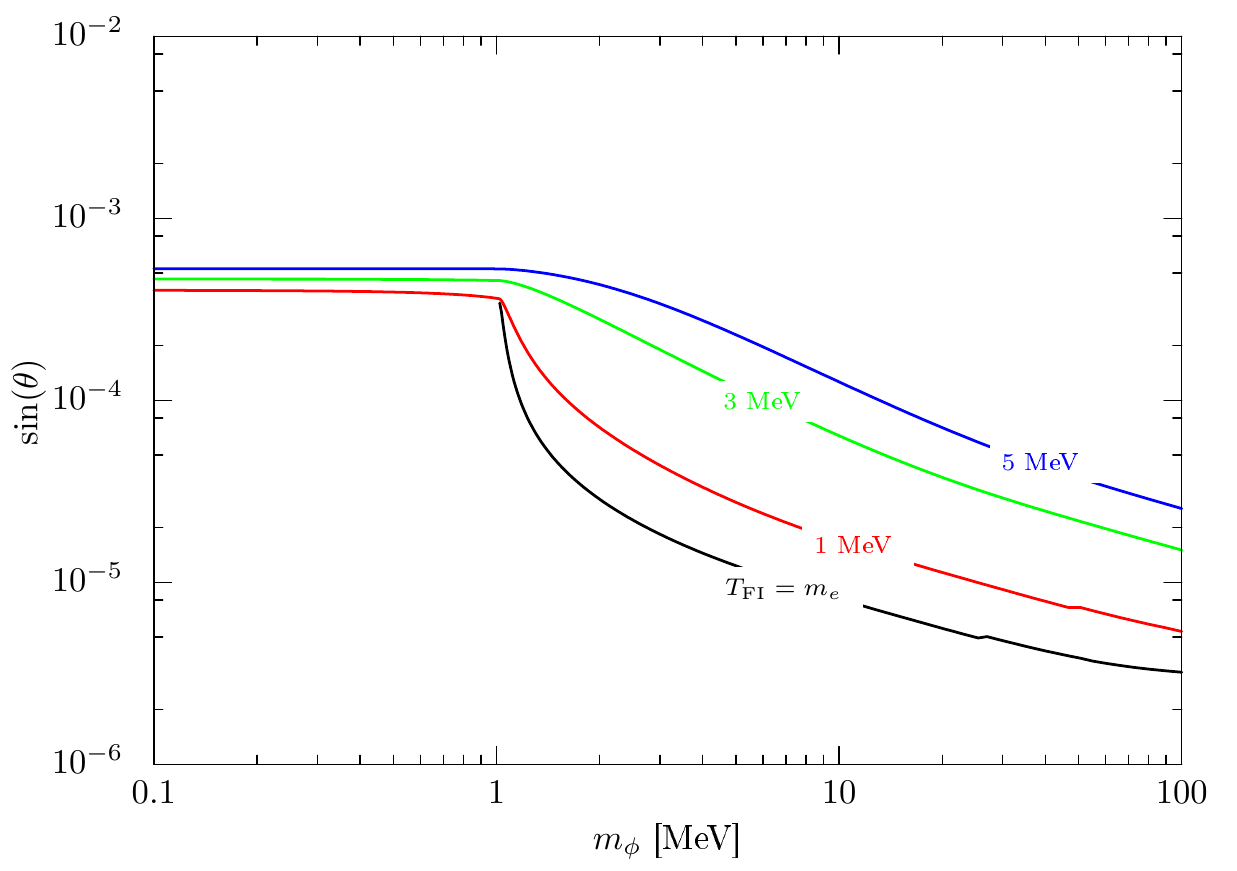}}
\caption{
(a): The UV decoupling temperature of $\phi$ from the SM sector as a function of the mixing angle.
The decoupling temperature hardly depends on the dark scalar mass as far as $ T_\mathrm{dec} \gtrsim m_\phi $.
(b): The contour plot of the freeze-in temperature 
on the $(m_\phi,\sin(\theta))$ plane.
For the definition of the freeze-in temperature, see the text.
}
\label{fig:decoupling freezein}
\end{figure}

In Fig.\,\ref{fig:dec_temp}, we show the UV decoupling temperature as a function of the mixing angle.
The decoupling temperature is almost independent of the dark scalar mass if the decoupling temperature is greater than around $100$\,MeV in the parameter space of our interest, because the thermalization occurs through the ultra-relativistic scattering at high temperature.
In Fig.\,\ref{fig:FI_temp}, we show the freeze-in temperature on the $(m_\phi, s_\theta)$ plane.
For $m_\phi \gtrsim 2 m_e$, the dominant process is the inverse decay of the dark scalar from the electron-positron pair.
The cosmological constraint is insensitive to the reheating temperature above the freeze-in temperature for given $(m_\phi, s_\theta)$.
At the late time, i.e., $T_\gamma < T_\mathrm{FI}$, 
the dark scalar abundance is insensitive 
to the reheating temperature 
as long as $T_R > T_\mathrm{FI}$.
Thus, in the parameter region of $T_\mathrm{FI} \gtrsim 2$\,MeV,
that is $T_\mathrm{FI}$ is higher than the neutrino-decoupling temperature,
the cosmological constraints 
are insensitive to the reheating temperature as long as it is above the freeze-in temperature.

If the reheating temperature is above the decoupling temperature, the dark scalar abundance scarcely depends on the reheating temperature.
In this case, the dark scalar abundance can be approximately given by
\begin{align}
    Y_\phi = \frac{n_\phi}{s_\mathrm{SM}} = \frac{45 \zeta(3)}{2 g_{*s}(T_\mathrm{dec}) \pi^4}\ .
\end{align}
When the decoupling temperature is below $\mathcal{O}(10)$\,GeV, the dark scalar abundance at the lower temperature does not 
depend on the details of the direct interaction to the Higgs boson, which cannot be parameterized by the mixing angle. 

On the other hand, for the small mixing angle, $s_\theta \lesssim 10^{-7}$, the dark scalar is never thermalized via the processes included in our analysis.
However, the abundance may be affected by the 
interactions to the Higgs and the weak gauge bosons which we have not included.
In this sense, our analysis always provides 
conservative constraints.
Our analysis is also robust for the cases 
that $s_\theta \gtrsim 10^{-7}$ or $T_R \ll \mathcal{O}(10)$\,GeV.

\section{Cosmological Constraints}\label{sec:Constraints}
The produced dark scalar $\phi$ has significant impacts on the cosmological history.
The effect on the cosmology strongly depends on the lifetime of the dark scalar.
\begin{enumerate}
    \item $\tau_\phi \lesssim 1 $\,sec\\
    The dark scalar can be thermalized by $T_\gamma = \mathcal{O}(1)$\,MeV.
If the dark scalar mass is less than $\mathcal{O}(1)$\,MeV, the dark scalar modifies the standard neutrino decoupling processes, which alters the effective number of neutrino generations $N_\mathrm{eff}$.
On the other hand, if the mass is much greater than $\mathcal{O}(1)$\,MeV, there are no effect on the standard cosmology with a temperature below $\mathcal{O}(1)$\,MeV, as the cosmic abundance of the dark scalar is strongly suppressed by the Boltzmann factor of $e^{-m_\phi/T}$.
 \item $\tau_\phi \gg 1 $\,sec\\
 If the mixing angle and the mass of the dark scalar $\phi$ are small, the lifetime can be larger than $\mathcal{O}(1)$\,sec.
 For the larger reheating temperature, the dark scalar can be produced at a high temperature.
 After the decoupling from the SM sector, the comoving number density is approximately conserved and the energy density is relatively enhanced compared to the SM sector in the late time Universe.
 The non-thermal contribution of the dark scalar decay affects the cosmic microwave background (CMB) and the BBN observations.
\end{enumerate}

Here, we discuss the CMB and BBN constraints on the dark scalar.

\subsection{CMB Constraints}
\subsubsection*{Effective Number of Relativistic Species: $N_\mathrm{eff}$}
We parameterize the effect of $\phi$ on the neutrino decoupling through the effective number of relativistic species, $N_\mathrm{eff}$, defined by
\begin{align}
\label{eq:Neff}
    N_{\mathrm{eff}} = \left. \frac{8}{7}\left(\frac{11}{4}\right)^{4/3}
\frac{\rho_{\nu}}{\rho_{\gamma}}\right|_{T=T_\mathrm{CMB}}\ .
\end{align}
In the standard cosmology $N^{(\mathrm{SM})}_\mathrm{eff} \simeq 3.044$~\cite{deSalas:2016ztq, Akita:2020szl, Froustey:2020mcq, Bennett:2020zkv}, this value is slightly larger than the case of the instantaneous decoupling, $N_\mathrm{eff} = 3$, because of the entropy transfer into the high-momentum neutrinos. 
We adopt the simplified method discussed in Refs.\,\cite{EscuderoAbenza:2020cmq}.

Since the dark scalar only dominantly couples to the visible sector, the dark scalar heats up only the visible sector,
Therefore the $N_\mathrm{eff}$ becomes smaller than the SM case, in the presence of the dark scalar.

\begin{figure}[tbp]
\centering{\includegraphics[width=0.7\textwidth]{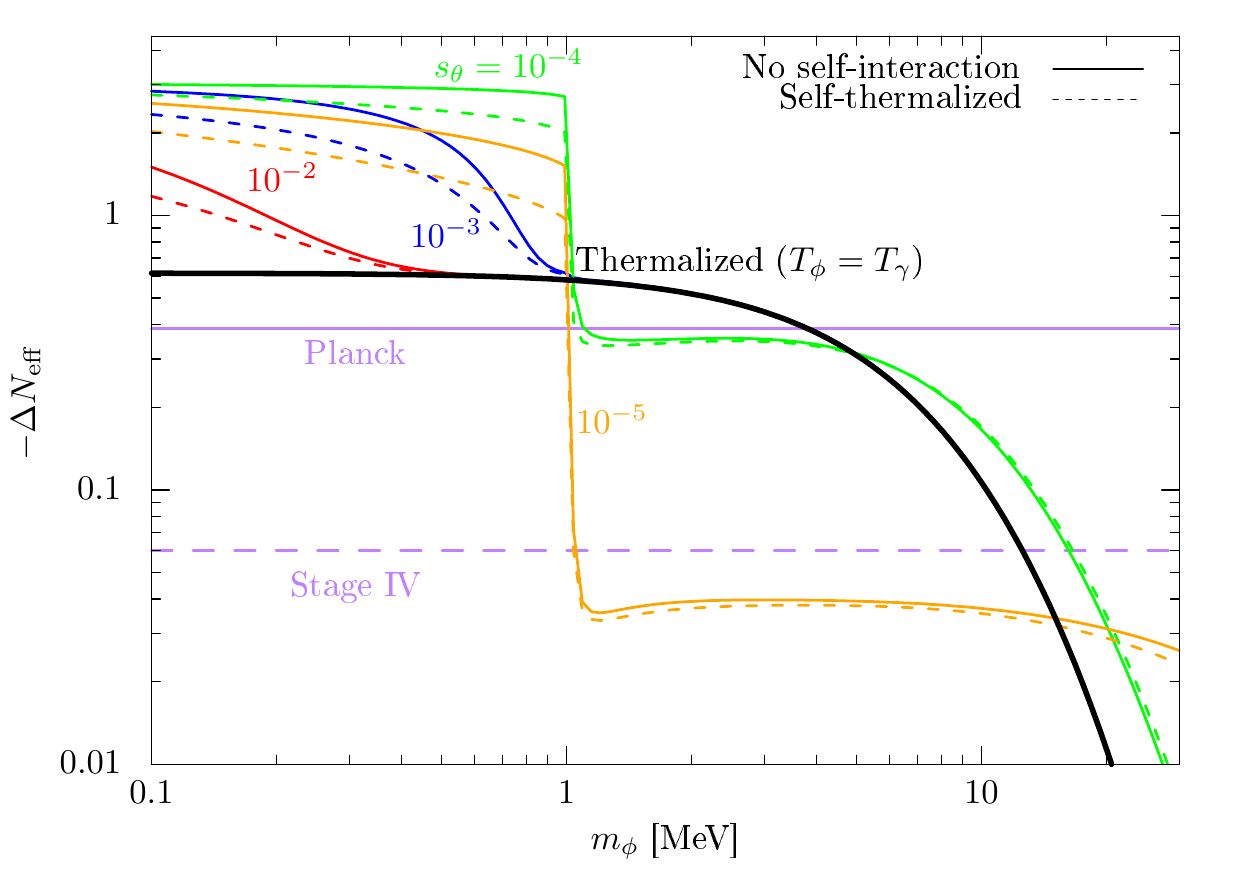}} 
\caption{
The $-\mathit{\Delta} N_\mathrm{eff} =- (N^\phi_\mathrm{eff} - N^\mathrm{SM}_\mathrm{eff})$ as a function of $m_{\phi}$.
Here we take $T_R = 30$\,MeV.
The purple solid line denotes the CMB constraint of $N_\mathrm{eff}>2.66$~\cite{Planck:2018vyg} 
by the Planck experiment
and the purple dashed line indicates the $2\sigma$ sensitivity, $|\mathit{\Delta}N_\mathrm{eff}| = 0.06$, of the CMB stage-IV experiments~\cite{CMB-S4:2016ple}.
}
\label{fig:Neff_thermal}
\end{figure}

In Fig.\,\ref{fig:Neff_thermal}, we show  $N_\mathrm{eff}$ for various values of $s_\theta$ as a function of $m_\phi$ for 
$T_R =30$\,MeV.
The black line shows $N_\mathrm{eff}$ for the case 
that the dark scalar is always thermalized 
with the visible sector.
As expected, the dark scalar reduces $N_\mathrm{eff}$ from the SM case. 
The Planck constraint $N_\mathrm{eff} > 2.66$ at 95\% C.L.~\cite{Planck:2018vyg} (the horizontal purple solid line) indicates the lower mass of the dark scalar is $3.8$\,MeV for the 
thermalized case.
The effect on $N_\mathrm{eff}$ can be larger than the thermalized case, when the lifetime of the dark scalar is longer than $\mathcal{O}(1)$\,sec.

\subsubsection*{Distortion}
For the late-time energy injection from the new physics, the photon background cannot keep thermal equilibrium and deviate from the black body radiation.
The CMB distortion parameters, $y_\mathrm{CMB}$ and $\mu_\mathrm{CMB}$, can be expressed in the following form \cite{Chluba:2011hw,Chluba:2013vsa,Chluba:2013pya,Chluba:2016bvg}
\begin{align}
    y_\mathrm{CMB} & = \frac{1}{4} \int_0^\infty dt \frac{Q(t)}{\rho_\gamma} \mathcal{J}_y (z)\ ,\\
     \mu_\mathrm{CMB} & =1.401 \int_0^\infty dt \frac{Q(t)}{\rho_\gamma} \mathcal{J}_\mu (z)\ .
   \end{align}
Here $1+z = (\rho_\gamma/\rho_{\gamma,0})^{1/4} $ with
$\rho_{\gamma}$ being the photon energy density and the subscript $0$ 
indicating the value of the present Universe. 
The term $Q(t)$ denotes the energy injection from the dark scalar, which is given by the integration of 
Eq.\,\eqref{eq: energy injection}, i.e., $Q(t) = C_{\phi\leftrightarrow \mathrm{vis}}(T_\gamma)$.
We adopt the model C of Ref.\,\cite{Chluba:2016bvg} as the window functions:%
\begin{align}
     \mathcal{J}_y (z) &= \frac{1}{1 + \left(\frac{1+z}{6\times 10^4}\right)^{2.58}} \theta(z - z_\mathrm{rec})\ , \\
     \mathcal{J}_\mu (z) &= e^{-(z/z_\mathrm{th})^{5/2}} 
     \left(
     1 - \exp\left(-\left(\frac{1 + z}{5.8\times10^4}\right)^{1.88}\right)
     \right)\ .
\end{align}
Here, we take $z_\mathrm{th} = 1.98 \times 10^6 $ and $z_\mathrm{rec} = 1000$, respectively.

\begin{figure}[tbp]
\centering{\includegraphics[width=0.7\textwidth]{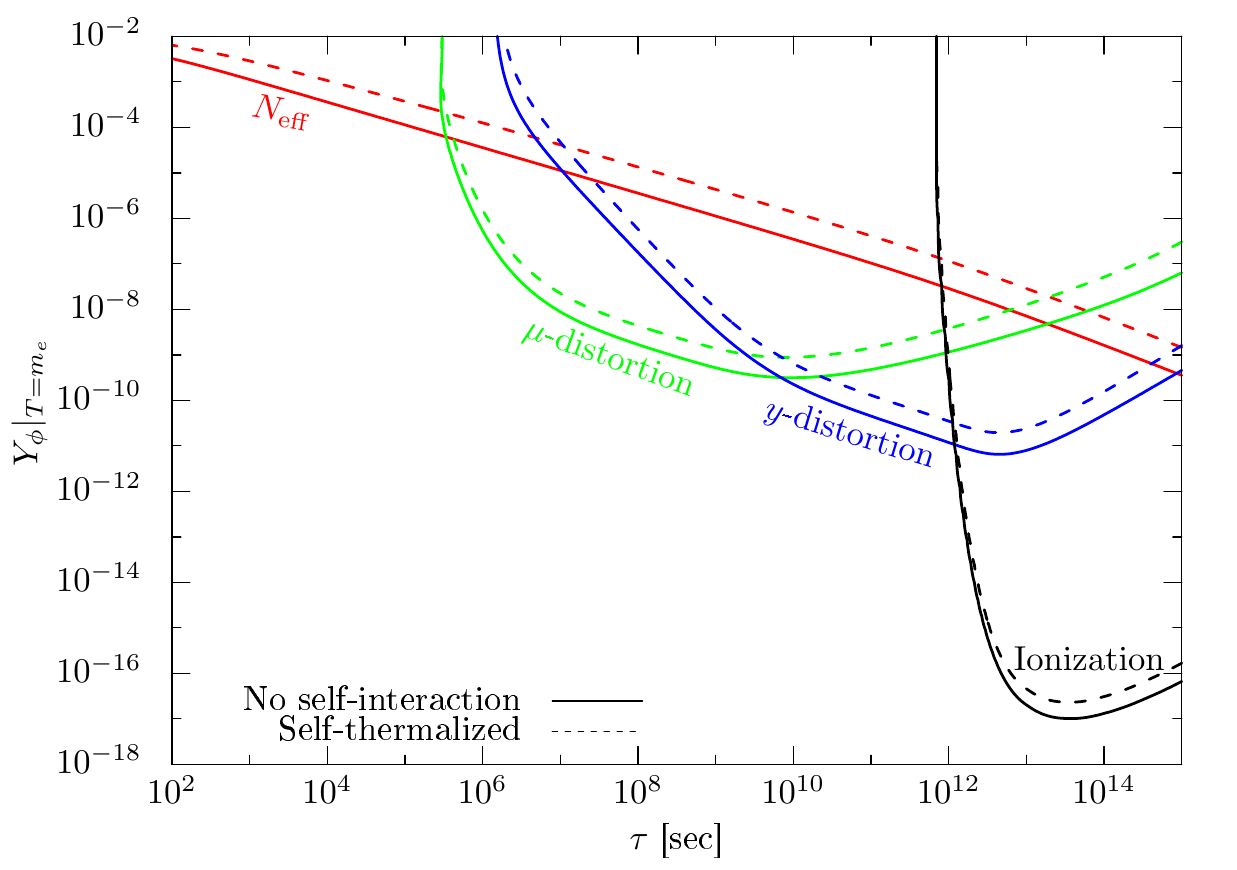}} 
\caption{
The upper limit on the dark scalar abundance 
from the CMB constraints for $m_\phi = 2$\,MeV.
The limits  roughly scale as $m_\phi^{-1}$.
}
\label{fig:Neff_non_thermal}
\end{figure}

The current constraints on the CMB distortion are given by COBE/FIRAS \cite{Fixsen:1996nj}:
\begin{align}
    |\mu_\mathrm{CMB}|_\mathrm{FIRAS} &< 9 \times 10^{-5}\ , \\
    |y_\mathrm{CMB}|_\mathrm{FIRAS} &< 1.5 \times 10^{-5}\ . 
    \label{eq: distortion constraints}
\end{align}
The future prospects of PIXIE \cite{Kogut:2011xw} are     
\begin{align}
    |\mu_\mathrm{CMB}|_\mathrm{PIXIE} &< 10^{-9}\ , \\
    |y_\mathrm{CMB}|_\mathrm{PIXIE} &< 2 \times 10^{-9}\ . 
\end{align}

\subsubsection*{Ionization}

The late time decay of the dark scalar after the recombination era, also affects the ionization history of the Universe, which leads to the modification of the CMB power spectrum~\cite{Chen:2003gz}. 
We study the ionization constraint with the method used in Ref.\,\cite{Slatyer:2016qyl}.
The constraint on the lifetime of the decaying dark scalar is approximately given by
\begin{align}
    \label{eq:ionization constraint}
    \tau \gtrsim f_X g_{\mathrm{eff}}(\tau) \times 2.6 \times 10^{25} \, \mathrm{sec}\, ,
\end{align}
where $f_X$ is the energy fraction of the decaying matter $X$.
The energy absorption efficiency function $g_{\mathrm{eff}}$ is provided in Ref.\,\cite{Slatyer:2016qyl}, which depends on the energy injection into the visible sector, the lifetime of the dark scalar, $\tau$, and the decay mode. 
The upper bound on $f_X$ is evaluated from the above inequality as
    \begin{align}
        \frac{1}{\tau_0} = \frac{g_{\mathrm{eff}}(\tau)}{\tau}f_X\, ,
    \end{align}
where $\tau_0 = 2.6\times 10^{25}$\, sec.

We note two points.
First, we evaluate $f_X$ just before the decay of the dark scalar. 
Thus, when we evaluate the ionization limit by using the abundance obtained by solving the Boltzmann equation, we multiply a rescale factor, $\exp(\Gamma_\phi t)|_{H=\Gamma_\phi} = e^{1/2}$, which accounts the reduction of energy density from the onset of the decay.
Second, the range of the table which we use to evaluate $g_{\mathrm{eff}}$ is $\tau_{\mathrm{min}}=10^{13}\, \mathrm{sec} < \tau < 10^{26} \, \mathrm{sec} = \tau_{\mathrm{max}}$.
When we evaluate the bound on $f_X$ in $\tau < 10^{13}\, \mathrm{sec}$ we extrapolate $g_{\mathrm{eff}}$ assuming the lower bound on $f_X$ is determined by 
    \begin{align}
        \frac{1}{\tau_0} = \frac{C(\tau_{\mathrm{min}})}{\tau}\exp(-t_*/\tau)f_X
    \end{align}
where $t_*$ is a parameter added by hand and $C(\tau_{\mathrm{min}})$ is a numerical constant calculated at $\tau = \tau_{\mathrm{min}}$.
Here, we determine the parameter $t_*$ and $C(\tau_{\mathrm{min}})$ by the smooth and continuous condition 
\begin{align}
    \left. \frac{g_{\mathrm{eff}}(\tau)}{\tau} \right|_{\tau = \tau_{\mathrm{min}}} &= \left. \frac{C(\tau_{\mathrm{min}})}{\tau}\exp(-t_*/\tau)\right|_{\tau = \tau_{\mathrm{min}}}\ ,\notag \\
    \left.\frac{d}{d\tau}\frac{g_{\mathrm{eff}}(\tau)}{\tau}\right|_{\tau = \tau_{\mathrm{min}}} &= \left.\frac{d}{d\tau}\frac{C(\tau_{\mathrm{min}})}{\tau}\exp(-t_*/\tau)\right|_{\tau = \tau_{\mathrm{min}}}\, .
\end{align}

In Fig.\,\ref{fig:Neff_non_thermal},
we show the CMB constraints on the
abundance of the long-lived dark scalar as a function of the lifetime $\tau$ for $m_\phi = 2$\,MeV. 
The solid lines are the constraints 
for the non-interacting case and 
the dashed lines are for the self-thermalized case.
Here, we neglect the scattering
of the dark scalar against the SM sector below $T = m_e$.

\subsection{BBN  Constraints}
The dark scalar also modifies the Standard BBN (SBBN) scenario.
As the current observation is consistent with the prediction of the SBBN, the contribution from the dark scalar can be constrained.

\subsubsection{Standard BBN}
Let us discuss the SBBN case.
To get the prediction of the SBBN, we make use of the program AlterBBN 2.2 \cite{Arbey:2011nf, Arbey:2018zfh} with slight modifications.
We modified the reaction rates of Deutron  processes d(p,$\gamma$)${}^3$He \cite{Tisma:2019acf,Mossa:2020gjc}, d(d,n)${}^3$He and d(d,p)${}^3$H \cite{Coc:2015bhi} to include the latest experimental results and correct typos in the original code.
We adopt the fitting functions given in Ref.\,\cite{Gariazzo:2021iiu}, with modifying the lower temperature region to remove numerical instability.
We adopt the following SBBN inputs: the neutron lifetime $879.4 \pm 0.6$\,sec~\cite{Zyla:2020zbs},  the baryon-to-photon ratio
$\eta_\mathrm{CMB}  = (6.105 \pm 0.055)\times 10^{-10}$\,\cite{Planck:2018vyg} and $N_\mathrm{eff} = 3.045$, and the SBBN prediction for the primordial mass fraction of ${}^4$He, $Y_\mathrm{p}$, and the ratios of primordial number density of D and ${}^3$He to hydrogen are
\begin{align}
    \left. Y_\mathrm{p} \right|_\mathrm{SBBN}= 0.2471 \pm 0.0029,~~\left.\frac{\mathrm{D}}{\mathrm{H}} \right|_\mathrm{SBBN}= (2.537 \pm 0.045)\times 10^{-5} ,~~\left.\frac{{}^3\mathrm{He}}{\mathrm{H}} \right|_\mathrm{SBBN}= (1.039 \pm 0.014) \times 10^{-5}.
\end{align}
Here the errors represent the sum of uncertainties from the neutron lifetime, nuclear reaction rates and the baryon-to-photon ratio $\eta_\mathrm{CMB}$, and we assume that the primordial $\eta_\mathrm{BBN} =\eta_\mathrm{CMB} $.
The predicted values are consistent with current observations:
\begin{align}
    \left. Y_\mathrm{p} \right|_\mathrm{obs}= 0.245 \pm 0.003,~~\left.\frac{\mathrm{D}}{\mathrm{H}} \right|_\mathrm{obs}= (2.547 \pm 0.025)\times 10^{-5} ,~~\left.\frac{{}^3\mathrm{He}}{\mathrm{H}} \right|_\mathrm{obs} < (1.1 \pm 0.2) \times 10^{-5}.
\end{align}
In the following analysis, we adopt $\chi^2$ analysis based on these three observations.
We define the $\chi^2$ variable as:
\begin{align}
    \chi^2 = \sum_{i,j} \delta x_j R^{-1}_{ij} \delta x_i,
\end{align}
where $R_{ij}$ is a covariant matrix including theory and observation errors,  $\delta x_1 =  Y^\mathrm{theory}_\mathrm{p} -  Y^\mathrm{obs}_\mathrm{p}$,
$\delta x_2 =  (\mathrm{D}/\mathrm{H})^\mathrm{theory}- (\mathrm{D}/\mathrm{H})^\mathrm{obs}$,
and $\delta x_3 =  \max (0, ({}^3\mathrm{He}/\mathrm{H})^\mathrm{theory}- ({}^3\mathrm{He}/\mathrm{H})^\mathrm{obs})$.

By using this likelihood analysis, we can estimate the baryon-to-photon ratio in the BBN era for the standard BBN case as
\begin{align}
    \eta^\mathrm{SBBN}_\mathrm{BBN} = (6.089 \pm 0.056) \times 10^{-10},
\end{align}
which is mainly determined with $\mathrm{D}/\mathrm{H}$.
This estimate is consistent with the more detailed analysis done in Ref.\,\cite{Yeh:2020mgl}.

\subsubsection{Effect from Dark Scalar}
In this work, we add the new scalar $\phi$ of which mass is less than $\mathcal{O}(10)$\,MeV.
The dark scalar has a significant impact on the prediction of the BBN for various aspects. 
An important contribution is the modification of $N_\mathrm{eff}$.
The $N_\mathrm{eff}$ modification at the BBN era and/or the energy density of the $\phi$ change the Hubble expansion rate at the BBN era and alter the SBBN prediction.
In addition, if $\phi$ decays after the BBN era, it provides additional entropy and reduces the baryon asymmetry at the CMB era, i.e., $\eta_\mathrm{BBN} > \eta_\mathrm{CMB}$.
Moreover, the late time decay of $\phi$ provides high energy particles. 
Such particles directly interact with the nucleons and alter the SBBN predictions.
For the particle of a mass  $\mathcal{O}(10)$\,MeV and lifetime greater than $\mathcal{O}(10^3)$\,sec, the photodissociation processes mainly affect the BBN.

We apply the photodissociation processes to the elements H, D, T, ${}^3$He and ${}^4$He for $T_\gamma < 10 $\,keV, well after the SBBN freeze-out. 
We adopt the fitting formula for the photodissociation rates of Ref.\,\cite{Cyburt:2002uv}.
We obey the procedure of Ref.\,\cite{Forestell:2018txr} for the electromagnetic cascade after the $\phi$ decay.
We consider the $\phi \to e^+e^-, \gamma\gamma$ decays and the final state radiation processes $e^+e^-\gamma$.
As for the final state radiation, we directly estimate the matrix element of the three-body decay processes, instead of the Altarelli-Parisi approximation formula used in Refs.\,\cite{Forestell:2018txr,Hufnagel:2018bjp}.

\begin{figure}[t]
\centering{\includegraphics[width=0.7\textwidth]{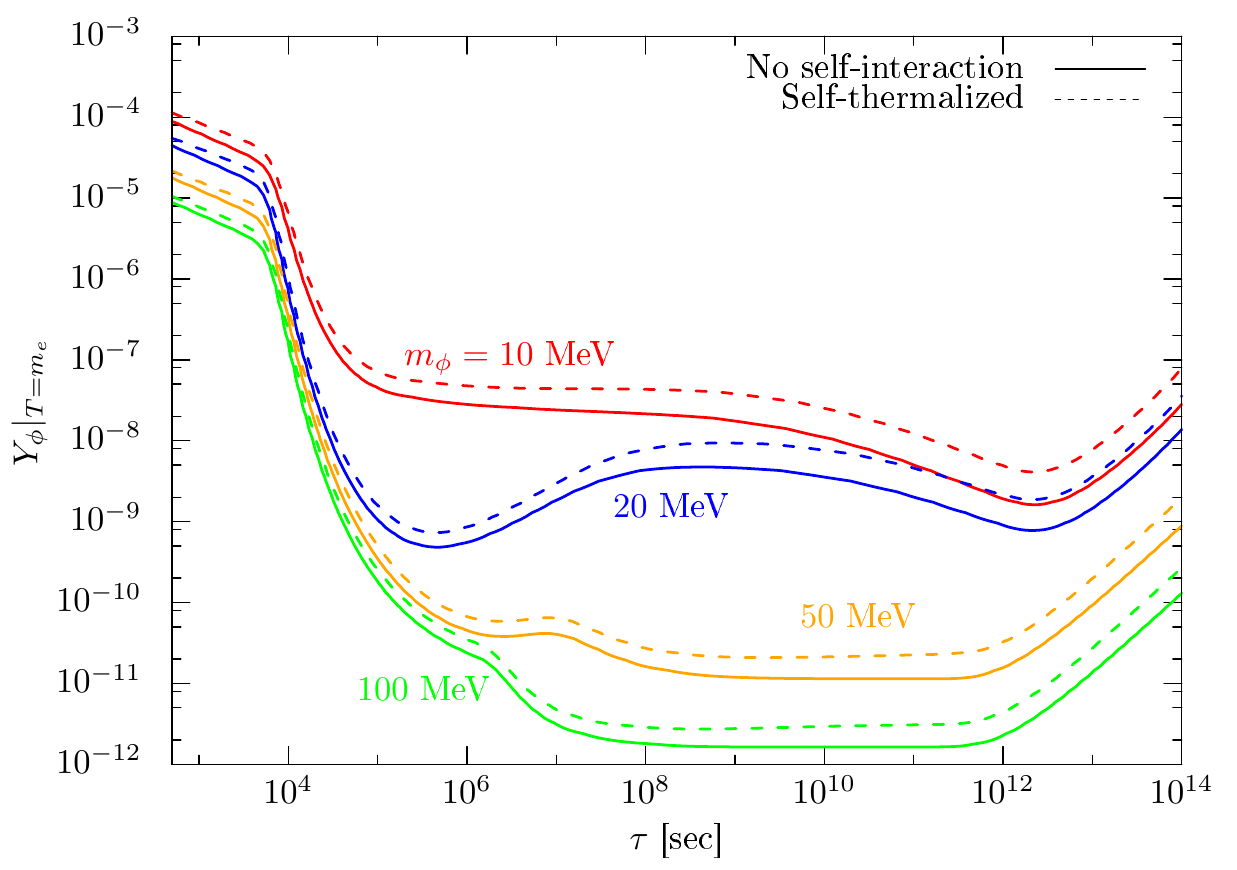}} 
\caption{
The upper limit on the abundance of the dark scalar at $T = m_e$ from BBN constraints.
We plot the bounds as functions of the lifetime of the dark scalar. 
}
\label{fig:BBN_example}
\end{figure}

In our analysis, we apply the BBN constraint for the dark scalar of which lifetime is longer than $500$\,sec.
In this case, only the late-time entropy production, which modifies $\eta$, and
the photodissociation affects the BBN.
In Fig.\,\ref{fig:BBN_example}, we show the BBN constraint for dark scalar abundance and lifetime.

\subsection{Effect of Freeze-out of Self-thermalization}
\begin{figure}[t]
\centering{\includegraphics[width=0.7\textwidth]{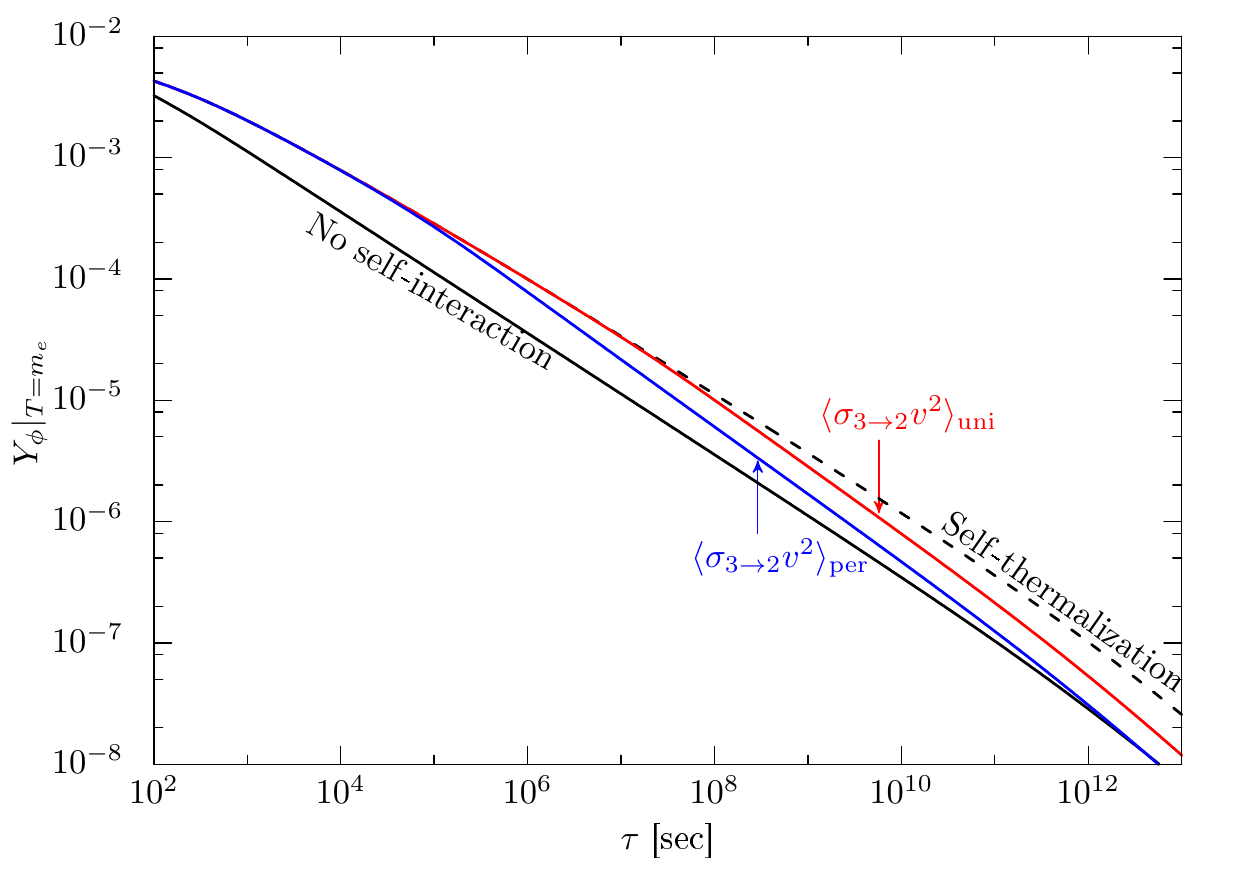}} 
\caption{The upper limits on the dark scalar abundance 
from the $N_\mathrm{eff}$ constraint 
for different choices of the self-scattering cross sections, $\langle\sigma_{3\to 2}v^2\rangle$. 
Here, we take $m_\phi = 2\,\mathrm{MeV}$.
This figure provides 
the closeup of the $N_\mathrm{eff}$ constraint in Fig.\,\ref{fig:Neff_non_thermal}.
}
\label{fig:Neff_cross}
\end{figure}

In the above discussion, it is assumed that the self-thermalization of the dark scalar is always maintained.
However, due to the finite cross section of the self-scattering, when the number density drops  as the Universe expands, the chemical equilibrium cannot be maintained.
As we have discussed, the limit is affected by the self-thermalization effect.
In fact, due to this freeze-out of self-thermalization (see Eq.\,\eqref{eq: 3-2 freeze out}), the constraint becomes severer, once we consider a realistic cross section of the self-scattering.
This phenomenon strongly affects the limit for long lifetimes and small number densities.
In Fig.\,\ref{fig:Neff_cross}, we show the $N_\mathrm{eff}$ constraint for several scattering cross sections, which provides the closeup of the 
$N_\mathrm{eff}$ constraint in Fig.\,\ref{fig:Neff_non_thermal}.
The black and solid line shows the case that there is no self-scattering of the dark scalar, which gives the strongest constraint.
The black and dashed line represents the case that the self-thermalization is always maintained.
The blue and red lines show the cases of the perturbative cross section in Eq.\,\eqref{eq:cross_per} with $\mu_\phi=m_\phi$ and $\lambda_s = 0$ and $s$-wave unitarity in Eq.\,\eqref{eq:cross_uni}, respectively.
In order to include the freeze-out effect, we simply set the EOS parameter $w=0$ when the freeze-out condition Eq.\,\eqref{eq: 3-2 freeze out} is satisfied. 

\begin{figure}[tbp]
\centering{\includegraphics[width=0.7\textwidth]{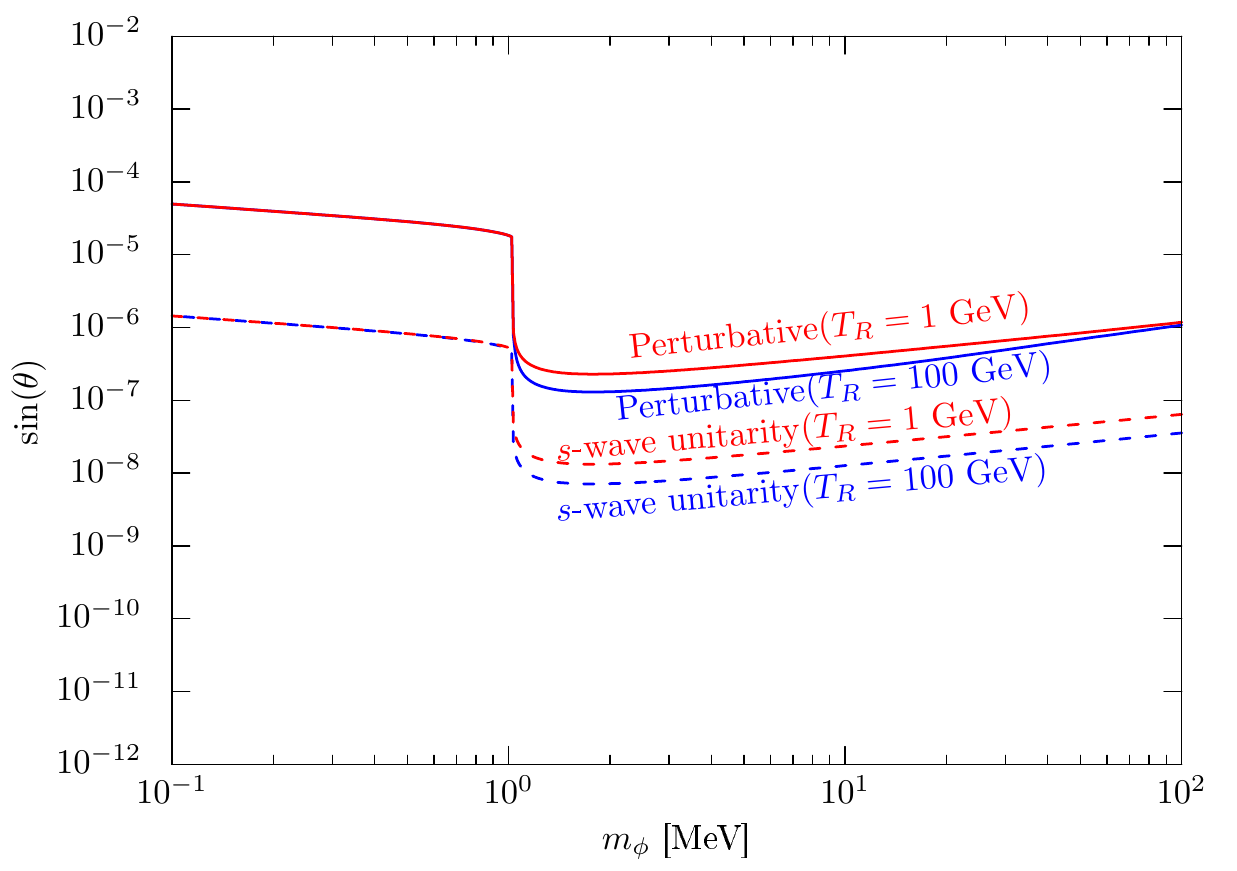}} 
\caption{
The parameter region in which the self-thermalization effect is valid.
Above the lines the self-thermalization is maintained
until the dark scalar decay time or the CMB era.
}
\label{fig:thermalization_bound}
\end{figure}

In Fig.\,\ref{fig:thermalization_bound}, we show the parameter region  where the self-thermalization is kept until the decay of the dark scalar.
Here, we again consider the perturbative cross section in Eq.\,\eqref{eq:cross_per} with $\mu_\phi=m_\phi$ and $\lambda_s = 0$ (solid lines) and $s$-wave unitarity in Eq.\,\eqref{eq:cross_uni} (dashed lines), respectively.
The self-thermalization does not freeze-out before the dark scalar decay or the CMB era in the regions above the lines.
For the smaller abundance or smaller mixing angle, the self-thermalization is unlikely maintained.
Therefore, if we consider the realistic cross section for the self-scattering, the cosmological constraint gets severer than the case that the self-thermalization is always realized.
In the following analysis, to get conservative constraints, we consider the case that the self-thermalization is always maintained.

\subsection{Result}
Finally, we show the cosmological constraints on the parameter region of the dark scalar.
Fig.\,\ref{fig:cosmological constraints} shows the constraints for different reheating temperatures.
The left columns represent the non-interacting case and right columns  the self-thermalized case.
For the self-thermalized case, we assume that the self-thermalization is always maintained.
For each case, we take the reheating temperature $T_R = 5$\,MeV
to $T_R =100$\,GeV from top to bottom row.
The contour lines show the $N_\mathrm{eff}$.
The red region is excluded by the CMB measurement of the $N_\mathrm{eff}$, $N_\mathrm{eff} = 2.99^{+0.34}_{-0.33}$~\cite{Planck:2018vyg}.
The CMB measurement also casts the $y$-distortion (yellow) and $\mu$-distortion (blue) constraints (Eq.\,\eqref{eq: distortion constraints}).
The light purple region is excluded by ionization after the recombination era, Eq.\,\eqref{eq:ionization constraint}.
The orange region is excluded by the X-ray constraints on the decaying dark matter~\cite{Essig:2013goa}.
We also show the BBN constraints as the green shaded region.

\begin{figure}[t]
	\centering
	\subcaptionbox{\label{fig:TR_constraint}No self-interaction.}{\includegraphics[width=0.47\textwidth]{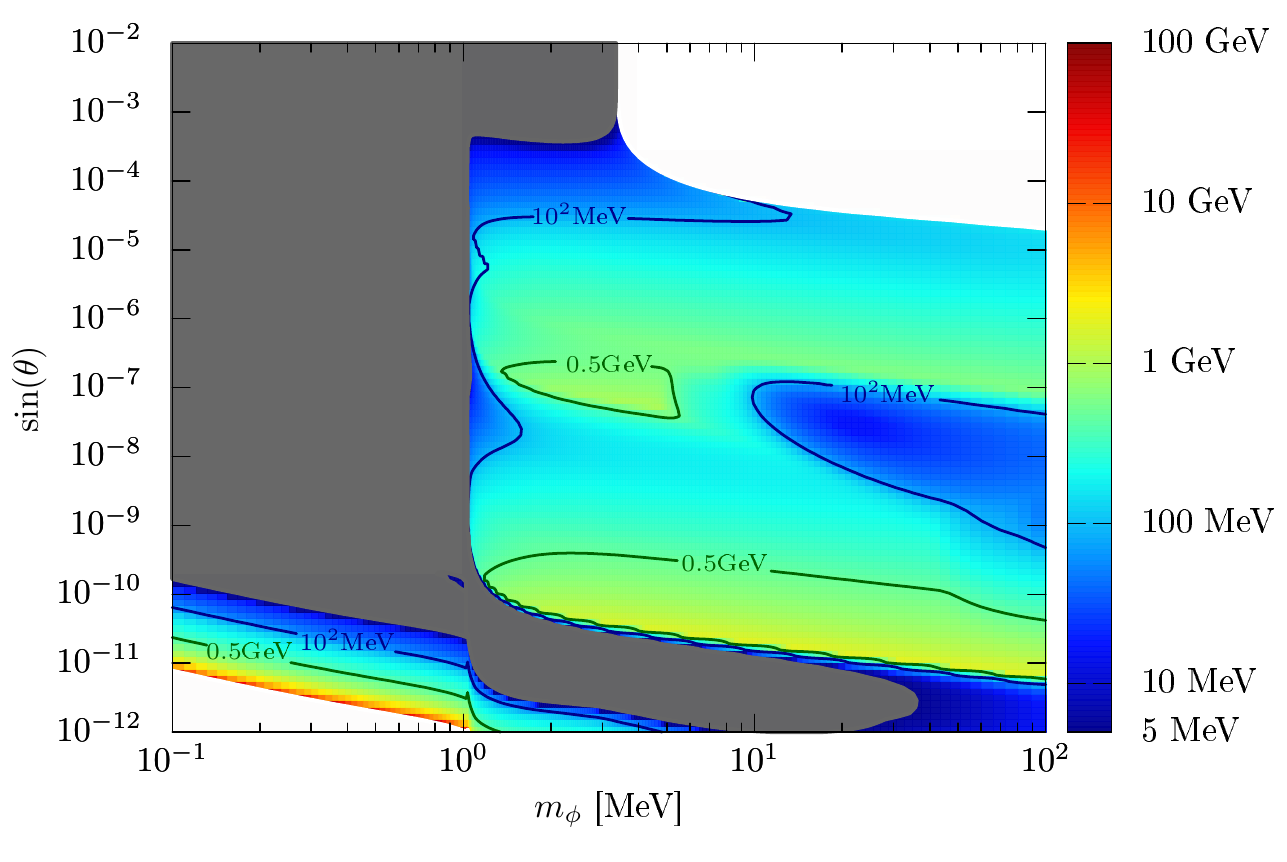}}
	\qquad
	\subcaptionbox{\label{fig:TR_constraint_self}Self-thermalization.}{\includegraphics[width=0.47\textwidth]{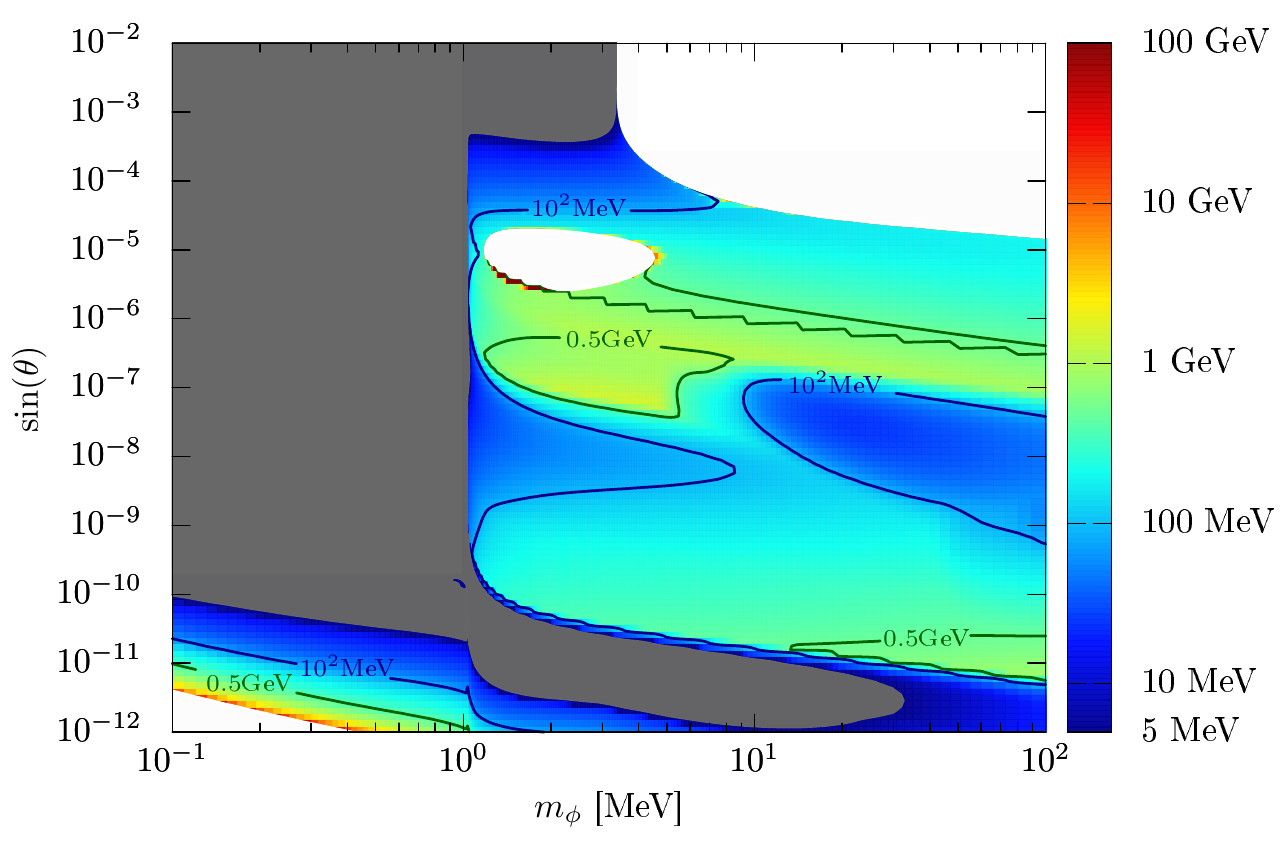}}
\caption{ 
The upper-bound on the reheating temperature from the cosmological constraints.
The gray regions are excluded for an arbitrary reheating temperature.
The white regions, on the other hand, are compatible with the cosmological constraints for an arbitrarily high reheating temperature.
}
\label{fig:reheat}
\end{figure}

In Fig.\,\ref{fig:reheat}, we show the constraint on the reheating temperature on the ($m_\phi$, $\sin(\theta)$) plane.
In the gray regions the upper bound on the reheating temperature is less than $5$\,MeV, which is in tension with the BBN.
On the other hand, the white regions are consistent with the cosmological constraints for any reheating temperature.
Note that for the self-thermalization case, there is a ``blind spot" for the dark scalar with mass $m_\phi$ around $2$\,MeV and the mixing angle $\sin(\theta)$ around $10^{-5}$.
In this region, $N_\mathrm{eff} \simeq 2.7$ and the lifetime of the dark scalar is around $\mathcal{O}(1)$\,min, which is marginally consistent with all the cosmological constraints in our analysis.
In this region, the decoupling temperature is around 
$100$\,MeV and therefore the constraint is valid even if the reheating temperature is much greater than the electroweak scale.
Moreover, as seen in Fig.\,\ref{fig:thermalization_bound}, the self-thermalization is always maintained before the dark scalar decay even for the perturbative cross section.
Therefore, our approximation for the self-scattering effect is also valid.

\begin{figure}[tbp]
	\centering

	{\includegraphics[width=0.45\textwidth]{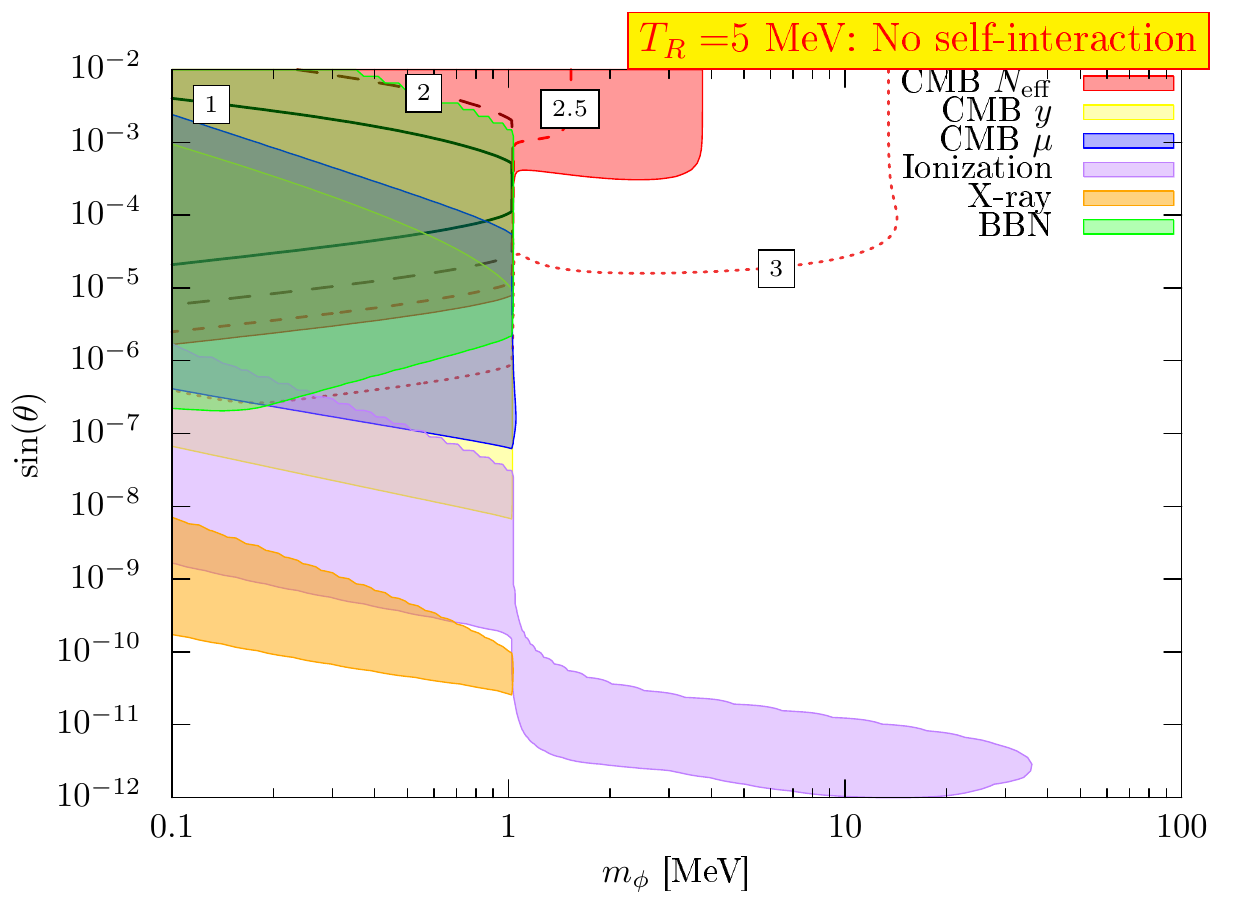}}
	{\includegraphics[width=0.45\textwidth]{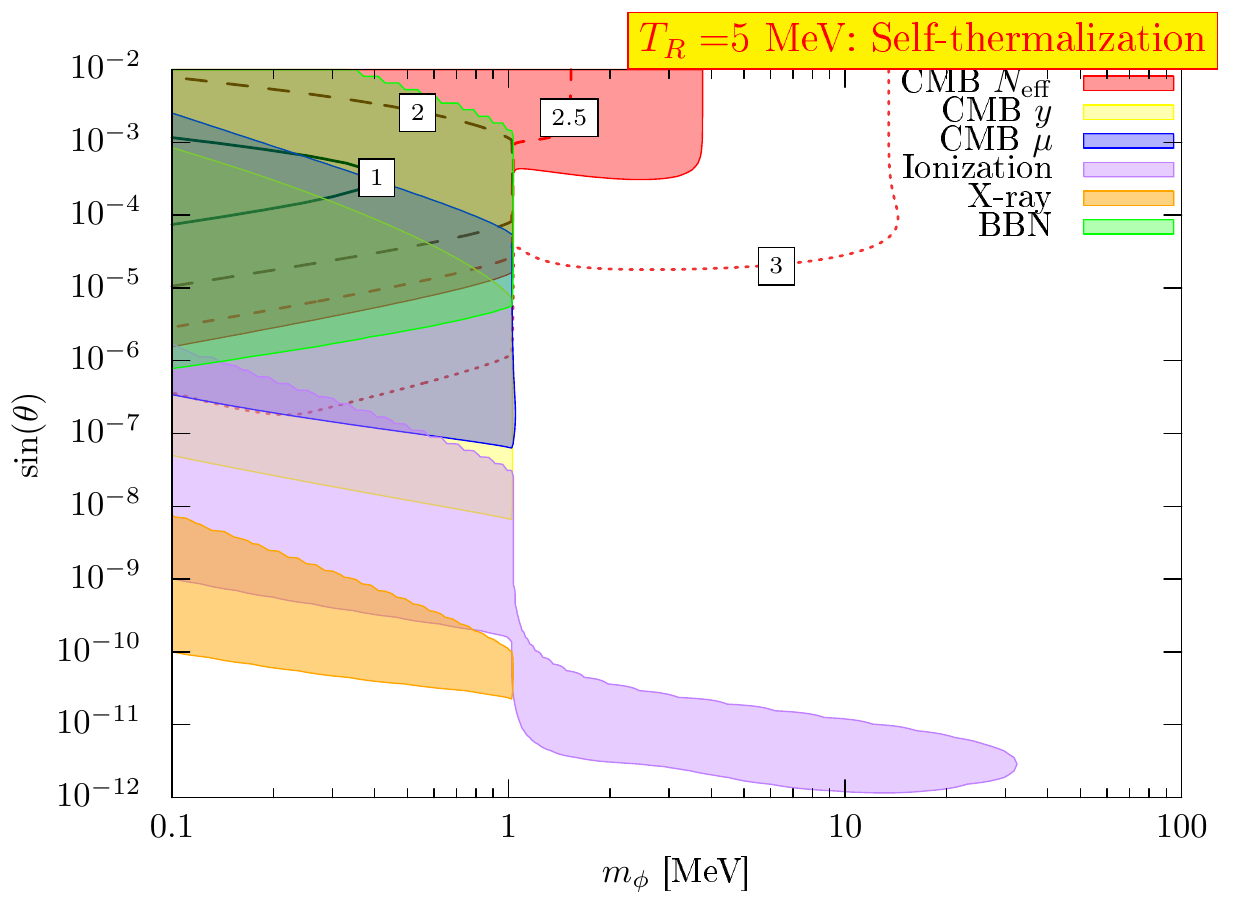}}

    \vspace{-.2cm}	
	{\includegraphics[width=0.45\textwidth]{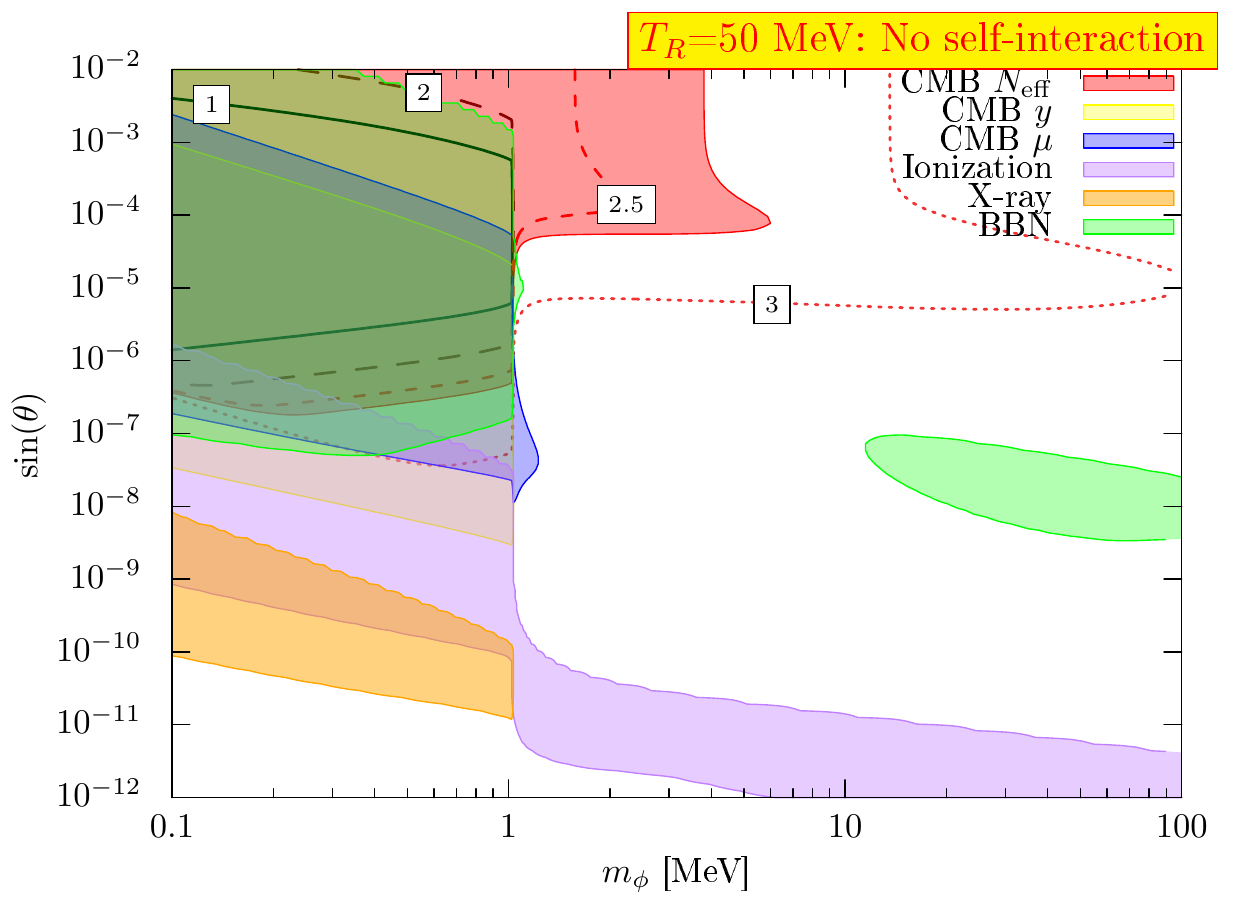}}
	{\includegraphics[width=0.45\textwidth]{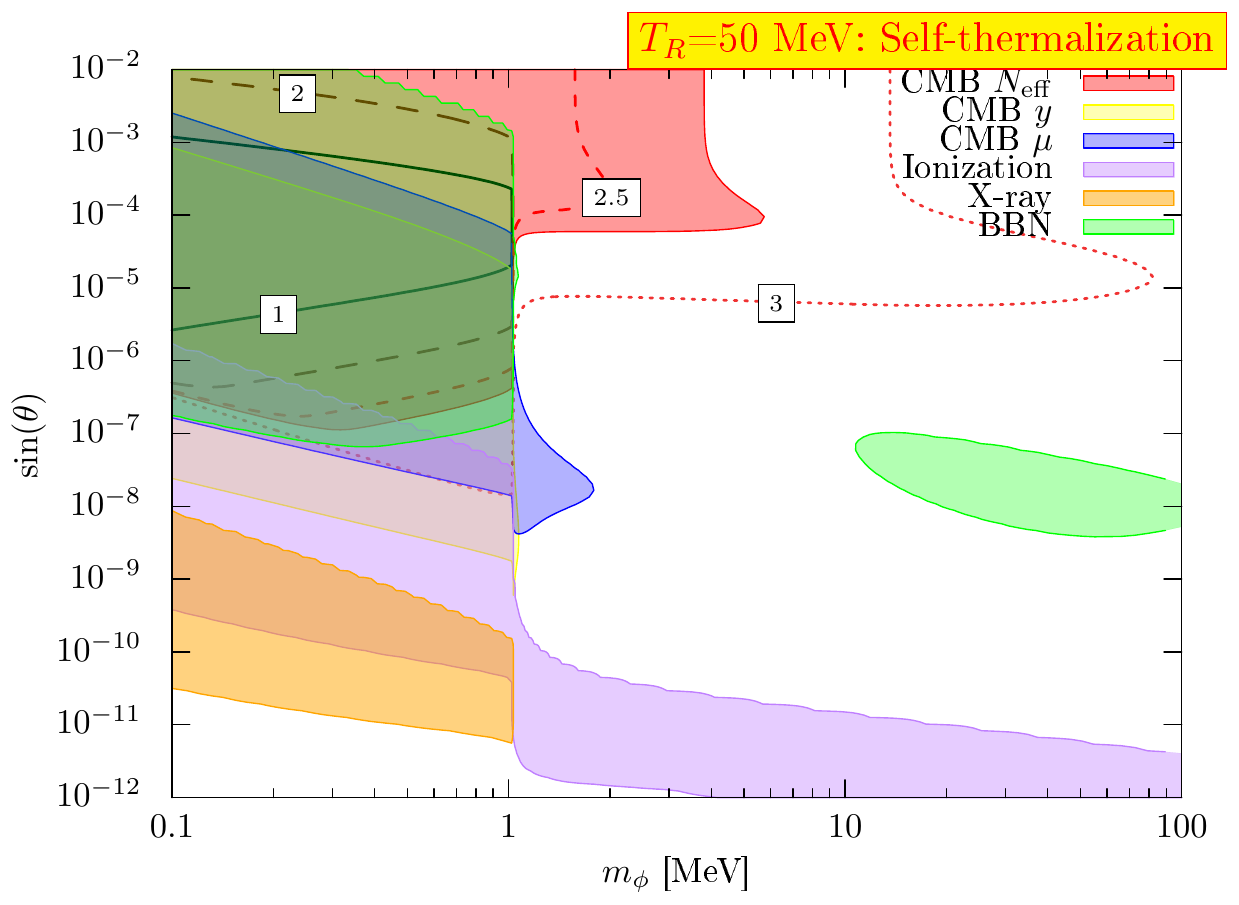}}
	
	    \vspace{-.2cm}	
	
	{\includegraphics[width=0.45\textwidth]{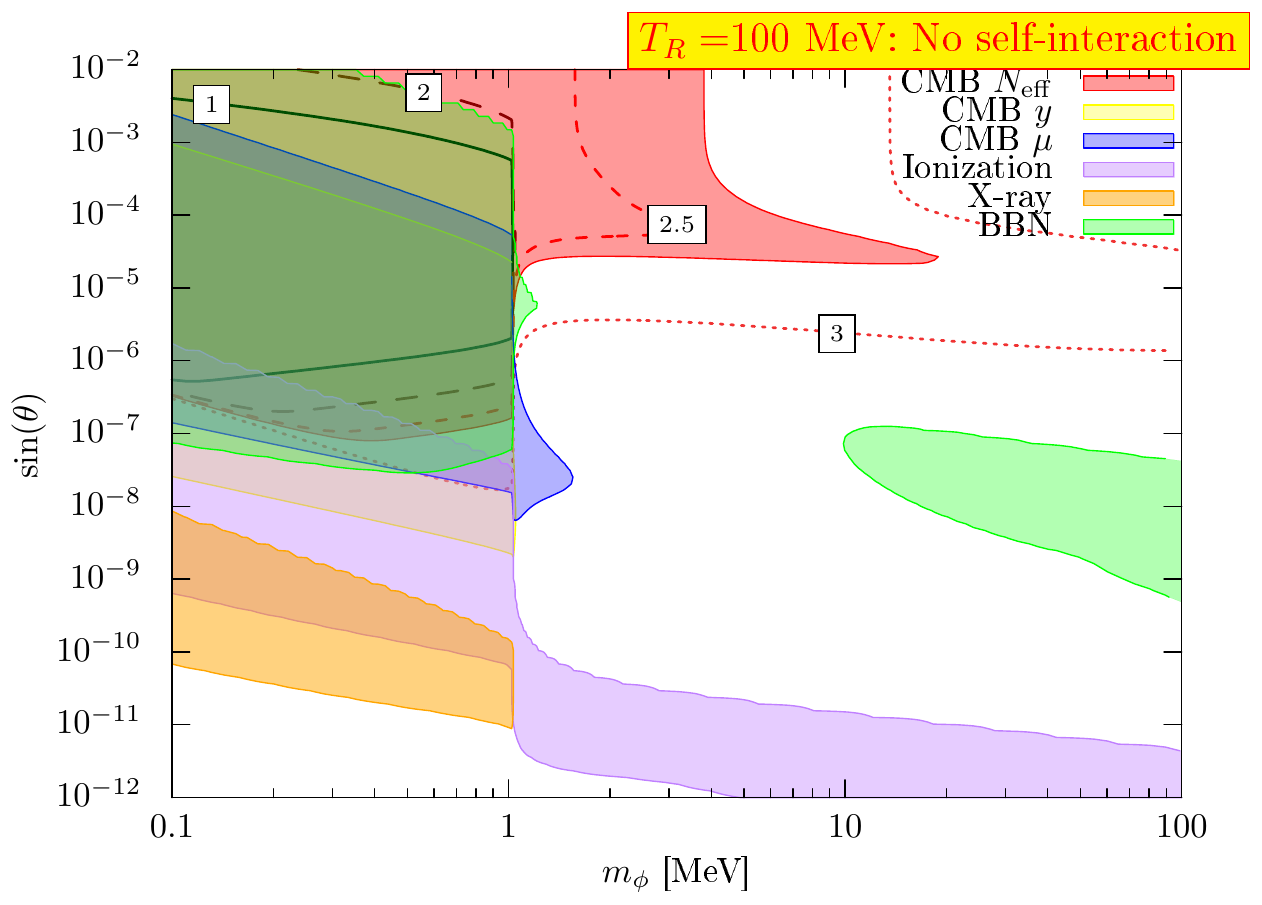}}
	{\includegraphics[width=0.45\textwidth]{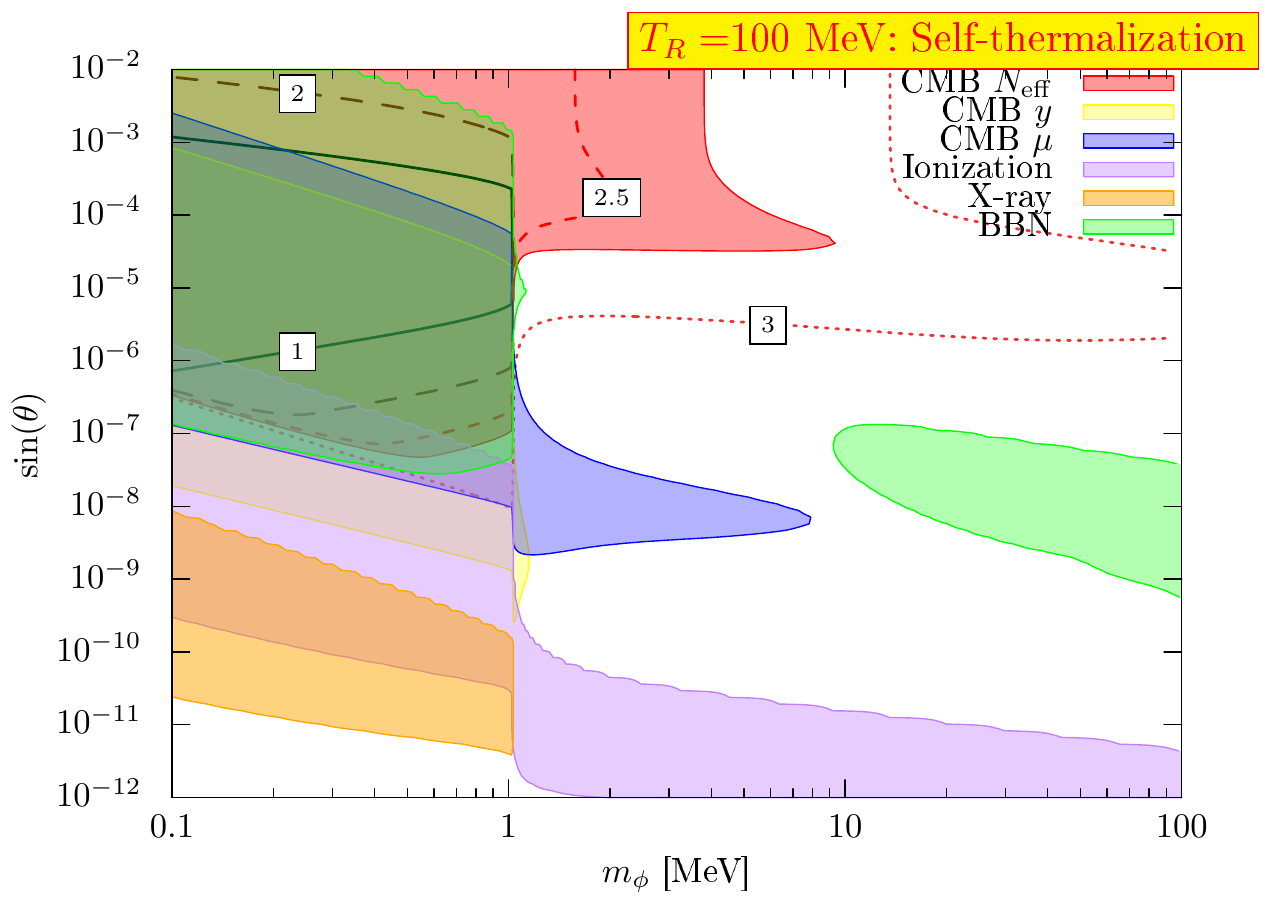}}
	
    \vspace{-.2cm}

	{\includegraphics[width=0.45\textwidth]{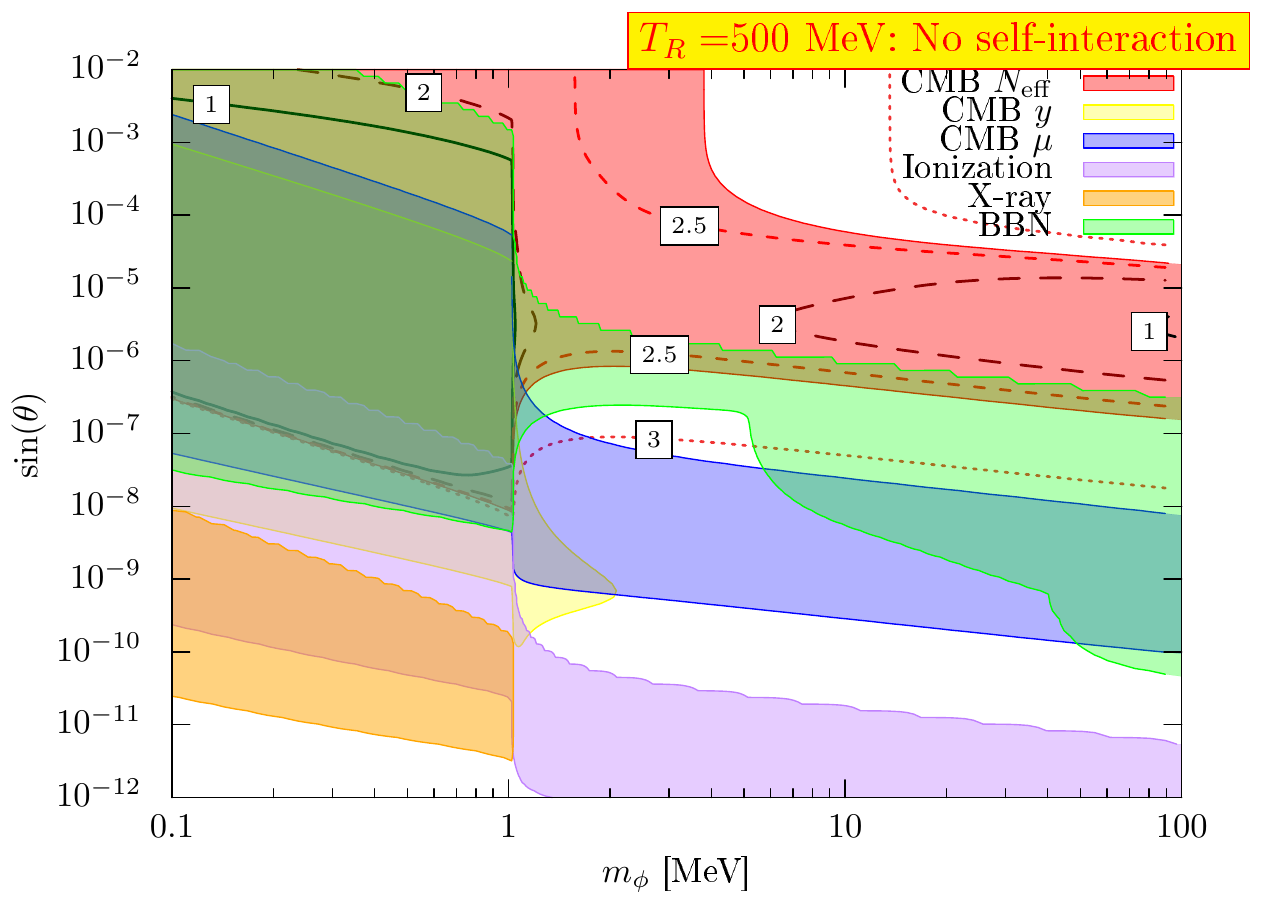}}
	{\includegraphics[width=0.45\textwidth]{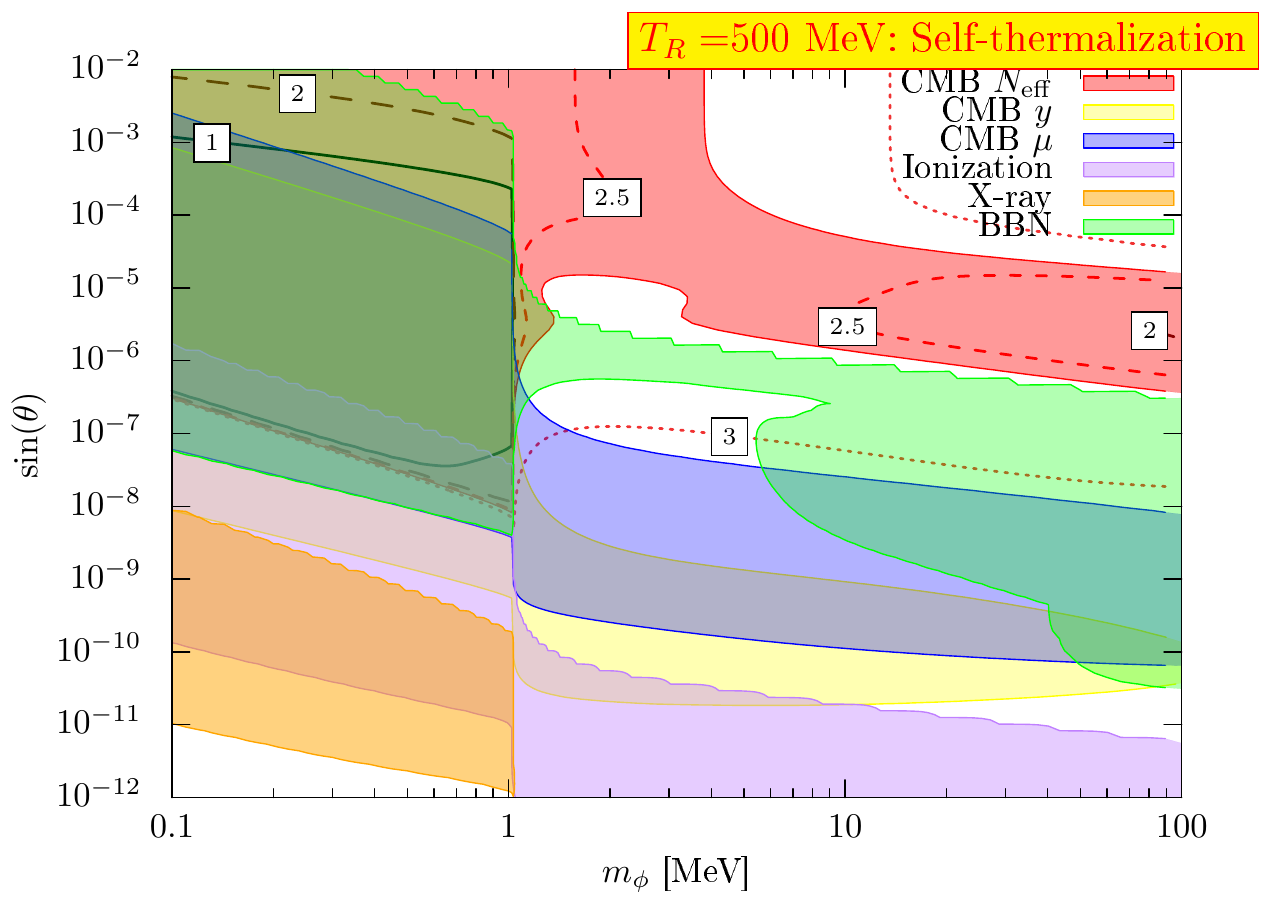}}	
	\caption{Cosmological constraints on the dark scalar for different reheating temperatures. \emph{(cont.)}}
	\label{fig:cosmological constraints}
\end{figure}
\begin{figure}[htbp] \ContinuedFloat

	\centering

	{\includegraphics[width=0.45\textwidth]{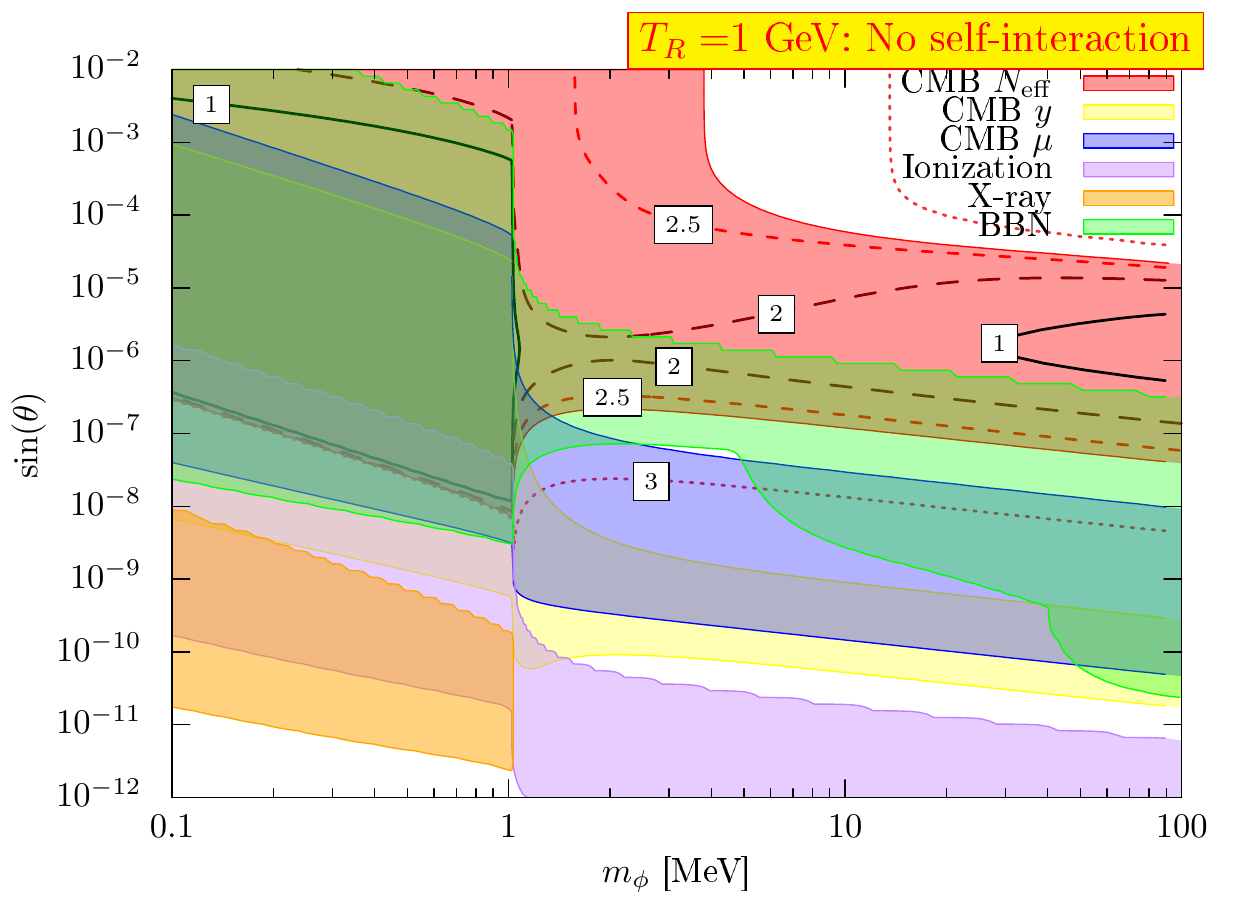}}
	{\includegraphics[width=0.45\textwidth]{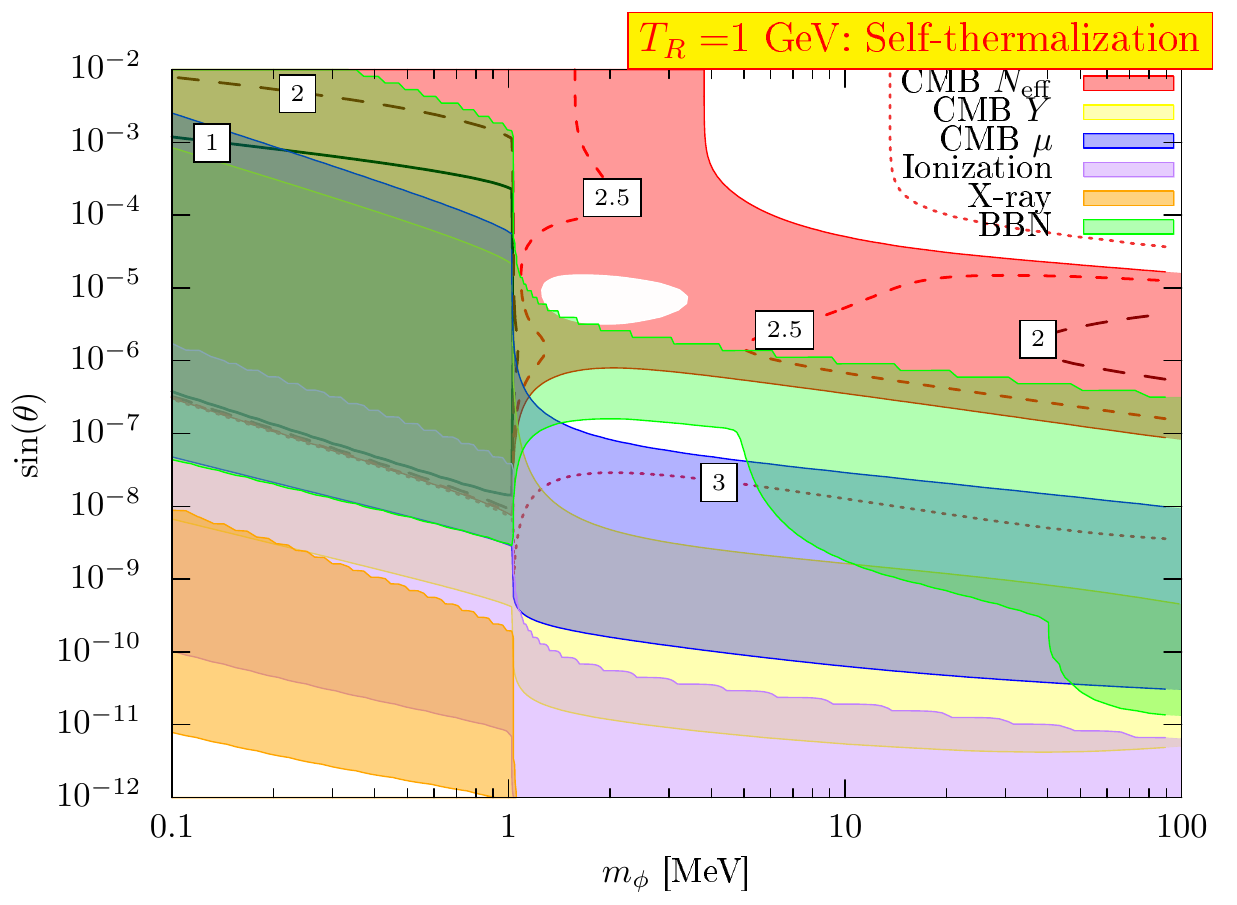}}

    \vspace{-.2cm}	
	{\includegraphics[width=0.45\textwidth]{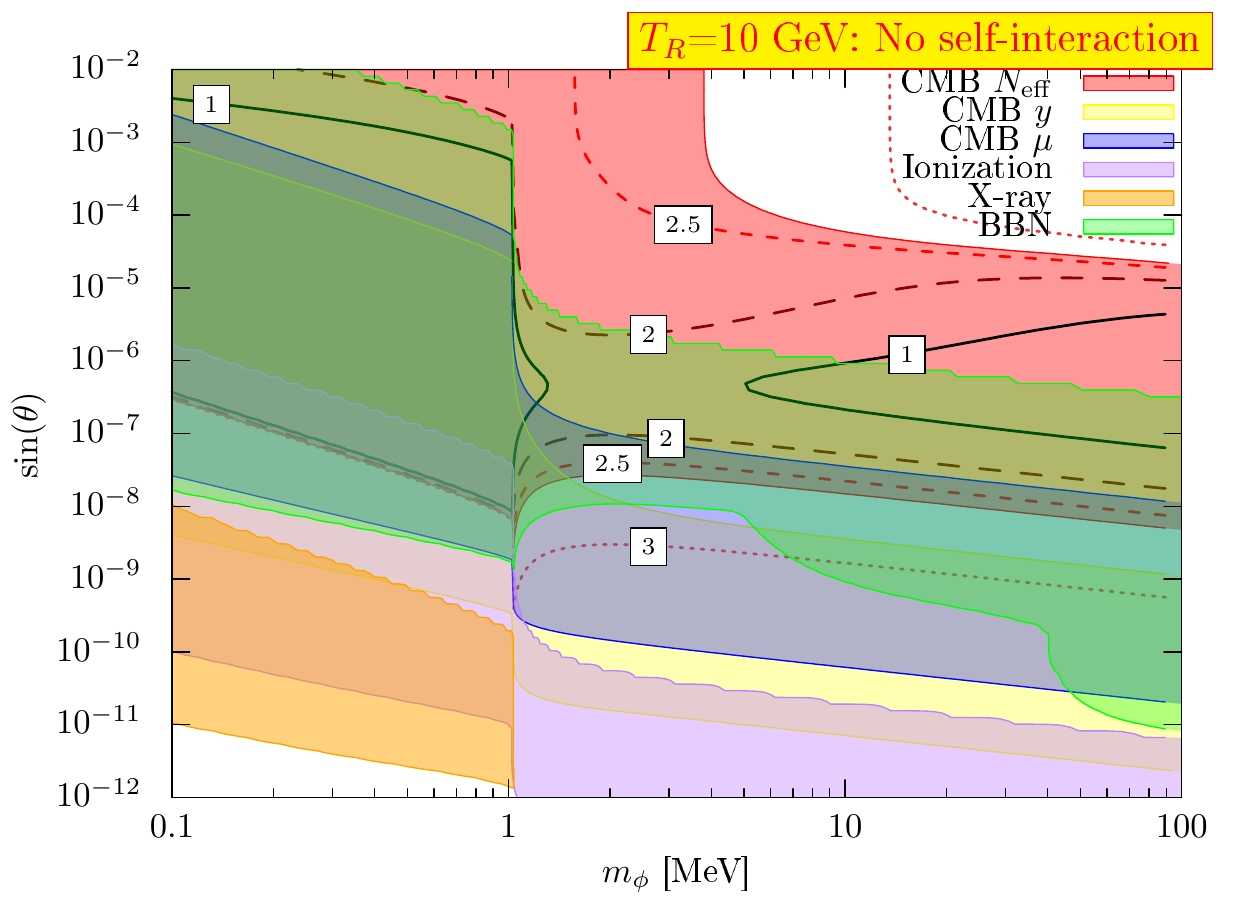}}
	{\includegraphics[width=0.45\textwidth]{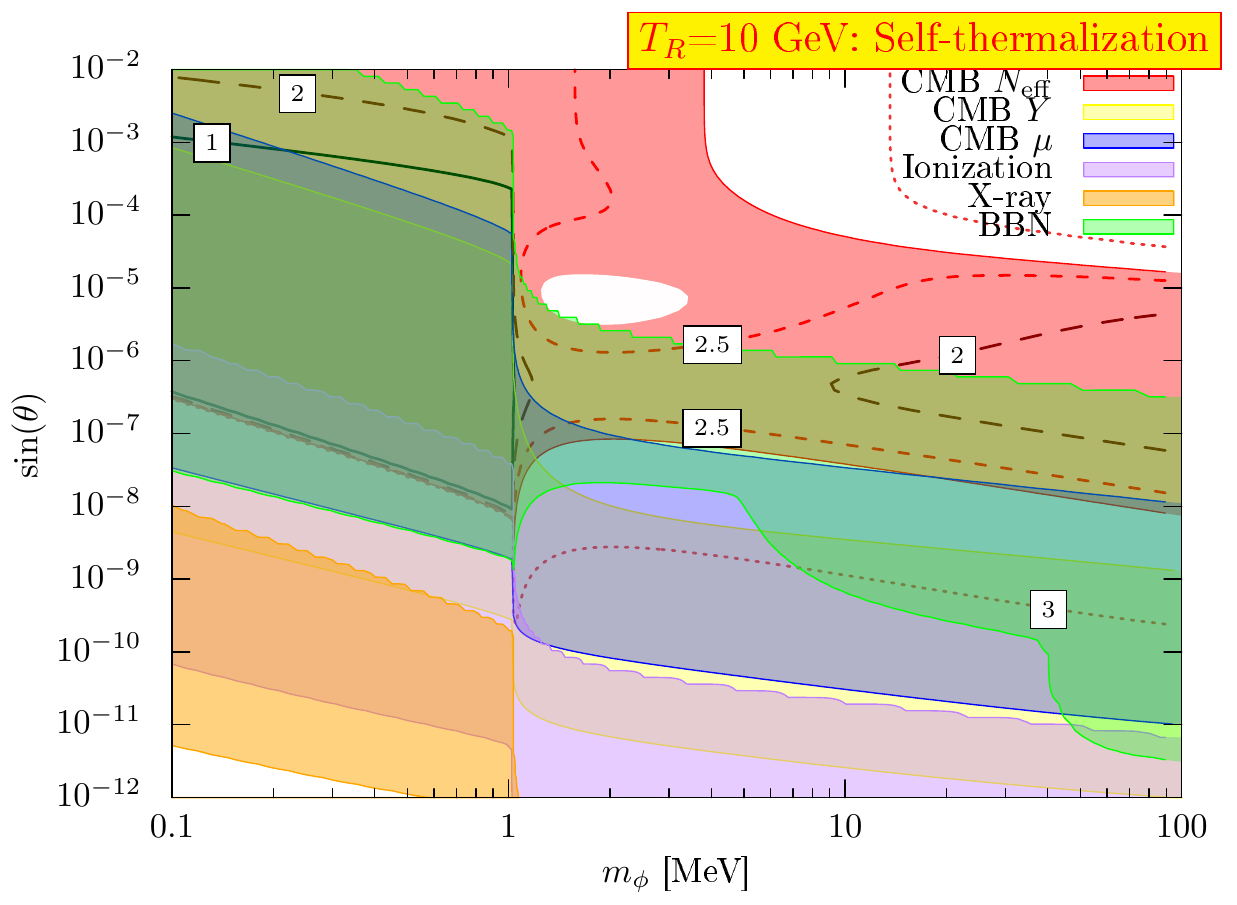}}
	
	    \vspace{-.2cm}	
	
	{\includegraphics[width=0.45\textwidth]{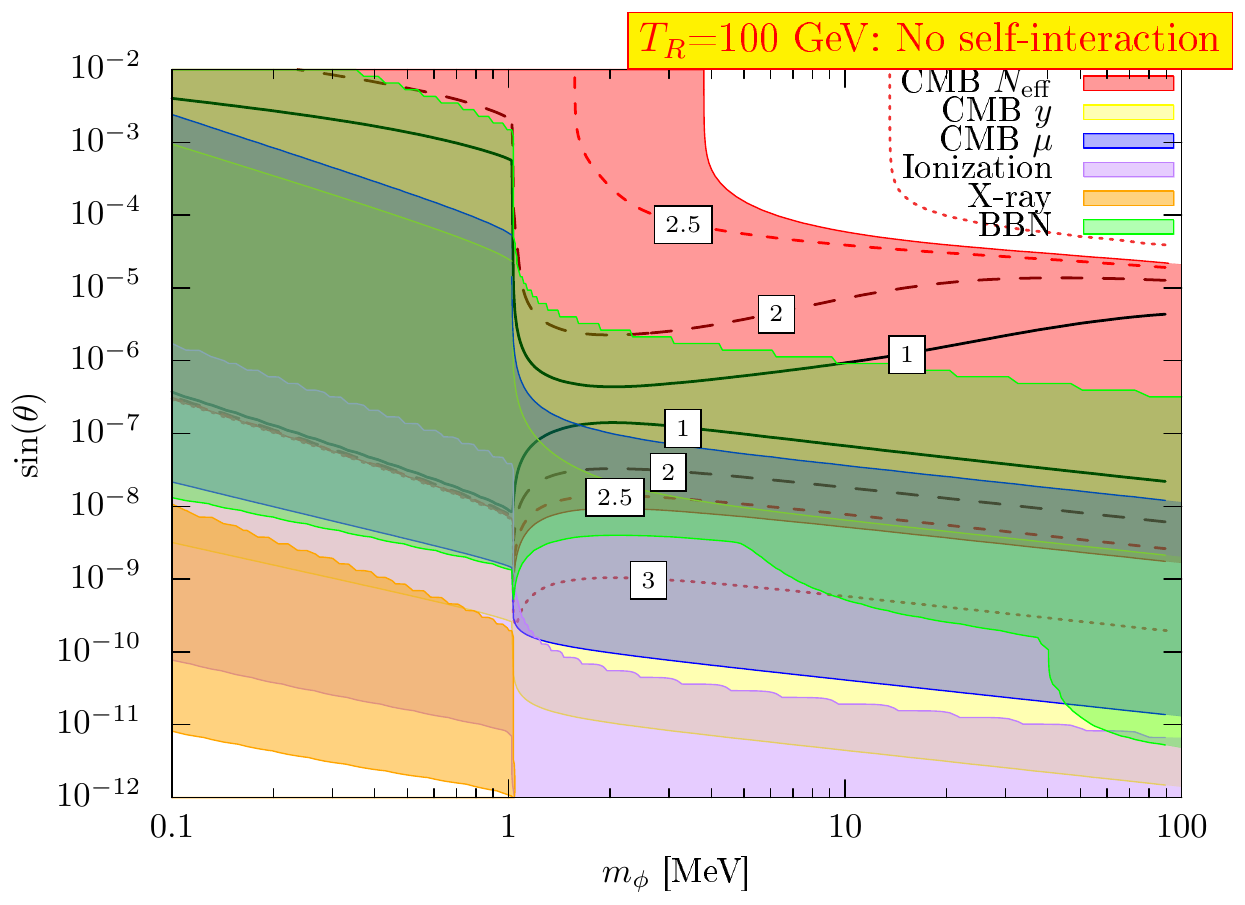}}
	{\includegraphics[width=0.45\textwidth]{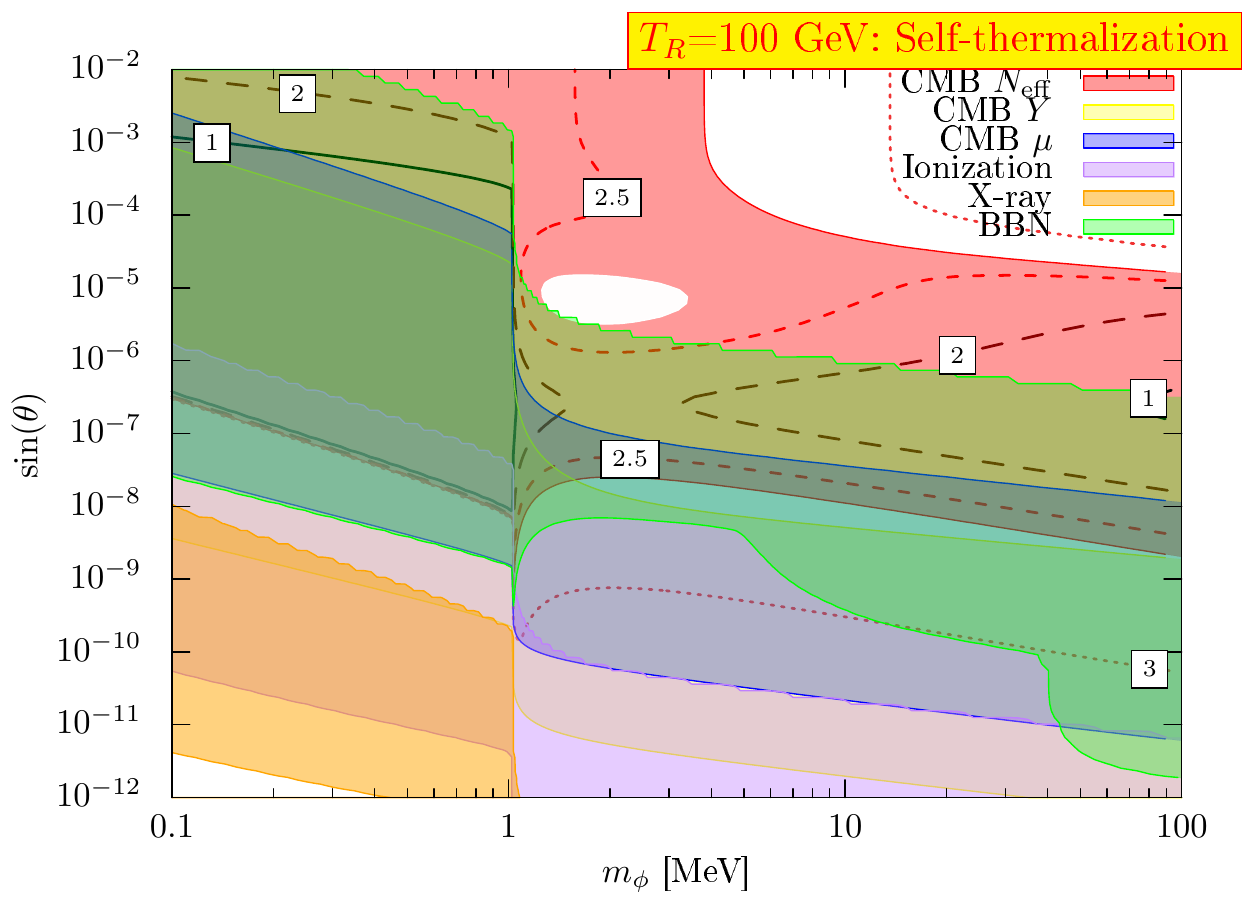}}

	\caption{Cosmological constraints on the dark scalar for different reheating temperatures.
	The red, yellow and blue regions correspond to the constraints by the CMB measurement of the $N_\mathrm{eff}$, $y$-distortion and $\mu$-distortion, respectively.
    The light purple region is excluded by ionization after the recombination era.
    The orange region is excluded by the X-ray constraints on the decaying dark matter.
    We also show the BBN constraints as the green shaded region.}
			\end{figure}

\begin{figure}[htbp]
	\centering
	\subcaptionbox{\label{fig:current}Current cosmological constraints.}{\includegraphics[width=0.47\textwidth]{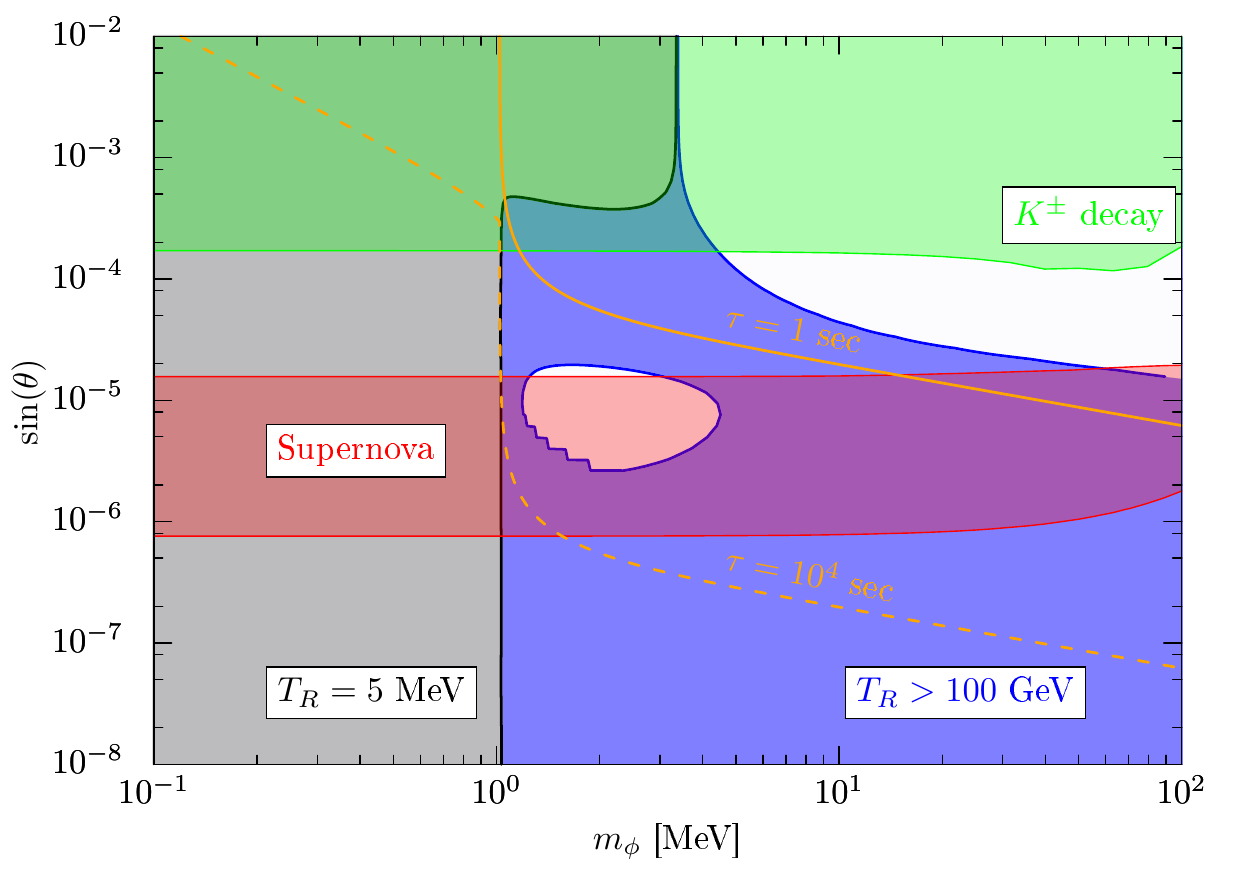}}
	\qquad
	\subcaptionbox{\label{fig:future}Future cosmological prospects.}{\includegraphics[width=0.47\textwidth]{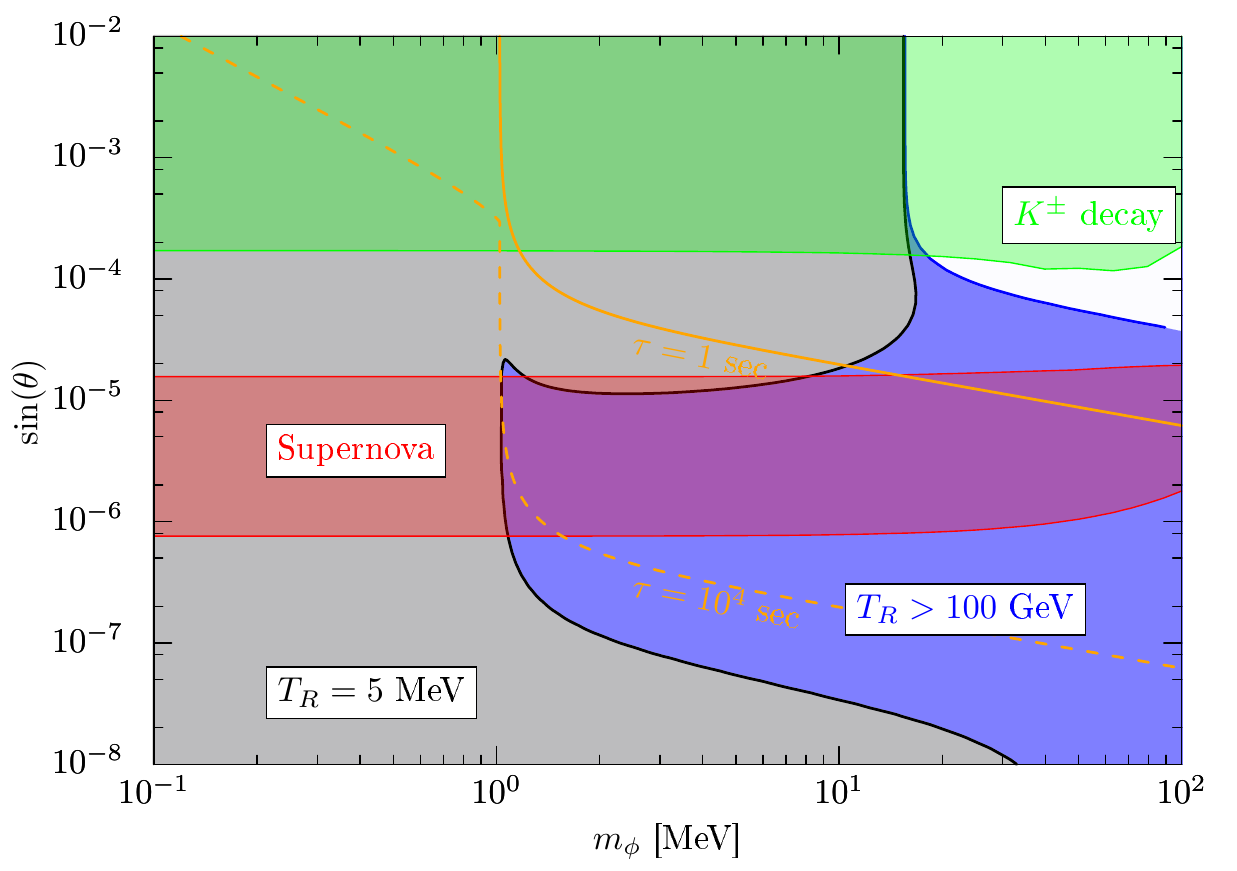}}
\caption{
The cosmological constraints on the dark scalar model.
The gray region shows the most conservative constraint, which is excluded even for an extremely low reheating temperature. 
The blue region shows the excluded region for the reheating temperature greater than $100$\,GeV.
The white region is cosmologically consistent region.
}
\label{fig:constraints}
\end{figure}

\section{Conclusion and Discussion}
\label{sec:Discussiions}
In this paper, we have discussed the cosmological constraints on the dark scalar which mixes with the Higgs boson.
We have paid particular attention to the 
cases with a low reheating temperature 
to derive the conservative constraints.
We also take account of the self-interaction of the dark scalar, which affects the time evolution of the dark scalar energy density.  

In Fig.\,\ref{fig:constraints}, we show the summary of the cosmological constraints and future prospects of the dark scalar parameter with the accelerator \cite{BNL-E949:2009dza,NA62:2021zjw} and astrophysical constraints \cite{Krnjaic:2015mbs}.
Here, we take $T_R = 100$\,GeV,
although the result does not change as long as the reheating temperature is higher than $\sim 10$\,GeV.
Such a high reheating temperature is the most common assumption in the applications of the dark scalar as a mediator between the 
dark sector and the SM sector.
In both panels, we show the weaker constraint among the non-interacting and the self-thermalized cases.

The blue region shows the excluded region for the reheating temperature greater than $\sim 10$\,GeV.
The gray region shows the most conservative constraint, where the reheating temperature is extremely small, i.e., $5$\,MeV, which is the almost lowest temperature for the BBN to work well.
The white region is cosmologically safe region regardless of the reheating temperature.
For a rather large mixing angle, $s_\theta \gtrsim 10^{-(3\mbox{-}4)}$, the  dark scalar is thermalized
when $T = \mathcal{O}(1)$\,MeV. In that region the cosmological constraints are insensitive to $s_\theta$ which put the lower limit on the dark scalar mass, $m_\phi \simeq 3.8$\,MeV for the current limit and $m_\phi\simeq 12$\,MeV can be tested by the future CMB stage-IV~\cite{CMB-S4:2016ple}.

For the current constraint, we find the blind spot of the cosmological constraint around $m_\phi\simeq 2$\,MeV and $s_\theta \simeq 10^{-5}$.
However, this parameter region just barely avoids the cosmological constraint, and is not completely safe.
This blind spot region may change drastically if we incorporate the BBN limit, which is not considered here, or if we change the statistical treatment of the CMB limit.
The analysis of these details is beyond the scope of this paper.

Let us discuss possible uncertainties and effects which are not included in this analysis.
First of all, in this analysis we only consider the interactions of the dark scalar described by the mixing angle, $\theta$, but in fact there are direct couplings with the Higgs boson.
Also, for the interaction with the QCD sector, there is indefiniteness in the treatment of hadrons and partons.
We also neglected the multiple $\phi$ production processes through the dark scalar self-interaction.
However, in most of the region we are considering, i.e., $s_\theta \gtrsim 10^{-7}$, the dark scalars will be thermalized, and these uncertainties will not loosen the constraints.

In this work, we focused on the dark scalar constraint from cosmology.
As a consequence, we obtain the conservative and robust constraints on the parameter region that have not been probed by either accelerator experiments or astrophysical observations.

\section*{Acknowledgments}
We thank M.~Kawasaki and  H.~Nakatsuka for helpful comments on the cosmological constraints of long-lived particles.
This work is supported by Grant-in-Aid for Scientific Research from the Ministry of Education, Culture, Sports, Science, and Technology (MEXT), Japan, 17H02878 (M.I. and S.S.), 18H05542 (M.I.), 18K13535, 19H04609, 20H01895, 20H05860 and 21H00067 (S.S.), and by World Premier International Research Center Initiative (WPI), MEXT, Japan. 
This work is also supported by the Advanced Leading Graduate Course for Photon Science (S.K.), the JSPS Research Fellowships for Young Scientists (S.K. and Y.N.) and International Graduate Program for Excellence in Earth-Space Science (Y.N.).

\appendix
\section{Reduction of Collision Integrals} \label{sec: reducecolint}
In this work we approximate collision integrals for the dark scalar $\phi$ in the similar way
as Ref.~\cite{Ibe:2019gpv}. Here let us summarize how to reduce the collision terms.
In the following, we use the symbol $\sum$ to indicate the summation over the degrees of freedom of the initial and final states (other than $\phi$).
\subsection{Decay and Inverse Decay}
The collision integral corresponding to the decay and the inverse decay $\phi\leftrightarrow 1+2$ is given by
\begin{align}
    \mathcal{C}[f_\phi] 
    =\frac{1}{2E_\phi}\int 
        &\frac{d^3p_2}{(2\pi)^32E_2}
        \frac{d^3p_{1}}{(2\pi)^32E_{1}}(2\pi)^4 \delta^4(p_\phi-p_{1}-p_{2}) \notag \\
   &\times \sum|\mathcal{M}_{\phi\leftrightarrow 1+2}|^2 
    (f_\phi(1\pm f_1)(1\pm f_2)- f_1 f_2(1+f_\phi))\ . \label{eq: coldecayexact}
\end{align}
Here, the plus (minus) sign corresponds to the case that particle 1 and 2 are bosons (fermions).
For our purpose, it is sufficient to consider the case in which the particle 1 and 2 have 
the same statistics.  In such a case, Eq.~\eqref{eq: coldecayexact} is reduced to
\begin{align}
    \mathcal{C}[f_\phi] 
    = \frac{1}{2E_\phi}\sum\abs{\mathcal{M}}^2[f_\phi F_1(E_\phi)-F_2(E_\phi)], 
\end{align}
where we defined $F_1, F_2$ as
\begin{align}
    F_1(E_\phi) &= \int\frac{d^3p_1}{(2\pi)^32E_1}\frac{d^3p_2}{(2\pi)^32E_2}
    (2\pi)^4\delta^4(p_\phi-p_1-p_2)(1\pm (f_1 + f_2))\ , \\
    F_2(E_\phi) &= \int \frac{d^3p_1}{(2\pi)^32E_1}\frac{d^3p_2}{(2\pi)^32E_2}
    (2\pi)^4\delta^4(p_\phi-p_1-p_2)f_1f_2\ .
\end{align}
Assuming that $f_{1, 2}$ is the Bose-Einstein or Fermi-Dirac distribution of the temperature $T$, 
we obtain the analytical expressions of these integrals\footnote{Although these integrals 
are no longer Lorentz invariant because of $f_{1,2}$ and contain the integration over an angular coordinate, we can perform it immediately by virtue of the delta function.}, 
\begin{align}
    F_1(E_\phi) 
    =\frac{p_1^0}{4\pi m_\phi}(1+\varphi_\mp(p_\phi, T))\ , \hspace{5pt}
    F_2(E_\phi) &= F_1(E_\phi)f_\phi^{\rm BE}(T)\ .
\end{align}
Here, $p^0_1$ is the final state momentum in the rest frame of $\phi$, 
\begin{align}
    p^0_1 = \frac{m_\phi}{2}\sqrt{\qty[1-\qty(\frac{m_1+m_2}{m_\phi})^2]\qty[1-\qty(\frac{m_1-m_2}{m_\phi})^2]}\ ,
\end{align}
and $\varphi_\mp(p_\phi, T)$ is given by
\begin{align}
    \label{eq: statistical factor}
    \varphi_\mp(p_\phi, T) 
    = \frac{m_\phi T}{p_\phi p^0_1}
    \log\left( 
    \frac{e^{E_\phi E^0_1/(T m_\phi)} \mp e^{-p_\phi p^0_1/(T m_\phi) }}{
    e^{E_\phi E^0_1/(T m_\phi)} \mp e^{ p_\phi p^0_1/(T m_\phi) }}
    \right)\ .
\end{align}
Note that  $\varphi_-$ ($\varphi_+$) represents the Bose-enhancement (Pauli-blocking) 
effect when the particle 1 and 2 are bosons (fermions).

Gathering the above results, we get the following formula
\begin{align}
    \mathcal{C}[f_\phi] 
    = \frac{1}{2E_\phi}\sum\abs{\mathcal{M}}^2
    \frac{p_1^0}{4\pi m_\phi}(1+\varphi_\mp(p_\phi, T))[f_\phi-f_\phi^{\rm BE}(T)]\ .
    \label{eq: coldecayformula}
\end{align}

\subsection{Scatterings}
The collision integral of the scattering process $1+2\leftrightarrow 3+\phi$ is given by
\begin{align}
    \mathcal{C}[f_\phi] 
    = \frac{1}{2E_\phi}\int 
        &\frac{d^3p_3}{(2\pi)^32E_3}
        \frac{d^3p_2}{(2\pi)^32E_2}
        \frac{d^3p_{1}}{(2\pi)^32E_{1}}(2\pi)^4 \delta^4(p_{3}+p_\phi-p_{1}-p_{2}) \notag \\
   &\times \sum|\mathcal{M}_{1+2\leftrightarrow 3+\phi}|^2 
    (f_3f_\phi(1\pm f_1)(1\pm f_2)- f_1 f_2(1\pm f_3)(1+f_\phi))\ . \label{eq: colscatexact}
\end{align}
We assume that the particle 1, 2, and 3 are in thermal equilibrium with each other and 1 and 2 obey 
the Maxwell-Boltzmann distribution. 
In this approximation, we observe that the distribution
functions in Eq.\,\eqref{eq: colscatexact} can be simplified as
\begin{align}
    &(f_3f_\phi(1\pm f_1)(1\pm f_2)- f_1 f_2(1\pm f_3)(1+f_\phi)) \notag \\
    &\to 
   (f^{\rm eq}_3f_\phi- f^{\rm MB}_1 f^{\rm MB}_2(1\pm f^{\rm eq}_3)(1+f_\phi)) \notag \\
   &= \frac{f^{\rm MB}_1 f^{\rm MB}_2(1\pm f^{\rm eq}_3)}{f^{\rm BE}_\phi}
      (f_\phi-f^{\rm BE}_\phi)\ .
   \label{eq: MBapprox}
\end{align}
To get the third line, we used the energy conservation imposed by the delta function in Eq.\,\eqref{eq: colscatexact}. Therefore, we can write the collision integral in the reduced form, 
\begin{align}
    \mathcal{C}[f_\phi] 
    = I_{1+2\leftrightarrow 3+\phi}
        \times
        \left(f_\phi(E_\phi)- \frac{1}{e^{E_\phi/T}-1}\right)\ ,
\end{align}
where
\begin{align}
 I_{1+2\leftrightarrow 3+\phi}  
 =  \frac{1}{2E_\phi f^{\rm BE}_\phi(E_\phi)}
 \int &
    \frac{d^3p_3}{(2\pi)^32E_3}
    \frac{d^3p_2}{(2\pi)^32E_2}
    \frac{d^3p_{1}}{(2\pi)^32E_{1}}(2\pi)^4 \delta^4(p_{3}+p_\phi-p_{1}-p_{2}) \cr
    &\times\sum|\mathcal{M}_{1+2\leftrightarrow 3+\phi}|^2
    f^{\rm MB}_1 f^{\rm MB}_2(1\pm f^{\rm eq}_3)\ .
\end{align}
This integral can be further reduced to the two-dimensional one, 
\begin{align}
 I_{1+2\leftrightarrow 3+\phi}  
    = \frac{1}{512\pi^3} \frac{T e^{-E_\phi/T}}{|\ps_\phi|E_\phi f_\phi(E_\phi)}
    \int & ds  \frac{1}{\sqrt{s} |\ps^{\mathrm{cms}}_{3\phi}|}
 \log\left[\frac{1\mp e^{-E_3^+/T}}{1\mp e^{-E_3^-/T}}\right]
    \int dt \sum|\mathcal{M}_{1+2\leftrightarrow 3+\phi}|^2\ .
    \label{eq: colintformula}
\end{align}
Here, $s$ and $t$ are the Mandelstam variables defined as
\begin{align}
    s = (p_1 + p_2)^2, \hspace{5pt} t = (p_1-p_\phi)^2, 
\end{align}
and $\ps^{\mathrm{cms}}_{3\phi}$ is the momentum
at the center of the mass frame, 
\begin{align}
|\ps^{\mathrm{cms}}_{3\phi}| = \frac{\sqrt{(s-(m_3+m_\phi)^2)(s-(m_3-m_\phi)^2)}}{2\sqrt{s}} \ . 
\end{align}
The kinematical range of the energy of the particle 3 is given by
\begin{align}
   & E_3^\pm  =\frac{{E}_\phi \left(s-m_3^2-m_\phi^2\right)\pm|\ps_\phi|\sqrt{\left(s-(m_3 + m_\phi)^2\right)\left(s-(m_3 - m_\phi)^2\right)}}{2m_\phi^2}\ .
\end{align}

\section{Explicit Form of Collision Terms} \label{sec: explicit colint}
Here, we present the reduced form of collision integrals for each process relevant for the production or decay of the dark scalar $\phi$ using the results of the appendix \ref{sec: reducecolint}.
\subsection{Decay into Electrons and Photon}
The dark scalar decays into the electron-positron pair via the Yukawa couplings in Eq.\,\eqref{eq:phi-fermion yukawa}. 
The corresponding matrix element is given by
\begin{align}
		\sum\abs{\mathcal{M}_{\phi\rightarrow e^- e^+}}^2
		&= 2s_\theta^2 y_e^2 (m_\phi^2-4m_e^2) 
		= \frac{16\pi m_\phi\Gamma_{\phi\rightarrow e^- e^+}}{\sqrt{1-(2m_e/m_\phi)^2}} \ .
\end{align}
Substituting this into Eq.\,\eqref{eq: coldecayformula}, we obtain the collision term for this process
\begin{align}
    \mathcal{C}[f_\phi]
	= -\frac{m_\phi\Gamma_{\phi\rightarrow e^- e^+}(1+\varphi_+(p_\phi, T))}{E_\phi}
	  (f_\phi-f_\phi^{\mathrm{eq}}(T))\ , 
\end{align}
where $\varphi_+(p_\phi, T)$ is the Fermi-Dirac version of Eq.\,\eqref{eq: statistical factor}.

Similarly, we can obtain the collision term for the decay into two photons 
$\phi\rightarrow\gamma\gamma$, 
\begin{align}
    \mathcal{C}[f_\phi]
	= -\frac{m_\phi\Gamma_{\phi\rightarrow \gamma\gamma}(1+\varphi_-(p_\phi, T))}{E_\phi}
	  (f_\phi-f_\phi^{\mathrm{eq}}(T))\ , 
\end{align}
where $\Gamma_{\phi\rightarrow \gamma\gamma}$ is given by Eq.\,\eqref{eq: two photon decay}, and
$\varphi_-(p_\phi, T)$ is the Bose-Einstein version of Eq.~\eqref{eq: statistical factor}.

\subsection{Production from Fermions}
The dark scalars are produced through the annihilation $f \bar{f}\rightarrow \phi V$  and the Compton-like 
scattering $fV\rightarrow f\phi$, where $V$ represents the photon or the gluon. 
The matrix elements squared for these processes are given by
\begin{align}
		\sum\abs{\mathcal{M}_{f \bar{f}\rightarrow \phi V}}^2/(16\pi\xi s_\theta^2 y_f^2)
		= &\frac{m_\phi^2t-st-t^2+m_f^2m_\phi^2-3s m_f^2-2t m_f^2+11m_f^4}{(m_f^2+m_\phi^2-s-t)^2} \notag \\
		&+ \frac{2(m_\phi^2 s +m_\phi^2 t -st -t^2-5m_\phi^2m_f^2-s m_f^2+2t m_f^2+7m_f^4)}{(t-m_f^2)(m_f^2+m_\phi^2-s-t)} \notag \\
		&+ \frac{m_\phi^2t-st-t^2-3m_\phi^2m_f^2+s m_f^2+6t m_f^2+3m_f^4}{(t-m_f^2)^2}\ , 
\end{align}

\begin{align}
		\sum\abs{\mathcal{M}_{fV\rightarrow f\phi}}^2/(16\pi\xi s_\theta^2 y_f^2)
		= &\frac{-st+m_f^2(2m_\phi^2-3s+t)-5m_f^4}{(s-m_f^2)^2} \notag \\
		  &+ 2\frac{-(s-m_f^2)(t-m_f^2)-m_f^2(-4m_\phi^2+s+t)-5m_f^4}{(s-m_f^2)(t-m_f^2)} \notag \\
		  &+ \frac{-st+m_f^2(2m_\phi^2+s-3t)-5m_f^4}{(t-m_f^2)^2}\ .
\end{align}
Here we define $\xi$ as 
\begin{align}
\xi = 
\begin{cases}
Q^2\times\displaystyle{\frac{e^2}{4\pi}} &\mathrm{for}~~ V=\gamma, \\
\tr(T^a T^a)\times
\displaystyle{\frac{g_3(\mu_\mathrm{RG} = m_f)^2}{4\pi}}  &\mathrm{for}~~V=g,
\end{cases}
\end{align}
where $e$ and $g_3$ are QED and QCD gauge couplings, respectively, and $Q$ and $T^a$ are the corresponding charges.
Applying the formula in Eq.\,\eqref{eq: colintformula}, we obtain the collision terms for these processes:
\begin{align}
         I_{f\bar{f} \leftrightarrow \phi V}(p_\phi, T_{\gamma})
         = &\frac{16\pi\xi s_\theta^2 y_f^2}{512\pi^3}
            \frac{T_{\gamma}e^{-E_\phi/T_{\gamma}}}{p_\phi E_\phi f^{\rm BE}_\phi(E_\phi, T_{\gamma})} \notag \\
           &\times\int_{\max\{m_\phi^2, 4m_f^2\}}^{\infty}ds \, 
		   2\log\qty(\frac{1-e^{-E_{V}^+/T_{\gamma}}}{1-e^{-E_{V}^-/T_{\gamma}}}) 
		   \notag \\
           &\times \left[4\sqrt{1-\frac{4m_f^2}{s}}(4m_f^2-m_\phi^2)
		   \qty{\frac{1}{s-m_\phi^2}+\frac{m_\phi^2}{(s-m_\phi^2)^2}} 
   			\right. \notag \\
           &\left.\hspace{20pt} + \log\qty(\frac{s - 2m_f^2 + \sqrt{s(s-4m_f^2)}}{2m_f^2}) \right. \notag \\
		   &\left.\hspace{30pt} \times 2\qty{1+\frac{2(m_\phi^2-4m_f^2)}{s-m_\phi^2}+\frac{2(m_\phi^2-4m_f^2)(m_\phi^2-2m_f^2)}{(s-m_\phi^2)^2}}\right]\ ,
\end{align}
\begin{align}
\label{eq: egamtoSe}
         I_{fV \leftrightarrow \phi f}(p_\phi, T_{\gamma})
         = &\frac{16\pi\xi s_\theta^2 y_f^2}{512\pi^3}
		 \frac{T_{\gamma}e^{-E_\phi/T_{\gamma}}}{p_\phi E_\phi f^{\rm BE}_\phi(E_\phi, T_{\gamma})} \notag \\
          &\times\int_{(m_f+m_\phi)^2}^{\infty} ds\, 
		  2\log\qty(\frac{1+e^{-E_f^-/T_{\gamma}}}{1+e^{-E_f^+/T_{\gamma}}}) \notag \\
		  &\times \left[-\frac{3}{2}+\frac{4m_f^2-m_\phi^2}{2s}
		  + \frac{m_f^2(m_\phi^2-m_f^2)}{2s^2} 
  		  - \frac{4(4m_f^2-m_\phi^2)}{s-m_f^2}\right. \notag \\
          &\left. \hspace{20pt} 
		  + \qty((s - (m_f + m_\phi)^2)(s - (m_f - m_\phi)^2))^{-1/2} \right. \notag \\
          &\left. \hspace{30pt} 
		  \times \log\qty(\frac{m_f^2-m_\phi^2+s+\sqrt{(s-(m_f+m_\phi)^2)(s-(m_f-m_\phi)^2)}}{2m_f\sqrt{s}}) 
  			\right. \notag \\
          &\left. \hspace{30pt} \times 
		  2\qty(s-2m_\phi^2+7m_f^2 + \frac{2(m_\phi^2-4m_f^2)(m_\phi^2-2m_f^2)}{s-m_f^2}) \right]\ .
\end{align}
\subsection{Production from Mesons}
For the energy scale below $4\pi f_\pi$, the production from the hadronic sector can be 
described by the effective Lagrangian~\eqref{eq:scalar-meson interaction}. 
As in the production from fermions, the relevant processes are the meson annihilation
$\pi^+\pi^-\rightarrow \phi\gamma$ and the Compton-like scattering $\pi\gamma\rightarrow\pi\phi$, 
whose matrix elements are given by
\begin{align}
	&\sum\abs{\mathcal{M}_{\pi^+\pi^- \rightarrow \gamma \phi}}^2
	/\qty(4\pi\alpha_\mathrm{QED} s_\theta^2\mu_\pi^2)
	= \frac{-st(s+t-m_\phi^2)+m_\pi^2(2st-(m_\pi^2-m_\phi^2)s - m_\phi^4)}
	{(t-m_\pi^2)^2(s+t-m_\pi^2-m_\phi^2)^2}\ , \\
	&\sum\abs{\mathcal{M}_{\pi\gamma \rightarrow \pi \phi}}^2
	/(4\pi\alpha_\mathrm{QED} s_\theta^2\mu_\pi^2)\notag \\
	&\hspace{40pt}= 
	\frac{-st(s+t-m_\phi^2)+m_\pi^2(2st-m_\phi^2(s+t)+m_\pi^2(s+t+m_\phi^2)-2m_\pi^4)}
			{(st-m_\pi^2(s+t)+m_\pi^4)^2}\ ,
\end{align}
where $\mu_\pi = 2\qty[(1+\kappa)m_\pi^2+\kappa m_\phi^2]/v$. 
With the formula \eqref{eq: colintformula}, the corresponding collision terms can be written as
\begin{align}
         I_{\pi^+\pi^- \leftrightarrow \gamma \phi}(p_\phi, T_{\gamma})
         = &\frac{4\pi\alpha_\mathrm{QED} s_\theta^2\mu_\pi^2}{512\pi^3}
            \frac{T_{\gamma}e^{-E_\phi/T_{\gamma}}}{p_\phi E_\phi f^{\rm eq}_\phi(E_\phi, T_{\gamma})} \notag \\
           &\times\int_{4m_\pi^2}^{\infty} ds\, 
		   2\log\qty(\frac{1-e^{-E_{\gamma}^+/T_{\gamma}}}{1-e^{-E_{\gamma}^-/T_{\gamma}}}) 
		   \notag \\
		   &\times \left[\frac{s-2m_\pi^2}{(s-m_\phi^2)^2} 
		   				\log\qty(\frac{s-2m_\pi^2+\sqrt{s(s-4m_\pi^2)}}{2m_\pi^2})
		   			- \frac{\sqrt{s(s-4m_\pi^2)}}{(s-m_\phi^2)^2}\right]\ ,
\end{align}
\begin{align}
         I_{\pi\gamma \rightarrow \pi \phi}(p_\phi, T_{\gamma})
         = &\frac{4\pi\alpha_\mathrm{QED} s_\theta^2\mu_\pi^2}{512\pi^3}
            \frac{T_{\gamma}e^{-E_\phi/T_{\gamma}}}{p_\phi E_\phi f^{\rm eq}_\phi(E_\phi, T_{\gamma})} \notag \\
		   &\times\int_{(m_\pi + m_\phi)^2}^{\infty} ds\, 
		   2\log\qty(\frac{1-e^{-E_\pi^+/T_{\gamma}}}{1-e^{-E_\pi^-/T_{\gamma}}}) 
		   \notag \\
		   &\times \left[\qty(1+\frac{2m_\pi^2-m_\phi^2}{s-m_\pi^2})
		   ((s-(m_\pi - m_\phi)^2)(s-(m_\pi+m_\phi)^2))^{-1/2} \right. \notag \\
		   &\left. \hspace{20pt}\times
				   \log\qty(\frac{s+m_\pi^2-m_\phi^2 + \sqrt{(s-(m_\pi - m_\phi)^2)(s-(m_\pi+m_\phi)^2)}}
								{2m_\pi\sqrt{s}}) \right. \notag \\
		   &\left. \hspace{40pt}- \frac{1}{s-m_\pi^2}\right]\ .
\end{align}
Note that these formulae apply to the kaon thanks to the SU(3) isospin symmetry.

\bibliographystyle{apsrev4-1}
\bibliography{ref}

\end{document}